\documentclass[11pt,letterpaper]{article}
\usepackage[margin=1in]{geometry}
\usepackage[T1]{fontenc}
\usepackage[utf8]{inputenc}
\usepackage{amsmath,amsthm,mathtools}  
\usepackage{newtxtext}                 
\usepackage[varvw]{newtxmath}          
\DeclareFontFamily{OMS}{cmcal}{\skewchar\font48}
\DeclareFontShape{OMS}{cmcal}{m}{n}{<-6>cmsy5<6-8>cmsy7<8-10>cmsy8<10->cmsy10}{}
\DeclareFontShape{OMS}{cmcal}{b}{n}{<-6>cmbsy5<6-8>cmbsy7<8-10>cmbsy8<10->cmbsy10}{}
\DeclareMathAlphabet{\mathcal}{OMS}{cmcal}{m}{n}
\SetMathAlphabet{\mathcal}{bold}{OMS}{cmcal}{b}{n}
\usepackage{booktabs}
\usepackage{array}
\usepackage{graphicx}
\usepackage{xcolor}
\usepackage{enumitem}
\usepackage{multicol}
\usepackage[titles]{tocloft}  
\usepackage{microtype}
\usepackage{algorithm}        
\usepackage{algpseudocode}
\makeatletter\def\theHALG@line{\theHalgorithm.\arabic{ALG@line}}\makeatother 
\usepackage{capt-of}          
\usepackage{hyperref}
\hypersetup{
  colorlinks=true,
  linkcolor=blue!60!black,
  citecolor=green!50!black,
  urlcolor=blue!70!black
}
\usepackage[capitalize,nameinlink]{cleveref}
\theoremstyle{plain}
\newtheorem{theorem}{Theorem}[section]
\newtheorem{proposition}[theorem]{Proposition}
\newtheorem{lemma}[theorem]{Lemma}
\newtheorem{corollary}[theorem]{Corollary}
\theoremstyle{definition}

\newtheorem{remark}[theorem]{Remark}

\newcommand{\Description}[1]{}

\newcommand{\vx}{\mathbf{x}}  
\newcommand{\Sbb}{\mathbb{S}^2}
\newcommand{\Rbb}{\mathbb{R}}
\newcommand{\Zbb}{\mathbb{Z}}
\newcommand{\EBHS}{E_{\mathrm{BHS}}}
\newcommand{\CBHS}{C_{\mathrm{BHS}}}
\newcommand{\Ezonal}{E_{\mathrm{zonal}}}
\newcommand{\Emult}{E_{\mathrm{mult}}}
\newcommand{\TZ}[1]{\texttt{Terraced}\-\texttt{Zonal}$_{#1}$}  
\newcommand{\TZs}[1]{\ensuremath{\texttt{TZ}_{#1}}}  
\newcommand{\intmark}{\raisebox{0.78ex}{\scalebox{0.62}{$\int$}}}  

\title{A Linear-Time Construction on $\mathbb{S}^2$ Achieving a Sublinear Energy Gap on Smale's Seventh Problem}

\author{
Rishabh Kabra$^{1,2}$ \quad
Wim Dehaen$^{3}$ \quad
Andrea Karlova$^{2}$ \quad \\
Mark Hamilton$^{1,4}$ \quad
Joao Carreira$^{1}$ \quad
Maks Ovsjanikov$^{1}$ \quad
Jonathan T. Barron$^{1}$ \\
\normalsize $^{1}$Google DeepMind \quad $^{2}$University College London \\
\normalsize $^{3}$University of Chemistry and Technology Prague \quad $^{4}$Massachusetts Institute of Technology
}
\date{}

\begin{document}

\pagenumbering{roman}
\maketitle
\thispagestyle{empty}
\vspace{-2.1em}

\begin{abstract}
Smale's seventh problem asks for a polynomial-time algorithm that places $N$ points on the unit sphere $\Sbb$ with \emph{logarithmic energy} (the sum of $-\log$ of all pairwise distances) within $\mathcal{O}(\log N)$ of the minimum. The minimal energy has an asymptotic expansion with explicit terms of order $N^2$ and $N\log N$ and a linear term whose constant is unknown. Brauchart, Hardin, and Saff conjectured this constant; the resulting expansion, $\EBHS(N)$, is known to be an upper bound on the minimum, up to $o(N)$. All previously analyzed constructions on $\Sbb$ have energy exceeding $\EBHS(N)$ by an amount that grows at least linearly in $N$, i.e. $\Omega(N)$, far from the $\mathcal{O}(\log N)$ accuracy sought by Smale.

We present \texttt{Terraced}\-\texttt{Zonal}, a deterministic construction that generates $N$ points in $\Theta(N)$ time and memory. Points are placed on latitude rings whose populations are held constant across bands (``terraces'') chosen by a one-dimensional dynamic program, so that adjacent rings within each terrace lock into a staggered triangular lattice. Under a locality assumption, we prove that the energy exceeds $\EBHS(N)$ by only $\Theta(N^{2/3})$, with an explicit constant; exact evaluation up to $N = 10^9$ confirms this rate. Under the same assumption, its distance from the true minimum is sublinear if the conjecture holds, and in any case at most about $6.24 \times 10^{-4}$ per point for large $N$, compared with at least $1.9 \times 10^{-3}$ per point for the previously analyzed constructions. To our knowledge, this is the first algorithmic construction with a sublinear excess over $\EBHS(N)$ of explicit order (proven under the locality assumption), a necessary step toward Smale's seventh problem. With one parameter changed, the same algorithm also gives an equal-weight integration rule on $\Sbb$ that runs just as fast and whose error decays at an empirical rate of $N^{-4}$ on smooth functions, versus $N^{-3/2}$ for every previous fast construction we tested.
\end{abstract}

\vfill
\begingroup
\small
\setlength{\cftbeforesecskip}{1.2pt plus 3pt minus 0.4pt}
\setlength{\columnsep}{2em}
\setlength{\multicolsep}{2pt}
\noindent{\normalsize\bfseries }\par\vspace{2pt}\nopagebreak
\makeatletter
\begin{multicols}{2}
\@starttoc{toc}
\end{multicols}
\makeatother
\endgroup

\newpage
\pagenumbering{arabic}

\section{Introduction}
\label{sec:intro}

Place $N$ points $X_N = \{\vx_1,\dots,\vx_N\}$ on the unit sphere $\Sbb \subset \Rbb^3$ so as to minimize their logarithmic energy
\begin{equation}
  \label{eq:log-energy}
  E(X_N) \;=\; \sum_{1 \le i < j \le N} \log \frac{1}{\|\vx_i - \vx_j\|}.
\end{equation}
Smale's seventh problem~\cite{Smale1998} asks for an algorithm that, in time polynomial in $N$, outputs $X_N$ with
\[
  E(X_N) - E_{\min}(N) \;\le\; c \log N, \qquad E_{\min}(N) \;=\; \min_{|Y|=N} E(Y).
\]
Shub and Smale showed that such points give well-conditioned polynomials~\cite{ShubSmale1993}, natural starting points for homotopy methods that solve polynomial equations~\cite{BCSS1998,BurgisserCucker2013} (Appendix~\ref{app:bg-polynomials}).

The first two terms of the minimal energy are classical, $E_{\min}(N) = \frac{1}{2}(\frac{1}{2} - \log 2) N^2 - \frac{1}{4} N \log N + \mathcal{O}(N)$~\cite{RSZ1994}. Brauchart, Hardin, and Saff~\cite{BHS2012} conjectured that the linear term is the one obtained when the points locally form a perfect triangular lattice, which gives the explicit benchmark
\begin{equation}
  \label{eq:bhs-expansion}
  \EBHS(N) \;=\; \frac{1}{2}\left(\frac{1}{2} - \log 2\right) N^2 \;-\; \frac{1}{4} N \log N \;+\; \frac{1}{2}\CBHS\, N,
\end{equation}
where $\frac{1}{2}\CBHS \approx -0.0278$ is the energy per point, at this order, of points forming a unit-density triangular lattice (Lemma~\ref{lem:kronecker}; $\CBHS$ is the regularized Epstein zeta constant of that lattice, with closed form~\eqref{eq:c-bhs} in Appendix~\ref{app:bg-minimum}). B\'etermin and Sandier~\cite{BeterminSandier2018} proved that the linear term exists, and with the lower bounds of~\cite{Lauritsen2021,BeltranLizarte2023} it is known that
\[
  E_{\min}(N) \;=\; \EBHS(N) - \delta N + o(N), \qquad 0 \le \delta < 6.24 \times 10^{-4},
\]
for some constant $\delta$ (Lemma~\ref{lem:emin-lower}); the conjecture is $\delta = 0$. We call $E(X_N) - \EBHS(N)$ the \emph{excess} of $X_N$. Since $\delta \ge 0$, any solution of Smale's problem must have excess at most $o(N)$, whether or not the conjecture holds. Configurations with excess at most $o(N)$ exist~\cite[\S\S6.3--6.4, \S7.3]{BeterminSandier2018}, but the existence proof is not an algorithm and gives no rate. The linear term is also the first order at which the arrangement of the points, and not merely their density, matters (Appendix~\ref{app:bg-riesz}): conjecturally, reaching $\EBHS(N)$ at this order requires the points to form a triangular lattice almost everywhere. Every explicit construction analyzed so far misses it, with excess $\Omega(N)$.

\begin{figure}[!t]
  \centering
  \includegraphics[width=0.9\textwidth]{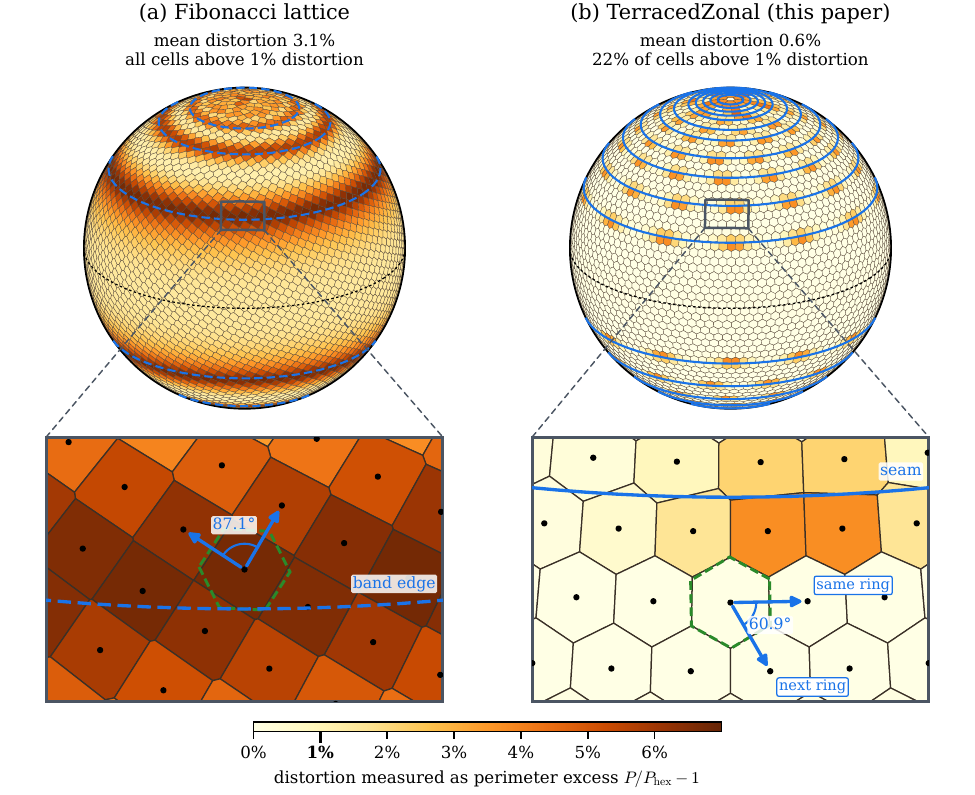}
  \caption{\textbf{Smale’s Seventh Problem} asks for an algorithm to place points on the unit
sphere $\Sbb$ to minimize the repulsive logarithmic energy $\log(1/\|\vx_i - \vx_j\|)$ over all pairs of points
$(\vx_i, \vx_j)$. Previous solutions like \textbf{(a)} the \textbf{Fibonacci lattice} \cite{SwinbankPurser2006} spread lattice defects over wide belts. Our method, \textbf{(b) TerracedZonal}, confines them to thin seams, achieving an overall sublinear $\Theta(N^{2/3})$ energy
excess over the benchmark $\EBHS(N)$. We visualize the $N = 5001$ case. Voronoi cells are shaded by how far each cell's perimeter exceeds that of a regular hexagon of the same area. Cells of the Fibonacci lattice shear toward squares. In each zoom, the arrows join a point to its nearest neighbor and to the nearest neighbor at an acute angle to that one; in a triangular lattice they meet at $60^\circ$, in a square lattice at $90^\circ$. In TerracedZonal, ring populations are constant on bands of rings called \emph{terraces} (Section~\ref{sec:terraces}), separated by seams (solid blue); everywhere except along the seams the points form a triangular lattice and the cells are nearly regular hexagons. Appendix~\ref{app:three-families} explains this measure. An interactive version of this figure and the Python code for \texttt{Terraced}\-\texttt{Zonal} are available at \url{https://terraced-zonal.github.io/}.}
  \label{fig:spiral-vs-terraced}
\end{figure}

\begin{figure}[!t]
  \centering
  \includegraphics[width=\textwidth]{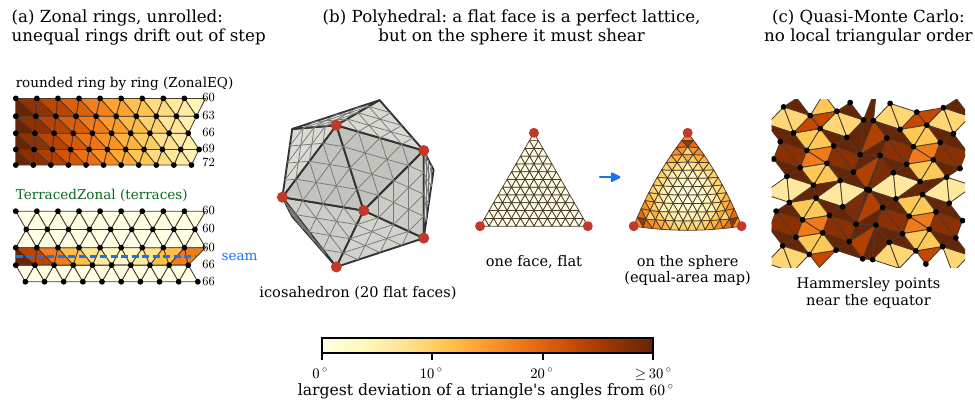}
  \caption{\textbf{Why previous constructions miss the triangular lattice, shown on the flat objects they start from.} Triangles join points whose cells share an edge: they form the Delaunay triangulation, the dual of the Voronoi tessellation of Figure~\ref{fig:spiral-vs-terraced}, and a cell is a regular hexagon exactly when its six surrounding triangles are equilateral. Each triangle is shaded by the largest deviation of its angles from $60^\circ$ (white: equilateral); Section~\ref{sec:existing} details the three families~\textbf{(a)--(c)}.}
  \Description{Three panels shaded from white (equilateral triangles) to dark brown (angles 30 degrees or more away from 60). Left: two strips of five rows of points. In the upper strip the rows have 60, 63, 66, 69, and 72 points and the triangles between them shade from dark on the left to pale on the right. In the lower strip, labeled TerracedZonal, the rows have 60, 60, 60, 66, and 66 points; all triangles are white except one orange row at the seam between 60 and 66, which is marked by a dashed blue line. Middle: a gray icosahedron (20 triangular faces) with a triangular grid on its faces and red dots at its 12 corners, then one flat face with a white triangular grid, an arrow, and the same face mapped to the sphere, with orange, distorted triangles near its corners. Right: irregular triangles between scattered points, mostly orange and dark brown.}
  \label{fig:families}
  \vspace{0.7em}
  \captionof{table}{Limiting energy excess per point, $(E - \EBHS)/N$, by family of Figure~\ref{fig:families} (full list in Table~\ref{tab:prior}, Appendix~\ref{app:tables}). $^{\ast}$Expected value of a random construction.}
  \label{tab:prior-compact}
  \vspace{2pt}
  \footnotesize
  \setlength{\tabcolsep}{4.2pt}
  \begin{tabular}{@{}lr@{\hspace{1.8em}}lr@{\hspace{1.8em}}lr@{}}
    \toprule
    \multicolumn{2}{@{}l}{\textbf{(a) Zonal / spiral}} &
    \multicolumn{2}{l}{\textbf{(b) Polyhedral}} &
    \multicolumn{2}{l@{}}{\textbf{(c) QMC / random}} \\
    \cmidrule(r{1.8em}){1-2}\cmidrule(r{1.8em}){3-4}\cmidrule{5-6}
    \texttt{Pearl}$^{\ast}$~\cite{GariboldiGigante2026} & $0.00315$ &
    \texttt{EqAreaIco}~\cite{HardinMichaelsSaff2016} & $0.00198$ &
    \texttt{Hammersley}~\cite{CuiFreeden1997} & $0.0655$ \\
    \texttt{ZonalEQ}~\cite{Leopardi2006} & $0.00326$ &
    \texttt{HEALPix}~\cite{Gorski2005} & $0.00791$ &
    \texttt{EllipticZeros}$^{\ast}$~\cite{ArmentanoBeltranShub2011} & $0.1244$ \\
    \texttt{FibLattice}~\cite{SwinbankPurser2006} & $0.00548$ &
    \texttt{Octahedral}~\cite{HolhosRosca2014} & $0.00908$ &
    \texttt{SphericalEnsemble}$^{\ast}$~\cite{Krishnapur2009,AlishahiZamani2015} & $0.2301$ \\
    \midrule
    \multicolumn{6}{@{}c@{}}{\textbf{This paper} (\texttt{Terraced}\-\texttt{Zonal}, coaxial terraces): \quad $(E - \EBHS)/N \;\approx\; 0.026\,N^{-1/3} \;\longrightarrow\; 0$} \\
    \bottomrule
  \end{tabular}
\end{figure}

\subsection{Why Existing Families Suffer an \texorpdfstring{$\Omega(N)$}{Omega(N)} Energy Excess}
\label{sec:existing}
Previously studied constructions on $\Sbb$ (surveyed in~\cite{HardinMichaelsSaff2016,BorodachovHardinSaff2019}; Appendix~\ref{app:related} gives a fuller account) fall into three families. Each has a geometric obstruction to forming a triangular lattice almost everywhere, which leaves an excess of at least a constant per point (Figures~\ref{fig:spiral-vs-terraced} and~\ref{fig:families}, Table~\ref{tab:prior-compact}).
\begin{enumerate}[label=\textbf{\arabic*.}]
  \item \textbf{Smooth zonal rings and spirals.} Equal-area ring partitions~\cite{RSZ1994,Zhou1995,Leopardi2006} and the \texttt{Diamond}~\cite{BeltranEtayo2020,BeltranEtayoLopez2023} and \texttt{Pearl}~\cite{GariboldiGigante2026} ensembles place points on latitude rings whose populations follow a smooth profile, so adjacent rings almost always differ in population and cannot interlock (Figure~\ref{fig:families}(a), top). \texttt{Pearl} has the smallest proven excess of any explicit construction, $0.00315\,N$ in expectation. Spirals such as the Fibonacci lattice~\cite{RSZ1994,Bauer2000,SwinbankPurser2006} suffer the analogous problem: their nearest-neighbor structure changes at $\Theta(\log N)$ band edges, each of which shears the lattice toward squares over a wide belt (Figure~\ref{fig:spiral-vs-terraced}(a)). \texttt{Terraced}\-\texttt{Zonal} belongs to this family but removes its main limitation (Section~\ref{sec:terraces}; Figure~\ref{fig:families}(a), bottom).
  \item \textbf{Polyhedral tessellations.} Icosahedral and other polyhedral grids~\cite{Goldberg1937,CasparKlug1962,HardinMichaelsSaff2016} put a perfect lattice on each flat face, but by Gauss's \emph{Theorema Egregium}~\cite{Gauss1828} any map of a flat face onto the sphere shears it by $\Theta(1)$ (Figure~\ref{fig:families}(b)). The equal-area icosahedral points of~\cite{HardinMichaelsSaff2016} pay $0.00198$ per point, the lowest measured excess of any previous construction.
  \item \textbf{Quasi-Monte Carlo and random points.} Low-discrepancy~\cite{Hammersley1960,CuiFreeden1997,WongLukHeng1997} and random~\cite{Muller1959,ArmentanoBeltranShub2011,AlishahiZamani2015} point sets are evenly spread on large scales but have no local triangular order (Figure~\ref{fig:families}(c)); even random points that repel each other pay $0.12$--$0.23$ per point.
\end{enumerate}

\subsection{Resolution via Terraces}
\label{sec:terraces}
\texttt{Terraced}\-\texttt{Zonal} places two points at the poles and the remaining $N-2$ points on coaxial latitude rings with integer populations $m_0,\dots,m_{K-1}$. The rings are spaced at $h_0 = \sqrt{2\pi\sqrt{3}/N}$, the row spacing of a triangular ($A_2$) lattice in which every point has its share $4\pi/N$ of the sphere's area, so about $K \approx \pi/h_0 \approx 0.952\,N^{1/2}$ rings fit between the poles (eq.~\eqref{eq:discretization}, Section~\ref{sec:stage1}). The construction rests on a single trade-off. Two adjacent rings with the \emph{same} population $m$, rotated by half a spacing ($\Delta\alpha = \frac{1}{2m}$) relative to each other, interlock into a perfect triangular lattice; rings with different populations cannot. The reason is in Lemma~\ref{lem:cyclotomic}: the part of the energy of two rings $a, b$ that depends on how their points line up is governed by $\rho_{ab} = (\zeta_a/\zeta_b)^{\operatorname{lcm}(m_a, m_b)}$, where $\zeta_a < \zeta_b$ are the stereographic radii of the two rings and $\operatorname{lcm}(m_a, m_b)$ is the least common multiple of their populations, the lowest azimuthal frequency present in both rings. Two equal rings at the triangular-lattice spacing have $\rho = e^{-\pi\sqrt{3}}$: in the conformal coordinate $\log\zeta + i\phi$ a ring of $m$ points is a row with spacing $2\pi/m$, so the triangular-lattice row spacing is $\frac{\sqrt{3}}{2}\cdot\frac{2\pi}{m} = \pi\sqrt{3}/m$, and $\operatorname{lcm}(m, m) = m$. Staggering them by half a spacing turns their phase-dependent interaction into $-m\log(1 + \rho)$, so they gain the binding energy per point
\begin{equation}
  \label{eq:c-step}
  c_{\mathrm{step}} \;:=\; \log\bigl(1 + e^{-\pi\sqrt{3}}\bigr) \;\approx\; 0.0043241.
\end{equation}
For unequal rings $\operatorname{lcm}(m_a, m_b) \ge 2\min(m_a, m_b)$, so $\rho \le e^{-2\pi\sqrt{3}} \approx 2 \times 10^{-5}$, and the binding is lost. But the ideal population $y_k \approx \frac{1}{2}\sqrt{2\pi\sqrt{3}\,N}\,\sin\theta_k$ of a ring at colatitude $\theta_k$ changes from ring to ring, so populations cannot be equal everywhere. \par
\begin{figure}[!t]
  \centering
  \includegraphics[width=\textwidth]{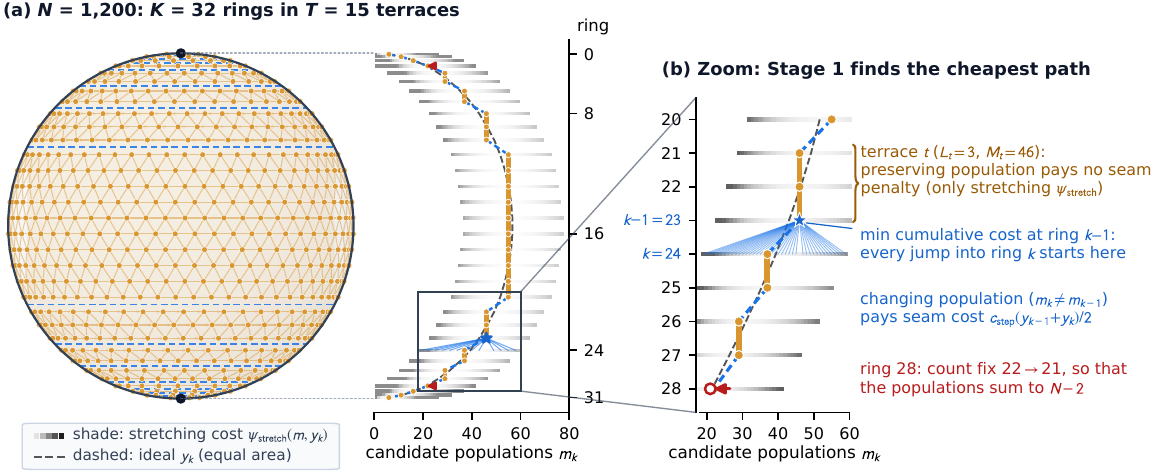}
  \caption{\textbf{\texttt{Terraced}\-\texttt{Zonal} at $N = 1{,}200$: terraces on $\Sbb$ and how Stage~1 chooses them.} \textbf{(a)}~Left: apart from the two poles, the points lie on $K = 32$ latitude rings grouped into $T = 15$ constant-population terraces; within each terrace, half-period staggered rings interlock into a triangular ($A_2$) lattice (solid bonds), broken only at the $T - 1 = 14$ seams (dashed blue). Right: Stage~1's vertical dynamic-programming trellis aligned with the sphere's heights $z_k$. Each row shows the candidate populations $m_{k,i}$ around the ideal equal-area population $y_k \propto \sin\theta_k$ (dashed), shaded by stretching cost $\psi_{\mathrm{stretch}}(m, y_k)$~\eqref{eq:stretch-cost}. \textbf{(b)}~Enlargement of rings $20$--$28$. Keeping $m_k = m_{k-1}$ pays no seam penalty, whereas changing population ($m_k \neq m_{k-1}$) pays $c_{\mathrm{step}}(y_{k-1}+y_k)/2$~\eqref{eq:seam-cost} and always starts from the minimum-cumulative-cost node of ring $k-1$ (blue star, shown for $k = 24$), giving $\mathcal{O}(W)$ work per ring~\eqref{eq:viterbi}. Red arrows ($k = 3$ in (a) and $k = 28$ in (b)): post-DP count repair adjusting $22 \to 21$ so $\sum_k m_k = N - 2$ (Appendix~\ref{app:alg-restore}).}
  \Description{Two panels. Panel a shows a sphere on the left with 32 horizontal rings of amber points grouped into 15 bands separated by dashed blue seam circles, alongside a vertical Viterbi trellis on the right aligned with the sphere's height axis, showing gray candidate nodes per ring, a dashed curve of ideal populations, and the cheapest path of amber vertical terraces and blue dashed seams. A box around rings 20 to 28 connects to Panel b, which enlarges those rows, brackets a three-ring terrace of 46 points per ring, marks the best node at ring 23 with a blue star and a fan of possible jumps into ring 24, and shows a red arrow at ring 28 adjusting 22 to 21 so populations sum to N minus 2.}
  \label{fig:concept}
\end{figure}

\smallskip\noindent
\begin{minipage}[c]{0.57\textwidth}
\setlength{\abovedisplayskip}{5pt}\setlength{\belowdisplayskip}{5pt}%
\texttt{Terraced}\-\texttt{Zonal} therefore holds the population constant across bands of consecutive rings (``terraces'') $\mathcal{T}_0,\dots,\mathcal{T}_{T-1}$ and lets it jump only between bands (Figure~\ref{fig:spiral-vs-terraced}(b)). The number of terraces is one more than the number of jumps (``seams''):
\begin{equation}
  \label{eq:T-def}
  T = 1 + \bigl\lvert\bigl\{1 \le k < K : m_k \neq m_{k-1}\bigr\}\bigr\rvert.
\end{equation}
Terrace $t$ consists of $L_t$ rings, each carrying $M_t$ points ($m_k = M_t$; lowercase refers to a ring, uppercase to a terrace; Figure~\ref{fig:notation}). A terrace of $L$ rings pays two costs (Figure~\ref{fig:concept}):
\end{minipage}\hfill
\begin{minipage}[c]{0.408\textwidth}
  \centering
  \setlength{\abovecaptionskip}{2pt}\setlength{\belowcaptionskip}{0pt}%
  \includegraphics[width=\linewidth]{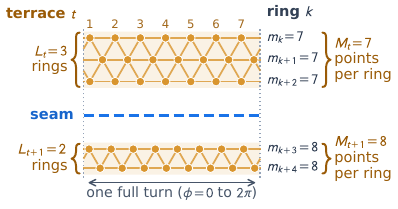}
  \captionof{figure}{Notation (two terraces, unrolled).}
  \label{fig:notation}
  \Description{Two small terraces unrolled into rows over one full turn from phi equals 0 to 2 pi. On the left, under terrace t, curly braces mark L sub t equals 3 rings above a dashed blue seam and L sub t plus 1 equals 2 rings below it. On the right, under ring k, the rows carry m sub k, m sub k plus 1, and m sub k plus 2 equal to 7 above the seam (grouped by a curly brace as M sub t equals 7 points per ring) and m sub k plus 3 and m sub k plus 4 equal to 8 below it (grouped as M sub t plus 1 equals 8).}
\end{minipage}
\par\smallskip
\begin{itemize}
  \item \textbf{Stretching inside the terrace.} Take the ring $u$ positions from the middle of the terrace. Its population differs from its ideal value $y_k = \Theta(N^{1/2})$ by about $u\,|y_{k+1} - y_k|$ points, where consecutive ideal populations differ by $\pi\sqrt{3}\,|\cos\theta_k| + \mathcal{O}(N^{-1/2}) = \Theta(1)$ away from the equator, so its local lattice is stretched by a relative strain $\Theta(u/N^{1/2})$. The elastic energy is quadratic in the strain, so this ring costs $\Theta(u^2/N^{1/2})$; averaged over the terrace, stretching costs $A\,L^2$ per ring with $A = \Theta(N^{-1/2})$ (eq.~\eqref{eq:A-theta}).
  \item \textbf{One seam per terrace.} Across a jump the two rings no longer interlock and lose their binding energy of $c_{\mathrm{step}} \approx 0.0043$ per point, i.e.\ $B = c_{\mathrm{step}}\,y_k = \Theta(N^{1/2})$ per seam. This binding energy $B$ is shared among the $L$ rings of the terrace, costing $B/L$ per ring (eq.~\eqref{eq:B-theta}).
\end{itemize}
Minimizing $A L^2 + B/L$ gives terraces of $L_* = (B/2A)^{1/3} = \Theta(N^{1/3})$ rings, hence $T \approx K/L_* = \Theta(N^{1/6})$ terraces, and an excess of $\Theta(N^{1/6})$ per ring, i.e.\ $\Theta(N^{1/2} \cdot N^{1/6}) = \Theta(N^{2/3})$ in total (Theorem~\ref{thm:energy}). Stage~1 of the algorithm finds the optimal terraces exactly with a one-dimensional dynamic program (Section~\ref{sec:stage1}).

\subsection{Main Results}
\label{sec:main-results}
Our two main contributions concern Smale's problem and are stated formally in Sections~\ref{sec:proof-complexity} and~\ref{sec:proof-energy}; the third is an empirical application.
\begin{enumerate}[label=(\roman*)]
  \item \textbf{Linear time (Proposition~\ref{thm:complexity}).} \texttt{Terraced}\-\texttt{Zonal} builds $N$ points in $\Theta(N)$ time and memory. The only search is a one-dimensional dynamic program over the $\Theta(N^{1/2})$ ring populations; phases and coordinates are then explicit.
  \item \textbf{Sublinear excess (Theorem~\ref{thm:energy}, Corollary~\ref{cor:emin}).} In conformal coordinates each terrace is a planar lattice whose energy per point is $\frac{1}{2}\CBHS$ plus a stretching cost that vanishes when the lattice is equilateral (Lemma~\ref{lem:kronecker}). Our accuracy analysis assumes locality~(A), stated precisely in Section~\ref{sec:variational}: up to lower-order terms, the energy is the sum of the per-ring stretching costs and per-seam costs that Stage~1 minimizes. Locality~(A) is plausible because rings interact only over very short ranges: inside a terrace the coupling between rings $d$ apart decays like $0.0043^d$ (Lemma~\ref{lem:cyclotomic}), so each ring effectively feels only its nearest neighbors, as in a crystal with grain boundaries. We do not prove locality~(A), but we test it against exact energies: its residual falls from $17.5\%$ of the excess at $N = 10^3$ to $1.0\%$ at $N = 10^8$, roughly halving with every decade of~$N$ (Appendix~\ref{app:locality-test}). Under locality~(A), balancing stretching against seams gives $E(X_N) - \EBHS(N) = C_1 N^{2/3} + o(N^{2/3})$ with an explicit constant $C_1 \approx 0.0262$, and exact evaluations up to $N = 10^9$ confirm the rate (Section~\ref{sec:verification}). Against the true minimum, this gives $E(X_N) - E_{\min}(N) = \delta N + o(N)$: sublinear if the Brauchart--Hardin--Saff conjecture holds, and at most $6.24 \times 10^{-4}\,N + o(N)$ in any case.
  \item \textbf{One algorithm, two problems (Section~\ref{sec:qmc}).} \texttt{Terraced}\-\texttt{Zonal} has a single parameter $Q \ge 0$, the number of height moments matched. Only the ring heights are tuned to the logarithmic energy; the terraces and the stagger depend only on the geometry. Setting $Q = 96$ adjusts the ring heights so that every polynomial in $z$ of degree at most $96$ is integrated exactly. The result, \TZ{96}, is an equal-weight integration rule whose error on smooth functions decays at an empirical rate of $N^{-4}$, versus $N^{-3/2}$ for every previous fast construction we tested, while its energy and running time stay essentially those of \TZ{0}. Our complexity and accuracy analysis concerns \TZ{0}; for \TZ{96} we report only empirical results.
\end{enumerate}
Appendix~\ref{app:glossary} collects the terms and symbols used throughout.

\section{The \texttt{Terraced}\-\texttt{Zonal} Algorithm}
\label{sec:algorithm}

Given any integer $N \ge 1$ (for $N=1,2$ the algorithm trivially returns the north pole or both poles $(0,0,\pm 1)$), \texttt{Terraced}\-\texttt{Zonal} places two polar points at $\vx_0 = (0,0,1)$ and $\vx_{N-1} = (0,0,-1)$ and constructs the remaining $N-2$ points in three deterministic stages summarized in Algorithm~\ref{alg:terraced-zonal} and detailed in Sections~\ref{sec:stage1}--\ref{sec:stage2}. The algorithm has one parameter $Q \ge 0$, the number of height moments matched in an optional final step (line~7; equivalently, the highest polynomial degree in $z$ integrated exactly when $Q \ge 1$). Its other constants are either fixed by the geometry and the energy or set with a wide safety margin; Table~\ref{tab:constants} in Appendix~\ref{app:constants} lists them all. With $Q = 0$ it is \TZ{0}, our construction for Smale's problem, which Sections~\ref{sec:proof-complexity}--\ref{sec:verification} analyze and which we also call simply \texttt{Terraced}\-\texttt{Zonal} (abbreviated $\TZs{0}$ in tables). With $Q = 96$ it is \TZ{96}, an equal-weight integration rule that we evaluate empirically in Section~\ref{sec:qmc}.

\begin{figure}[t]
  \centering
  \begin{minipage}{0.85\textwidth}
  \begin{algorithm}[H]
  \small
  \caption{\TZ{Q}: linear-time terraced zonal construction on $\Sbb$ \\(each stage is spelled out line by line in Algorithms~\ref{alg:stage1} and~\ref{alg:stage23}, Appendix~\ref{app:algorithm})}
  \label{alg:terraced-zonal}
  \raggedright
  \begin{algorithmic}[1]
    \setlength{\itemsep}{1pt}
    \Require Point count $N \ge 3$
    \Statex \hspace*{-\algorithmicindent}\phantom{\textbf{Require:} }Matched height moments $Q \ge 0$ ($Q = 0$ for Smale's problem, $Q = 96$ for integration)
    \Ensure Configuration $X_N = \{\vx_0,\dots,\vx_{N-1}\} \subset \Sbb$\par\vspace{4pt}
    \Statex \textbf{Stage 1: how many points on each ring} (Section~\ref{sec:stage1}) \hfill $\triangleright$ $\mathcal{O}(N)$ time
    \State Place $K \approx 0.952\,N^{1/2}$ rings at spacing $h_0 = \sqrt{2\pi\sqrt{3}/N}$ between two polar caps; ring $k$ ideally holds $y_k \propto$ the area of its strip \Comment{\eqref{eq:discretization}--\eqref{eq:y-ideal}}
    \State Choose integers $m_k \in \operatorname{round}(y_k) \pm w$, $w = \Theta(N^{1/3})$, minimizing $\sum_{k} \psi_{\mathrm{stretch}}(m_k, y_k) + \sum_{k} c_{\mathrm{step}}\,\tfrac{y_{k-1} + y_k}{2}\,\mathbf{1}_{\{m_k \neq m_{k-1}\}}$ (stretching inside terraces plus one seam cost per population change), exactly, by a dynamic program with $\mathcal{O}(w)$ work per ring \Comment{\eqref{eq:stretch-cost}--\eqref{eq:viterbi}$^\dagger$}
    \State \textbf{while} $\sum_k m_k \neq N - 2$: move a terrace boundary by one ring, or change a whole terrace's population by $\pm 1$, at the least stretching cost per point (if neither is possible, change one end ring) \Comment{Appendix~\ref{app:alg-restore}}\par\vspace{4.5pt}
    \Statex \textbf{Stage 2: height and rotation of each ring} (Section~\ref{sec:stage2}) \hfill $\triangleright$ $\mathcal{O}(N^{1/2}\log N)$ time
    \State $z_k \gets$ the heights minimizing the zonal energy, in closed form \Comment{\eqref{eq:z-zonal}$^\dagger$}
    \State $\alpha_0 \gets 0$; \ $\alpha_k \gets \bigl(\alpha_{k-1} + 1/(2\operatorname{lcm}(m_{k-1}, m_k))\bigr) \bmod 1$, a half-spacing stagger inside each terrace \Comment{\eqref{eq:phase-locking}}
    \State Apply $4$ fixed-point corrections to $z$ from the multipole forces of rings at most $6$ apart, keeping each only if the rings stay ordered \Comment{\eqref{eq:multipole-grad}--\eqref{eq:z-fixed-point}$^\dagger$}
    \State \textbf{if} $Q > 0$: adjust $z$ so that $P_1(z), \dots, P_{\min(K-1, Q)}(z)$ have their exact sphere averages \Comment{\eqref{eq:zonal-moment-constraint}\intmark}\par\vspace{4.5pt}
    \Statex \textbf{Stage 3: points} (Appendix~\ref{sec:stage3}) \hfill $\triangleright$ $\Theta(N)$ time
    \State Poles $(0,0,\pm 1)$; on ring $k$, $m_k$ equally spaced points at height $z_k$, rotated by $\alpha_k$ \Comment{\eqref{eq:coord-synthesis}}
    \State \Return $X_N$
  \end{algorithmic}
  \vspace{4pt}\noindent $^\dagger$\,Specific to the logarithmic energy; \intmark\,specific to integration ($Q > 0$; Section~\ref{sec:qmc}); all other steps depend only on the geometry.
\end{algorithm}
  \end{minipage}
  \Description{Numbered pseudocode in three stages. Stage 1 places the rings and their ideal populations, chooses integer populations by an exact dynamic program that charges stretching inside terraces and a seam cost for every population change, and then adjusts terraces until the populations sum to N minus 2. Stage 2 computes closed-form ring heights and phase offsets, then applies four corrections to the heights. Stage 3 places the two poles and the points of each ring.}
\end{figure}

\subsection{Stage 1: 1D Viterbi Potts DP and Marginal-Cost Refinement}
\label{sec:stage1}

\paragraph{1. Ring Discretization.}
Following Rakhmanov--Saff--Zhou~\cite{RSZ1994} and Leopardi~\cite{Leopardi2006}, but with the equilateral ($A_2$) ring spacing $h_0 = \sqrt{2\pi\sqrt{3}/N}$ instead of the square-cell spacing $\sqrt{4\pi/N}$, we place $K \approx 0.952\,N^{1/2}$ rings between two polar caps of area $4\pi/N$. The ideal real-valued population $y_k$ of ring $k$ is $N - 2$ times the relative area of its equal-colatitude strip, and $y_k^{\mathrm{base}} = \max(1, \operatorname{round}(y_k))$ (Appendix~\ref{app:alg-discretization}, eqs.~\eqref{eq:discretization}--\eqref{eq:y-ideal}).

\paragraph{2. Unbiased Third-Order Elastic Stretching Cost and Potts Seam Penalty.}
To segment the rings into piecewise-constant terraces, we formulate a 1D Potts model~\cite{Potts1952} (or $\ell_0$-jump-penalized piecewise-constant regression~\cite{WinklerLiebscher2002}) over a candidate window of half-width
\begin{equation}
  \label{eq:window}
  w = \max\!\left(20,\; \operatorname{round}\!\bigl(0.65\,N^{1/3}\bigr)\right), \qquad W = 2w + 1 = \Theta(N^{1/3}).
\end{equation}
For each ring $k \in \{0,\dots,K-1\}$ and offset index $i \in \{0,\dots,W-1\}$, the candidate integer population is $m_{k,i} = y_k^{\mathrm{base}} + (i - w)$, and each candidate is assigned the node stretching cost
\begin{equation}
  \label{eq:stretch-cost}
  \psi_{\mathrm{stretch}}(m, y) \;:=\; c_{\mathrm{stretch}}\,\frac{(m - y)^2}{y^{1/3}\,m^{2/3}} \;\;\text{for } m \ge 1, \qquad \psi_{\mathrm{stretch}}(m, y) \;:=\; +\infty \;\;\text{for } m < 1.
\end{equation}
The value $+\infty$ excludes the candidates with $m_{k,i} < 1$. These occur only on rings near the poles, where $y_k^{\mathrm{base}} \le w$; every ring keeps the candidate $m_{k,w} = y_k^{\mathrm{base}} \ge 1$, so all minima in the dynamic program below are finite. The constant $c_{\mathrm{stretch}} = \frac{1}{4}$ is the stretching stiffness of the planar lattice energy at the $A_2$ point $\tau_0 = e^{i\pi/3}$~\cite{Epstein1903,BeterminSandier2018} (derived in Lemma~\ref{lem:kronecker}); the mixed denominator $y^{1/3} m^{2/3}$ is explained by Lemma~\ref{lem:unbiased}. The transition penalty between ring $k-1$ and ring $k$ is the Potts grain-boundary seam cost~\cite{Potts1952}
\begin{equation}
  \label{eq:seam-cost}
  \psi_{\mathrm{seam}}(m_{k-1}, m_k;\,k) \;=\; c_{\mathrm{step}}\,\frac{y_{k-1} + y_k}{2}\,\mathbf{1}_{\{m_k \neq m_{k-1}\}},
\end{equation}
where $c_{\mathrm{step}}$~\eqref{eq:c-step} is the binding energy per point lost at a seam.

\paragraph{3. Exact Minimization by Dynamic Programming.}
Stage~1 chooses each $m_k$ from its window to minimize
\[
  \sum_{k=0}^{K-1}\underbrace{\psi_{\mathrm{stretch}}(m_k, y_k)}_{\text{per ring (node cost)}}
  + \sum_{k=1}^{K-1}\underbrace{c_{\mathrm{step}}\frac{y_{k-1}+y_k}{2}\,\mathbf{1}_{\{m_k \neq m_{k-1}\}}}_{\text{once per jump (edge cost)}}
  \;=\;
  \sum_{t=0}^{T-1}\underbrace{\sum_{k \in \mathcal{T}_t}\psi_{\mathrm{stretch}}(M_t, y_k)}_{\text{terrace } t \text{ strain } (\approx\, A L_t^3)}
  + \sum_{\substack{1 \le k < K \\ m_k \neq m_{k-1}}}\underbrace{c_{\mathrm{step}}\frac{y_{k-1}+y_k}{2}}_{\approx\, B \text{ per seam}},
\]
where the right-hand side groups the sum over maximal constant runs $\mathcal{T}_0,\dots,\mathcal{T}_{T-1}$ ($m_k = M_t$), recovering the collective terrace trade-off of Section~\ref{sec:terraces}. Because the objective decomposes into single-ring node costs and pairwise jump indicators (Figure~\ref{fig:concept}), it is a first-order Potts problem~\cite{Potts1952,WinklerLiebscher2002} that the Viterbi algorithm~\cite{Bellman1957,Viterbi1967} solves exactly without tracking terrace length $L_t$ in its state. Writing $D[k,i]$ for the least cost of rings $0,\dots,k$ ending in candidate $m_{k,i}$ (with $D[0,i] = \psi_{\mathrm{stretch}}(m_{0,i}, y_0)$),
\begin{equation}
  \label{eq:viterbi}
  D[k,i] = \psi_{\mathrm{stretch}}(m_{k,i}, y_k) + \min\Bigl(D[k-1,\, i+s_k],\ \min_{j} D[k-1, j] + c_{\mathrm{step}}\frac{y_{k-1}+y_k}{2}\Bigr),
\end{equation}
where $s_k = y_k^{\mathrm{base}} - y_{k-1}^{\mathrm{base}}$, so that candidate $i+s_k$ on ring $k-1$ has the same population as candidate $i$ on ring $k$ (the first term is dropped when $i+s_k$ falls outside the window). Because the seam cost does not depend on \emph{which} populations differ, the inner minimum is computed once per ring, so each ring costs $\mathcal{O}(W)$ rather than $\mathcal{O}(W^2)$; this is the Potts case of the min-convolution speed-up of Felzenszwalb and Huttenlocher~\cite[\S3 and \S3.1]{FelzenszwalbHuttenlocher2006}. Backtracking from $\arg\min_i D[K-1, i]$ yields $m_0,\dots,m_{K-1}$.

\paragraph{4. Restoring the Point Count.}
The unconstrained optimum need not satisfy $\sum_k m_k = N - 2$. A greedy loop repairs the residual $\mathrm{diff} = (N-2) - \sum_k m_k$ by sliding a seam by one ring---so a boundary ring switches population from $M_{t+1}$ to $M_t$ (or vice versa)---or shifting a whole terrace by $\pm 1$, neither of which creates a new seam, choosing the move of least marginal stretching cost per point. Each move strictly decreases $|\mathrm{diff}|$, so the loop ends after at most $|\mathrm{diff}_0|$ iterations, where $\mathrm{diff}_0$ is the initial residual (Appendix~\ref{app:alg-restore}); in practice it needs at most $10$ (Table~\ref{tab:runtime}).

\subsection{Stages 2--3: Latitudes, Phases, and Coordinates}
\label{sec:stage2}

Given the integer ring populations $m_0,\dots,m_{K-1}$ summing to $N - 2$:
\begin{enumerate}[label=\textbf{\arabic*.}]
  \item \textbf{Zonal Equilibrium Latitudes:}
  Let $m_{\mathrm{above},k} = 1 + \sum_{j < k} m_j$ and $m_{\mathrm{below},k} = 1 + \sum_{j > k} m_j$ be the numbers of points strictly above and below ring $k$ (including the two poles). The heights minimizing the zonal energy are
  \begin{equation}
    \label{eq:z-zonal}
    z_k^{\mathrm{zonal}} \;=\; \frac{m_{\mathrm{below},k} - m_{\mathrm{above},k}}{N - 1}.
  \end{equation}
  Equivalently, if all $N$ points were spaced evenly in height from pole to pole (which by Archimedes' theorem is equal area), $z_k^{\mathrm{zonal}}$ is the mean height of the $m_k$ points assigned to ring $k$; Remark~\ref{rem:stationarity} shows that~\eqref{eq:z-zonal} is the exact minimizer.
  \item \textbf{Lowest-Common-Harmonic Phase Locking:}
  Set $\alpha_0 = 0$ and for $k = 1,\dots,K-1$:
  \begin{equation}
    \label{eq:phase-locking}
    \begin{gathered}
    \Lambda_{k-1,k} = \operatorname{lcm}(m_{k-1}, m_k) = \frac{m_{k-1} m_k}{\gcd(m_{k-1}, m_k)}, \\
    \alpha_k = \left(\alpha_{k-1} + \frac{1}{2\,\Lambda_{k-1,k}}\right) \bmod 1.
    \end{gathered}
  \end{equation}
  \item \textbf{Multipole Latitude Relaxation:}
  Starting from $z^{\mathrm{zonal}}$, four fixed-point steps~\eqref{eq:z-fixed-point} adjust each ring's latitude to balance the exact multipole forces between rings (the forces that depend on how the points of two rings line up; Lemma~\ref{lem:cyclotomic}) within a band of $6$ neighboring rings; a step is accepted only if the latitudes remain strictly ordered (Appendix~\ref{app:alg-relax}).
\end{enumerate}

Finally, \textbf{Stage 3} places the two poles and, on each ring $k$, the points $\vx_{k,j} = (r_k\cos\phi_{k,j}, r_k\sin\phi_{k,j}, z_k)$ with $\phi_{k,j} = 2\pi(j/m_k + \alpha_k)$ and $r_k = \sqrt{1 - z_k^2}$ (Appendix~\ref{sec:stage3}).

\section{Complexity Analysis}
\label{sec:proof-complexity}

\begin{proposition}[Complexity]
\label{thm:complexity}
For $N \ge 3$, \TZ{0} constructs $X_N \subset \Sbb$ in $\Theta(N)$ time and $\Theta(N)$ memory. Stage~1 takes $\Theta(N^{5/6})$ time and memory for the dynamic program plus $\mathcal{O}(N)$ time for the point-count repair, Stage~2 takes $\mathcal{O}(N^{1/2}\log N)$ time and $\Theta(N^{1/2})$ memory, and Stage~3 takes $\Theta(N)$ time and memory.
\end{proposition}

\emph{Proof idea.} There are $K = \Theta(N^{1/2})$ rings and $W = \Theta(N^{1/3})$ candidates per ring. Because the seam cost~\eqref{eq:seam-cost} does not depend on which populations differ, each step of the dynamic program costs $\Theta(W)$ rather than $\Theta(W^2)$, so Stage~1 takes $\Theta(KW) = \Theta(N^{5/6})$; the point-count repair adds at most $\mathcal{O}(N)$ by Lemma~\ref{lem:initial-residual}, and in practice needs at most $10$ iterations. Stage~2 is a constant number of passes over the rings plus $\mathcal{O}(K)$ gcds, and Stage~3 writes the $N$ points. The full proof is in Appendix~\ref{app:proof-complexity}, and Table~\ref{tab:runtime} reports measured runtimes for $N = 10^3, \dots, 10^8$.

\section{Accuracy Analysis}
\label{sec:proof-energy}

This section states the accuracy results and sketches their proofs; full proofs are in Appendix~\ref{app:proofs-energy}. Section~\ref{sec:cyclotomic} writes the energy exactly as a sum over pairs of rings; Section~\ref{sec:kronecker} shows that inside a terrace this reproduces the BHS constant up to a quadratic stretching cost; and Section~\ref{sec:variational} balances that cost against the seams.

\subsection{The Exact Energy of Ring Configurations}
\label{sec:cyclotomic}

We call $X_N$ a \emph{ring configuration} if it consists of the two poles and $K$ coaxial latitude rings, ring $k$ carrying $m_k$ equally spaced points at height $z_k$, rotated by a phase $\alpha_k$; every output of \texttt{Terraced}\-\texttt{Zonal} is one. Lemma~\ref{lem:cyclotomic} below is our main tool. It splits the energy of a ring configuration into a part that depends only on the ring heights and populations, and a part that depends on how the rings are rotated relative to each other. It gives the second part in closed form, as a sum over pairs of rings whose terms decay like $\rho_{ab} = (\zeta_a/\zeta_b)^{\operatorname{lcm}(m_a, m_b)}$, where $\zeta_k = \tan(\theta_k/2)$ is the radius of ring $k$ after stereographic projection from the south pole onto the equatorial plane. So only neighboring rings with equal populations interact noticeably. This is why terraces help (Section~\ref{sec:terraces}), why the energy should be local (Section~\ref{sec:variational}), and why it can be evaluated exactly in $\Theta(N^{1/2})$ time.

The closed form comes from one identity. For a point $\vx_a$ on ring $a$ and a point $\vx_b$ on ring $b$, with $z_a > z_b$ and azimuthal offset $\Delta\phi$, the chordal distance factors as
\[
  \|\vx_a - \vx_b\|^2 \;=\; (1 + z_a)(1 - z_b)\,\bigl|1 - v\bigr|^2, \qquad v = (\zeta_a/\zeta_b)\,e^{i\Delta\phi},
\]
which is the chordal metric in stereographic coordinates (cf.\ Appendix~\ref{app:bg-polynomials}) and is verified directly in Appendix~\ref{app:proof-cyclotomic}. The prefactor depends only on the heights of the two rings. Over the $m$ points of a regular polygon, the product of the factors $|1 - v|$ collapses by the cyclotomic identity $\prod_{\ell=0}^{m-1}(1 - v\,\omega^{\ell}) = 1 - v^m$, $\omega = e^{2\pi i/m}$:

\begin{lemma}[Exact Energy of Coaxial Regular Polygons]
\label{lem:cyclotomic}
Let $X_N$ be a ring configuration with heights $1 > z_0 > \dots > z_{K-1} > -1$ and phases as in~\eqref{eq:coord-synthesis}, and let $\zeta_k = \tan(\theta_k/2)$ and $\xi_k := \log\zeta_k$ be the radius and log-radius of ring $k$ after stereographic projection from the south pole. Then
\begin{equation}
  \label{eq:exact-decomposition}
  E(X_N) \;=\; \Ezonal(m, z) \;+\; \Emult(m, z, \alpha),
\end{equation}
where the zonal energy $\Ezonal(m, z)$, the energy with every ring smeared into a uniform circle, depends only on the populations and heights and is the explicit $\mathcal{O}(K)$ sum~\eqref{eq:e-zonal-exact}, and the multipole energy, which depends on how the points of different rings line up, is
\begin{equation}
  \label{eq:e-mult-exact}
  \Emult(m, z, \alpha) \;=\; -\frac{1}{2}\sum_{0 \le a < b < K} \gcd(m_a, m_b)\,\log\bigl(1 - 2\rho_{ab}\cos\chi_{ab} + \rho_{ab}^2\bigr),
\end{equation}
with $\Lambda_{ab} = \operatorname{lcm}(m_a, m_b)$, $\rho_{ab} = (\zeta_a / \zeta_b)^{\Lambda_{ab}} = e^{\Lambda_{ab}(\xi_a - \xi_b)} \in (0, 1)$, and $\chi_{ab} = 2\pi \Lambda_{ab}(\alpha_a - \alpha_b)$.
\end{lemma}

The proof (Appendix~\ref{app:proof-cyclotomic}) applies the cyclotomic identity twice, once over each ring of a pair. Truncating~\eqref{eq:e-mult-exact} to $8$--$10$ neighboring rings evaluates $E(X_N)$ to machine precision (Remark~\ref{rem:cost}); this is what makes our exact verification up to $N = 10^9$ possible. Because~\eqref{eq:e-mult-exact} is explicit in the log-radii $\xi_k$, its derivatives give closed-form forces between rings, which Stage~2 uses to relax the ring heights (Appendix~\ref{app:alg-relax}).

\subsection{Each Terrace Is a Planar Lattice}
\label{sec:kronecker}

Inside a terrace every ring has the same population $m$, and adjacent rings are staggered by half a spacing. In the conformal coordinates $(\phi, \xi = \log\tan(\theta/2))$, where the rings unroll into horizontal rows as in Figure~\ref{fig:notation}, the points therefore form a planar lattice whose only free parameter is its \emph{aspect ratio} $\beta_k$, the spacing between rings relative to the spacing between points on a ring. At the zonal heights, $\beta_k = \frac{\sqrt{3}}{2}(m_k/y_k)^2$, so the lattice is equilateral exactly when the ring has its ideal population.

\begin{lemma}[The Energy of a Terrace Is a Planar Lattice Energy]
\label{lem:kronecker}
Inside a terrace with half-spacing stagger, the energy per point beyond the first two terms of $\EBHS(N)$ is the energy $\psi_{\mathrm{terrace}}(\beta_k)$ of the planar lattice with aspect ratio $\beta_k$, and
\[
  \psi_{\mathrm{terrace}}(\beta_k) \;=\; \frac{1}{2}\CBHS + \Phi_{\mathrm{strain}}(\beta_k),
  \qquad
  m_k\,\Phi_{\mathrm{strain}}(\beta_k) \;=\; c_{\mathrm{stretch}}\,\frac{(m_k - y_k)^2}{y_k} + \mathcal{O}\!\left(\frac{|m_k - y_k|^3}{y_k^2}\right),
\]
with $c_{\mathrm{stretch}} = \frac{1}{4}$, where $\Phi_{\mathrm{strain}}(\beta_k) \ge 0$ vanishes if and only if the lattice is equilateral ($\beta_k = \frac{\sqrt{3}}{2}$, i.e.\ $m_k = y_k$).
\end{lemma}

\emph{Proof sketch.} The multipole series inside a terrace is the product expansion of the Dedekind eta function at $\tau_k = \frac{1}{2} + i\beta_k$, so by Kronecker's first limit formula $\psi_{\mathrm{terrace}}$ is half the regularized Epstein zeta constant of that lattice, a classical quantity minimized by the equilateral lattice~\cite{Ryshkov1973,Montgomery1988}. The Chowla--Selberg formula gives its value $\frac{1}{2}\CBHS$ at $\tau_0 = e^{i\pi/3}$, the order-$3$ modular symmetry fixing $\tau_0$ makes it stationary there, and Ramanujan's differential equation for $E_2$ gives the stiffness $c_{\mathrm{stretch}} = \frac{1}{4}$. Lemma~\ref{lem:kronecker-full} states the result precisely, and Appendix~\ref{app:proof-kronecker} proves it.

\subsection{Locality and the \texorpdfstring{$\Theta(N^{2/3})$}{Theta(N\^{2/3})} Excess}
\label{sec:variational}

Lemma~\ref{lem:kronecker} describes an idealized terrace: an infinite stack of equal, staggered rings with constant aspect ratio. A real terrace differs in two ways: its aspect ratio drifts slowly from ring to ring, and it ends at seams. Because rings $d$ apart couple only through $\rho \approx e^{-\pi\sqrt{3}\,d} \approx 0.0043^d$ (Lemma~\ref{lem:cyclotomic}), we expect the energy to be \emph{local}, as for a crystal with grain boundaries: each ring contributes the stretching cost $\psi_{\mathrm{stretch}}$ of Stage~1, and each seam removes the binding of every ring pair that straddles it. A seam separates $d$ pairs at distance $d$, so it costs
\begin{equation}
  \label{eq:c-seam}
  c_{\mathrm{seam}} \;:=\; \sum_{d \ge 1} d\,\log\bigl(1 - (-1)^d e^{-\pi\sqrt{3}\,d}\bigr) \;\approx\; 0.0042867
\end{equation}
per point, slightly less than the nearest-neighbor value $c_{\mathrm{step}} \approx 0.0043241$ that Stage~1 uses. We do not prove this locality; we assume it:
\begin{enumerate}[label=\textbf{(A)}]
  \item \emph{Locality.} For the populations $m$ produced by Stage~1, the heights $z^{\mathrm{zonal}}$ and the phases~\eqref{eq:phase-locking},
  \[
    E(X_N) - \EBHS(N) \;=\; \sum_{k} \psi_{\mathrm{stretch}}(m_k, y_k) \;+\; c_{\mathrm{seam}} \sum_{k:\, m_k \neq m_{k-1}} \tfrac{1}{2}(y_{k-1} + y_k) \;+\; o(N^{2/3}),
  \]
  i.e.\ the energy excess is Stage~1's own cost~\eqref{eq:stretch-cost}--\eqref{eq:seam-cost}, with seam weight $c_{\mathrm{seam}}$ instead of $c_{\mathrm{step}}$.
\end{enumerate}
Exact energies support locality~(A): its residual falls from $17.5\%$ of the excess at $N = 10^3$ to $1.0\%$ at $N = 10^8$ (Appendix~\ref{app:locality-test}).

\begin{theorem}[Sublinear $\Theta(N^{2/3})$ BHS Energy Excess]
\label{thm:energy}
Assume locality~(A). For the configurations $X_N \subset \Sbb$ produced by \TZ{0} with the latitudes $z^{\mathrm{zonal}}$ of~\eqref{eq:z-zonal}, i.e.\ before the four relaxation steps of Stage~2,
\begin{equation}
  \label{eq:main-energy-bound}
  E(X_N) - \EBHS(N) \;=\; C_1\,N^{2/3} + o(N^{2/3}),
\end{equation}
where $C_1 \approx 0.026225$ is an explicit constant~\eqref{eq:c1-closed-form}.
\end{theorem}

\emph{Proof sketch.} By locality~(A), the excess is the one-dimensional cost that Stage~1 minimizes, with seams priced at $c_{\mathrm{seam}}$ instead of $c_{\mathrm{step}}$. Integrating the per-ring trade-off $A L^2 + B/L$ of Section~\ref{sec:terraces} over the sphere shows that the minimum of such a cost is proportional to $N^{2/3}$ (Lemma~\ref{lem:continuum}), with one third spent on stretching and two thirds on seams. Stage~1 attains this minimum for the seam price $c_{\mathrm{step}}$, and re-pricing its seams at $c_{\mathrm{seam}}$ gives $C_1$. The full proof is in Appendix~\ref{app:proof-energy-thm}; there we also report that the four relaxation steps, which the theorem does not cover, lower the exact energy further by $1.2$--$1.9\%$.

\paragraph{Accuracy relative to the true minimum.}
Theorem~\ref{thm:energy} measures \texttt{Terraced}\-\texttt{Zonal} against $\EBHS(N)$. Known bounds on the minimal energy $E_{\min}(N) = \min_{|Y|=N} E(Y)$ turn it into a statement about the true minimum.

\begin{corollary}[Accuracy Relative to the True Minimum]
\label{cor:emin}
Assume locality~(A), and let $X_N$ be the configurations satisfying Theorem~\ref{thm:energy}. Then:
\begin{enumerate}[label=(\roman*)]
  \item $E(X_N) - E_{\min}(N) \;\le\; 6.24 \times 10^{-4}\,N + C_1 N^{2/3} + o(N^{2/3})$;
  \item $E(X_N) - E_{\min}(N) = o(N)$ if and only if the Brauchart--Hardin--Saff conjecture holds;
  \item if moreover $E_{\min}(N) = \EBHS(N) + \mathcal{O}(\log N)$, as conjectured in~\cite{BHS2012,HardinMichaelsSaff2016}, then $E(X_N) - E_{\min}(N) = C_1 N^{2/3} + o(N^{2/3})$.
\end{enumerate}
\end{corollary}

The proof (Appendix~\ref{app:proof-emin}) combines Theorem~\ref{thm:energy} with an explicit lower bound on $E_{\min}(N)$, valid for every $N$, from the work of Beltr\'an and Lizarte~\cite{BeltranLizarte2023} (Lemma~\ref{lem:emin-lower}), and with~\cite[Theorem~1.5]{BeterminSandier2018}. Combined with exact energies, the lower bound already gives $E(X_N) - E_{\min}(N) < 0.00298\,N$ at $N = 10^3$ and $< 0.00065\,N$ at $N = 10^9$.

\section{Numerical Verification (\texorpdfstring{$N = 10^2 \dots 10^9$}{N = 10\^2 ... 10\^9})}
\label{sec:verification}

\begin{figure}[!t]
  \centering
  \includegraphics[width=\textwidth]{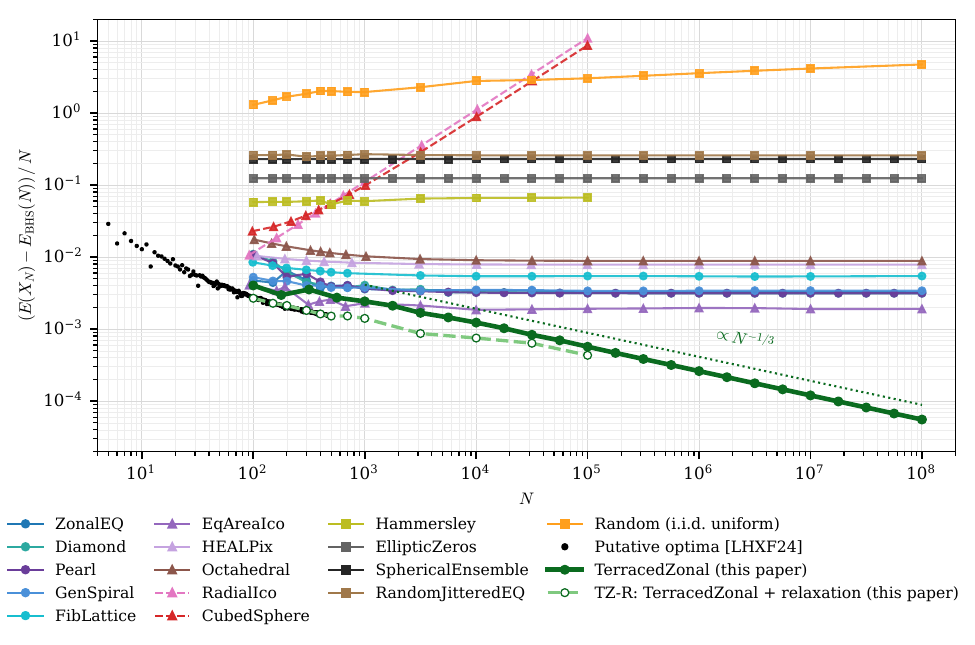}
  \caption{\textbf{\TZ{} is the only explicit construction whose per-point excess vanishes as $N \to \infty$.} Per-point excess over the BHS benchmark, $(E(X_N) - \EBHS(N))/N$, of the published explicit constructions of Table~\ref{tab:prior}, of \texttt{Terraced}\-\texttt{Zonal} (thick green), of \texttt{TZ-R}, a numerically optimized variant seeded by \texttt{Terraced}\-\texttt{Zonal} (dashed), and of the putative optima of~\cite{Lai2024} (black dots, $N \le 500$). Every previous construction levels off at a positive constant or grows. \texttt{Terraced}\-\texttt{Zonal} falls as $N^{-1/3}$ (dotted guide), drops below all of them near $N = 2 \times 10^3$, and at $N = 10^8$ is $34\times$ below the best of them. Appendix~\ref{app:tables} explains how each curve was computed.}
  \Description{One log-log plot of per-point excess against N from 5 to 10 to the 8. Random points sit near 1 to 5 and grow slowly; jittered points, the spherical ensemble, zeros of random elliptic polynomials, and Hammersley points are flat near 0.26, 0.23, 0.124, and 0.066; the gnomonic RadialIco and CubedSphere grow steeply. The remaining published constructions, including the Diamond and Pearl ensembles, level off between about 0.002 and 0.009. Black dots for the putative optima at N up to 500 fall from about 0.03 to 0.0015. The thick dark green TerracedZonal curve falls along a straight line of slope minus one third, from about 0.004 at N equals 100 to about 0.00006 at N equals 10 to the 8, below every other curve beyond N of about 2000. A dashed light green curve for TZ-R lies on the black dots up to N equals 500 and below the TerracedZonal curve up to N equals 10 to the 5.}
  \label{fig:all-constructors}
\end{figure}

\paragraph{Exact evaluation up to $N = 10^9$.}
We compute $E(X_N)$ exactly with Lemma~\ref{lem:cyclotomic} (validated against brute-force sums in Appendix~\ref{app:tables}) from $N = 10^3$ to $10^9$. The normalized excess $(E(X_N) - \EBHS(N))/N^{2/3}$ stays nearly constant, between $0.0243$ and $0.0266$, and is $0.0257$ at $N = 10^9$ (Table~\ref{tab:exact-1e9}), as Theorem~\ref{thm:energy} predicts; it lies slightly below $C_1 \approx 0.0262$ because the theorem does not include the four relaxation steps of Stage~2, which lower the excess by a further $1.2$--$1.9\%$ (Appendix~\ref{app:proof-energy-thm}). Figure~\ref{fig:all-constructors} compares \texttt{Terraced}\-\texttt{Zonal} with the published constructions, the putative optima~\cite{Lai2024}, and \texttt{TZ-R}, a numerically optimized variant seeded by \texttt{Terraced}\-\texttt{Zonal} (not covered by Proposition~\ref{thm:complexity}).

\paragraph{Stress test at adversarially chosen $N$.}
At $20$ values $N \in [349, 84{,}410]$ chosen to target five failure modes (Appendix~\ref{app:stress}), such as rounding resonances in $K$, $(E(X_N) - \EBHS(N))/N^{2/3}$ stays in $[0.0223, 0.0269]$, below \texttt{ZonalEQ} and \texttt{GenSpiral} at every one.

\section{An Application to Spherical Integration}
\label{sec:qmc}

As noted in Algorithm~\ref{alg:terraced-zonal} ($^\dagger$), Stage~1's terrace structure $(m_k)$ and Stage~2's half-spacing stagger $(\alpha_k)$ enforce local triangular ($A_2$) order purely from geometry; only the ring heights $z_0,\dots,z_{K-1}$ are tuned specifically to the logarithmic potential. Moreover, because the closed-form heights $z^{\mathrm{zonal}}$ of~\eqref{eq:z-zonal} sit at a stationary point of the zonal logarithmic energy ($\nabla_z \Ezonal = 0$), small adjustments to the $K = \Theta(N^{1/2})$ heights perturb $\Ezonal$ only at second order. This leaves $K$ nearly free one-dimensional degrees of freedom that can be directed toward a second fundamental task on $\Sbb$: \emph{equal-weight numerical integration} (quasi-Monte Carlo~\cite{BrauchartDick2012,BrauchartSaffSloanWomersley2014}), where the surface average
\begin{equation}
  \label{eq:sphere-integral}
  \bar{f} \;=\; \frac{1}{4\pi}\int_{\Sbb} f(\vx)\,d\sigma(\vx)
  \qquad\text{is approximated by the unweighted mean}\qquad
  I(f; X_N) \;=\; \frac{1}{N}\sum_{i=1}^N f(\vx_i).
\end{equation}
Unlike classical cubature rules $\sum_{i=1}^N w_i f(\vx_i)$, which assign unequal weights $w_i$ at special point counts $N$, equal weights $1/N$ are required whenever each point represents one sample, ray, or particle of equal weight.

\paragraph{Why logarithmic-energy points stall at $\mathcal{O}(N^{-3/2})$---even on $f(\vx) = z^2$.}
One might expect configurations of very low logarithmic energy to integrate smooth functions very accurately. Surprisingly, Brauchart, Saff, Sloan, and Womersley~\cite{BrauchartSaffSloanWomersley2014} observed that even putative global minimizers of $E(X_N)$ stall at integration error $\Theta(N^{-3/2})$, regardless of how smooth $f$ is beyond three derivatives. On \texttt{Terraced}\-\texttt{Zonal}, an elementary check reveals why without any specialized harmonic analysis:
\begin{itemize}
  \item \textbf{North--South functions of height $z$ alone.} Consider the quadratic $f(x,y,z) = z^2$, whose exact average over $\Sbb$ is $\frac{1}{2}\int_{-1}^1 z^2\,dz = \frac{1}{3}$. Because all $m_k$ points on ring $k$ share the same height $z_k$ (and the two poles sit at $z = \pm 1$), the integration error depends only on the ring heights:
  \[
    \frac{1}{N}\sum_{i=1}^N z_i^2 - \frac{1}{3}
    \;=\; \frac{2}{N} + \sum_{k=0}^{K-1} \frac{m_k}{N}\,z_k^2 - \frac{1}{3}.
  \]
  Because Stage~2 places $z_k$ to balance electrostatic repulsion rather than surface area, this difference does not vanish: it equals $+1.2 \times 10^{-5}$ at $N = 1{,}200$ and decays only as $\Theta(N^{-3/2})$. By Taylor's theorem, any point set that mis-integrates $z^2$ by $\Theta(N^{-3/2})$ inherits an $\Omega(N^{-3/2})$ error floor on generic smooth functions (such as $e^z$ or a smooth Gaussian hill).
  \item \textbf{East--West variation around each latitude circle.} Write $\vx = (\sqrt{1-z^2}\cos\varphi, \sqrt{1-z^2}\sin\varphi, z)$. Around latitude circle $k$, \texttt{Terraced}\-\texttt{Zonal} places $m_k$ equally spaced points at longitudes $\varphi_{k,j} = 2\pi(j/m_k + \alpha_k)$ ($j = 0,\dots,m_k - 1$). Around any circle of $m_k$ equally spaced points, every trigonometric wave $\cos(m\varphi)$ and $\sin(m\varphi)$ whose frequency $m \ge 1$ is not a multiple of $m_k$ sums to \emph{exact zero}:
  \begin{equation}
    \label{eq:azimuthal-cancel}
    \sum_{j=0}^{m_k - 1} e^{\mathrm{i} m \varphi_{k,j}} \;=\; 0 \qquad (m_k \nmid m).
  \end{equation}
  Near the poles, where $m_k$ is small (empirically $m_0 = m_{K-1} = 6$), the latitude circle itself has small radius $\sqrt{1 - z_k^2} = \Theta(m_k N^{-1/2})$, so smooth functions vary very little around it.
\end{itemize}
Equally spaced points on coaxial latitude circles therefore cancel East--West oscillations automatically; the $\Theta(N^{-3/2})$ integration bottleneck comes solely from one-dimensional functions of height $z$.

\begin{figure}[!t]
  \centering
  \includegraphics[width=\textwidth]{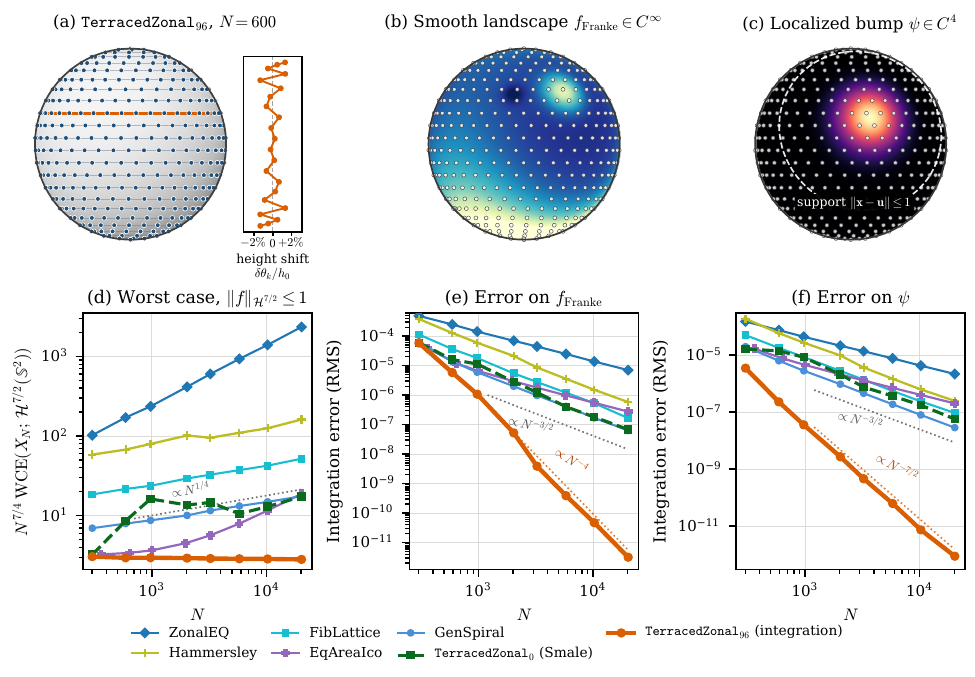}
  \caption{\textbf{Spherical integration by plain averaging: with \TZ{96}, the error on smooth functions decays empirically as $N^{-4}$, versus $N^{-3/2}$ for previous fast constructions.} Top: \textbf{(a)}~\TZ{96} at $N = 600$; right: how far each ring moves in latitude when its height is adjusted for integration, always less than $2\%$ of the nominal ring spacing $h_0$; \textbf{(b,\,c)}~the two test functions, a smooth function $f_{\mathrm{Franke}}$ and a $C^4$ bump $\psi$ (zero outside the dashed cap). Bottom: integration error, compared with \TZ{0} and the constructions of Figure~\ref{fig:all-constructors}. \textbf{(d)}~Worst-case error over a smoothness class, scaled so that the optimal rate is flat: only \TZ{96} stays flat. \textbf{(e,\,f)}~Errors on (b) and (c): every other construction falls only as $N^{-3/2}$, whereas \TZ{96} falls as $N^{-4}$ and $N^{-7/2}$. Details in Section~\ref{sec:qmc} and Appendix~\ref{app:qmc}.}
  \Description{Two rows of three panels. Top row: (a) a sphere with 22 horizontal rings of equally spaced points, one highlighted in orange, beside a narrow plot of ring colatitude shifts within plus or minus two percent; (b) a sphere shaded from dark blue to pale yellow for the Franke test function, with points overlaid; (c) a mostly black sphere with a glowing bump inside a dashed circle marking its support. Bottom row, log-log plots against N from 300 to about 20,000 with a shared legend: (d) N to the seven fourths times the worst-case error, rising for all baselines and TerracedZonal 0 but flat near 2.9 for TerracedZonal 96; (e) error on the Franke function, with baselines falling along N to the minus three halves and TerracedZonal 96 falling along N to the minus 4 to 3 times 10 to the minus 12; (f) error on the bump, with TerracedZonal 96 falling along N to the minus seven halves below 10 to the minus 12.}
  \label{fig:qmc}
\end{figure}

\paragraph{The integration setting $Q = 96$.}
To remove this one-dimensional bottleneck without disturbing the terraces, \TZ{Q} with $Q > 0$ runs Stages~1--3 exactly as \TZ{0} does, but line~7 of Algorithm~\ref{alg:terraced-zonal} adjusts the relaxed heights $z = (z_0,\dots,z_{K-1})$ so that the first $Q' := \min(K - 1, Q)$ moments in height $z$---equivalently, every one-dimensional polynomial in $z$ of degree $1 \le \ell \le Q'$---match their exact sphere averages. Expressed in the basis of Legendre polynomials $P_1(z),\dots,P_{Q'}(z)$ (which span all polynomials of degree $\le Q'$ modulo constants and satisfy $\frac{1}{2}\int_{-1}^1 P_\ell(z)\,dz = 0$), this condition is the system of $Q'$ equations in $K$ variables
\begin{equation}
  \label{eq:zonal-moment-constraint}
  r_\ell(z) \;:=\; \frac{1}{N}\Bigl(P_\ell(1) + P_\ell(-1) + \sum_{k=0}^{K-1} m_k P_\ell(z_k)\Bigr) \;=\; 0,
  \qquad \ell = 1,\dots,Q'.
\end{equation}
Starting from the heights of \TZ{0}, line~7 solves~\eqref{eq:zonal-moment-constraint} with a few Gauss--Newton steps, each costing $\mathcal{O}(N^{1/2})$ arithmetic operations for fixed $Q$ (Appendix~\ref{app:qmc}); we do not analyze \TZ{Q} further for $Q > 0$. We use $Q = 96$ throughout (Appendix~\ref{app:qmc} explains the choice), i.e.\ \TZ{96}, which in our measurements runs as fast as \TZ{0} ($\approx 1.0$~s for both at $N = 10^7$).

The resulting height shifts are small (below $2\%$ of the nominal ring spacing $h_0$ at $N = 600$; Figure~\ref{fig:qmc}(a)) and leave the energy almost unchanged (Appendix~\ref{app:qmc}).
\paragraph{Integration benchmarks.}
Figure~\ref{fig:qmc}(d--f) compares \TZ{96} across eight values of $N$ from $300$ to $20{,}252$, roughly doubling at each step (the largest is an icosahedral count, so all constructions, including \texttt{EqAreaIco}, share that $N$), against \TZ{0} and the deterministic equal-weight constructions of Figure~\ref{fig:all-constructors} on three benchmarks (exact formulas and tabular values in Appendix~\ref{app:qmc}):
\begin{enumerate}[label=(\alph*)]
  \item \textbf{Smooth global landscape ($C^\infty$).} On the spherical Franke function $f_{\mathrm{Franke}}$~\cite{Renka1988} (Figure~\ref{fig:qmc}(b)), averaged over random rotations, every prior equal-weight construction stalls along $\mathcal{O}(N^{-3/2})$ ($6.3 \times 10^{-8}$ for \texttt{GenSpiral} and $6.7 \times 10^{-8}$ for \texttt{Terraced}\-\texttt{Zonal} at $N = 20{,}252$), whereas the error of \TZ{96} falls as $N^{-4}$, to $3.2 \times 10^{-12}$, over $20{,}000\times$ smaller.
  \item \textbf{Localized bump of finite smoothness ($C^4$).} On a compactly supported Wendland bump $\psi$~\cite{Wendland1995} (Figure~\ref{fig:qmc}(c)), \TZ{96} tracks the $\mathcal{O}(N^{-7/2})$ smoothness rate down to $8.9 \times 10^{-13}$ at $N = 20{,}252$, over $30{,}000\times$ below the best prior equal-weight baseline.
  \item \textbf{Adversarial worst-case error over a function class.} To confirm that these gains do not depend on a particular test function, Figure~\ref{fig:qmc}(d) shows the worst-case error $\mathrm{WCE}(X_N; \mathcal{H}^{7/2}(\Sbb))$ over all functions of unit $\mathcal{H}^{7/2}(\Sbb)$ norm~\cite{BrauchartSaffSloanWomersley2014}, which has a closed form (Appendix~\ref{app:qmc}); the best possible rate for $N$-point sets is $N^{-7/4}$~\cite{HesseSloan2005}. Scaled by $N^{7/4}$, it grows for every prior equal-weight construction but stays flat in $[2.82, 3.06]$ for \TZ{96}.
\end{enumerate}

\section{Conclusion and Future Work}
\label{sec:conclusion}

We introduced \texttt{Terraced}\-\texttt{Zonal}, a deterministic construction that places $N$ points on $\Sbb$ in $\Theta(N)$ time. By holding ring populations constant across bands chosen by a one-dimensional dynamic program, it lets coaxial latitude rings lock into a local triangular lattice almost everywhere, which reduces the excess of the logarithmic energy over $\EBHS(N)$ from $\Omega(N)$, for every previously analyzed explicit construction, to $\Theta(N^{2/3})$. We proved this rate under locality~(A) and confirmed it by exact evaluation up to $N = 10^9$. Even if both locality~(A) and the Brauchart--Hardin--Saff conjecture hold, however, $\Theta(N^{2/3})$ remains polynomially far from $\mathcal{O}(\log N)$, and there is no reason to expect $N^{2/3}$---the exponent of balancing one-dimensional stretching against full-circle coaxial seams---to be the intrinsic algorithmic limit on $\Sbb$. Smale's seventh problem therefore remains open.

One natural attempt to go further is to avoid the polar regions, where the per-point excess grows as $|\cot\theta|^{2/3}$, by covering $\Sbb$ with three orthogonal belts of terraced rings. This variant (Appendix~\ref{app:multi-axis}) lowers the excess inside each belt by about $60\%$, but where the belts meet, one-dimensional phase locking cannot prevent close pairs, and the total excess grows by an order of magnitude. Going below $\Theta(N^{2/3})$ will likely require breaking the global rotational symmetry without creating such seam collisions, for instance by relaxing points locally around isolated defects.

Finally, the only geometric input to Stage~1 is the length of each latitude circle, so the same approach should carry over to other axially symmetric surfaces, such as spheroids, and to other energies, such as Riesz $s$-energies. Extending it to higher-dimensional spheres $\mathbb{S}^d$ is a natural open question.

\paragraph{AI methodology.}
This research relied on Gemini~4 Argon, which performed most of the day-to-day technical and expository work under the authors' direction and verification; see the \hyperref[sec:ai-methodology]{AI Methodology} statement (page~\pageref{sec:ai-methodology}) for details.

\clearpage
\appendix
\addtocontents{toc}{\protect\setcounter{tocdepth}{1}}

\section{Background on the Logarithmic Energy}
\label{app:background}
This appendix collects background that the introduction states only briefly.

\subsection{The Minimal Energy}
\label{app:bg-minimum}
The bounds of Rakhmanov, Saff, and Zhou~\cite{RSZ1994,RSZ1995} show that the minimal logarithmic energy is $\frac{1}{2}(\frac{1}{2} - \log 2) N^2 - \frac{1}{4} N \log N + \mathcal{O}(N)$~\cite{BorodachovHardinSaff2019}. Using the renormalized-energy theory of Sandier and Serfaty~\cite{SandierSerfaty2012}, B\'etermin and Sandier~\cite{BeterminSandier2018} proved that the $\mathcal{O}(N)$ term is $C_{\log} N + o(N)$ for some constant $C_{\log}$. Brauchart, Hardin, and Saff~\cite{BHS2012} conjectured that $C_{\log} = \frac{1}{2}\CBHS$; B\'etermin and Sandier showed that $\frac{1}{2}\CBHS$ is an upper bound for $C_{\log}$ and that the conjecture is equivalent to the optimality of the triangular lattice for the renormalized energy. Their proof of the upper bound builds configurations that locally follow the triangular lattice~\cite[\S\S6.3--6.4, \S7.3]{BeterminSandier2018}, so configurations with $E(X_N) \le \EBHS(N) + o(N)$ exist; that construction is not an algorithm and gives no rate. Lauritsen~\cite{Lauritsen2021} (see also Beltr\'an and Lizarte~\cite{BeltranLizarte2023}) showed that $C_{\log} \ge \frac{1}{2}(\log 2 - \frac{3}{4})$, which lies within $6.24 \times 10^{-4}$ of $\frac{1}{2}\CBHS$. The benchmark~\eqref{eq:bhs-expansion} of Section~\ref{sec:intro} uses the conjectured value, and the constant $\delta$ of Section~\ref{sec:intro} is $\delta = \frac{1}{2}\CBHS - C_{\log}$.
The Brauchart--Hardin--Saff constant $\CBHS$~\cite{BHS2012} is the regularized Epstein zeta constant~\cite{Epstein1903,ChowlaSelberg1967} of the unit-density planar triangular ($A_2$) Bravais lattice~\cite{Ryshkov1973,Montgomery1988,BeterminZhang2015}:
\begin{equation}
  \label{eq:c-bhs}
  \CBHS \;=\; 2 \log 2 + \frac{1}{2}\log\frac{2}{3} + 3 \log\frac{\sqrt{\pi}}{\Gamma(1/3)} \;\approx\; -0.05560530494339252.
\end{equation}

\paragraph{Normalization.} Much of the literature, including~\cite{BeterminSandier2018,Lauritsen2021,BeltranLizarte2023}, uses the double-counted energy $\sum_{i \neq j}\log(1/\|\vx_i - \vx_j\|)$, which is twice~\eqref{eq:log-energy}. In that convention the linear coefficient of the minimal energy is $2C_{\log}$, and the known bounds read $\log 2 - \frac{3}{4} \le 2C_{\log} \le \CBHS$~\cite[eq.~(1.1)]{BeltranLizarte2023}. In our single-counted convention they read $\frac{1}{2}(\log 2 - \frac{3}{4}) \le C_{\log} \le \frac{1}{2}\CBHS$, an interval of width $\frac{1}{2}(\CBHS - \log 2 + \frac{3}{4}) \approx 6.24 \times 10^{-4}$ (the constant $\Delta_{\log}$ of Lemma~\ref{lem:emin-lower}).

\subsection{Connection to Polynomials}
\label{app:bg-polynomials}
Equation~\eqref{eq:log-energy} is the $s = 0$ member of the Riesz $s$-energy family~\cite{BorodachovHardinSaff2019}, with pair potential $K_0(d) = \log(1/d)$, as opposed to $K_s(d) = d^{-s}$ for $s > 0$. The logarithmic case is singled out by its connection to polynomials~\cite{ShubSmale1993,Smale1998}. After a rotation so that no $\vx_i$ is the north pole, let $z_i = (x_{i,1} + \mathrm{i}\,x_{i,2})/(1 - x_{i,3}) \in \mathbb{C}$ be the stereographic image of $\vx_i$. Since $\|\vx_i - \vx_j\| = 2|z_i - z_j| / \sqrt{(1+|z_i|^2)(1+|z_j|^2)}$, the energy of $X_N$ is a function of the roots of $f(z) = \prod_{i=1}^N (z - z_i)$:
\begin{equation}
  \label{eq:energy-roots}
  E(X_N) \;=\; -\frac{1}{2}\log\bigl|\operatorname{Disc}(f)\bigr| \;+\; \frac{N-1}{2}\sum_{i=1}^{N}\log\bigl(1+|z_i|^2\bigr) \;-\; \binom{N}{2}\log 2,
\end{equation}
where $\operatorname{Disc}(f) = \prod_{i<j}(z_i - z_j)^2$ (this identity is~\cite[Prop.~1]{ArmentanoBeltranShub2011}; see also~\cite[eq.~(16)]{MichelenYakir2025}). Low energy therefore means roots that are well separated relative to their magnitudes. Shub and Smale were motivated by homotopy methods for solving polynomial equations, which deform a starting polynomial with known roots into the target polynomial and track the roots along the way; the cost depends on how well conditioned the roots remain~\cite{ShubSmale1993I,BCSS1998,BurgisserCucker2013}. They showed that if $E(X_N)$ is within $\mathcal{O}(\log N)$ of the minimum, then the roots of $f$ are well conditioned, with a condition number bounded by a polynomial in $N$~\cite{ShubSmale1993}. Smale's seventh problem~\cite{Smale1998} asks for an efficient algorithm that produces such points, giving one well-conditioned polynomial of each degree. Armentano, Beltr\'an, and Shub~\cite{ArmentanoBeltranShub2011} used~\eqref{eq:energy-roots} in the other direction: the stereographic images of the zeros of a random elliptic polynomial form a rotation-invariant random configuration whose expected energy they computed exactly. It lies $\Theta(N)$ above the minimum (Section~\ref{sec:existing}), with fluctuations of order only $\sqrt{N}$~\cite{MichelenYakir2025}.

\subsection{The Logarithmic Energy Among Riesz Energies}
\label{app:bg-riesz}
The known asymptotic expansions of the minimal Riesz $s$-energy on $\Sbb$~\cite{BHS2012,BorodachovHardinSaff2019,BeterminSandier2018} (Table~\ref{tab:riesz-regimes}) show why. For $s > 2$ the energy is dominated by nearest-neighbor interactions already at leading order, and for $0 < s < 2$ a continuum term of order $N^2$ leads while the arrangement of the points first matters at order $N^{1+s/2}$. For the logarithmic energy, the arrangement first matters only at order $N$, after the continuum $N^2$ term and the $N\log N$ term. Because Smale's problem demands accuracy $\mathcal{O}(\log N)$, any candidate construction must reproduce the true order-$N$ term, which in particular requires $E(X_N) - \EBHS(N) \le o(N)$; it is the first order at which the arrangement of the points, and not merely their density, matters.

\begin{table}[!htbp]
  \centering
  \caption{Asymptotics of the minimal Riesz $s$-energy of $N$ points on $\Sbb$~\cite{BHS2012,BorodachovHardinSaff2019,BeterminSandier2018}. The borderline case $s = 2$ is omitted.}
  \label{tab:riesz-regimes}
  \small
  \setlength{\tabcolsep}{9pt}
  \renewcommand{\arraystretch}{1.15}
  \begin{tabular}{@{}lll@{}}
    \toprule
    Regime & Leading term & First arrangement-dependent term \\
    \midrule
    $s > 2$ (hypersingular) & $\Theta(N^{1+s/2})$, local & leading order \\
    $0 < s < 2$ & $\Theta(N^2)$, continuum & order $N^{1+s/2}$ \\
    $s = 0$ (logarithmic) & $\frac{1}{2}\bigl(\frac{1}{2} - \log 2\bigr)N^2$, continuum & order $N$, after $-\frac{1}{4}N\log N$ \\
    \bottomrule
  \end{tabular}
\end{table}

\section{Extended Related Work: Three Families of Constructions}
\label{app:related}
This appendix expands Section~\ref{sec:existing}. Previously studied constructions on $\Sbb$ (surveyed in~\cite{HardinMichaelsSaff2016,BorodachovHardinSaff2019}) fall into three families, each with a geometric obstruction that leaves an excess $E(X_N) - \EBHS(N) = \Omega(N)$ (measured values in Table~\ref{tab:prior}, Appendix~\ref{app:tables}).
\begin{enumerate}[label=\textbf{\arabic*.}]
  \item \textbf{Smooth zonal rings, spirals, and helices.} Equal-area collar partitions~\cite{RSZ1994,Zhou1995,Leopardi2006}, the \texttt{Diamond} ensemble~\cite{BeltranEtayo2020,BeltranEtayoLopez2023}, and the recent \texttt{Pearl} ensemble~\cite{GariboldiGigante2026}, whose expected excess provably approaches $0.00315\,N$, the smallest proven for an explicit construction, place points on rings whose populations $m_k \approx \operatorname{round}(y_k)$ follow the smooth equal-area profile $y(\theta) = \frac{1}{2}\sqrt{2\pi\sqrt{3}\,N}\,\sin\theta$. Since $y_{k+1} - y_k = \pi\sqrt{3}\cos\theta_k + \mathcal{O}(N^{-1/2})$, adjacent rings almost always differ in population, and such rings cannot interlock (Figure~\ref{fig:families}(a), top). By Lemma~\ref{lem:cyclotomic}, the part of the energy of two rings $a, b$ that depends on how their points line up involves $\rho_{ab} = (\zeta_a/\zeta_b)^{\operatorname{lcm}(m_a, m_b)}$, where $\zeta_a < \zeta_b$ are their stereographic radii. Two equal rings at the $A_2$ (equilateral triangular lattice) spacing have $\rho = e^{-\pi\sqrt{3}}$ and, staggered by half a spacing, gain the binding energy $c_{\mathrm{step}} \approx 0.0043$ per point of~\eqref{eq:c-step}. For unequal rings $\operatorname{lcm}(m_a, m_b) \ge 2\min(m_a, m_b)$, so $\rho \le e^{-2\pi\sqrt{3}} \approx 2 \times 10^{-5}$, and the binding is lost at almost every ring. Spirals~\cite{RSZ1994,Bauer2000,SwinbankPurser2006,Gonzalez2010,Keinert2015} suffer the analogous problem: in the Fibonacci lattice the nearest-neighbor structure changes at $\Theta(\log N)$ band edges, each of which shears the lattice toward squares over a wide belt (Figure~\ref{fig:spiral-vs-terraced}(a); \cite[Sec.~2, eq.~(8) and Fig.~5]{HardinMichaelsSaff2016}). \texttt{Terraced}\-\texttt{Zonal} also belongs to this family, but it eases the main limitation of the family: by holding the population constant across terraces of $\Theta(N^{1/3})$ rings, it lets the rings interlock everywhere except at the $\Theta(N^{1/6})$ seams between terraces (Figure~\ref{fig:families}(a), bottom; Section~\ref{sec:terraces}).
  \item \textbf{Polyhedral tessellations.} Icosahedral and other polyhedral grids~\cite{Goldberg1937,CasparKlug1962,Sadourny1968} fold a triangular lattice onto $\Theta(1)$-sized faces, with $12$ five-fold disclinations at the vertices~\cite{Bowick2000} (Figure~\ref{fig:families}(b)). Radial projections~\cite{Teanby2006,Ronchi1996,Nair2005} are not equidistributed, and by Gauss's \emph{Theorema Egregium}~\cite{Gauss1828} any area-preserving map~\cite{Snyder1992,HardinMichaelsSaff2016} of a flat face onto a spherical triangle shears it by $\Theta(1)$, which costs strain energy proportional to $N$: $0.00198$ per point for the equal-area icosahedral points of~\cite{HardinMichaelsSaff2016}.
  \item \textbf{Quasi-Monte Carlo and random points.} Low-discrepancy~\cite{Hammersley1960,CuiFreeden1997,WongLukHeng1997}, jittered~\cite{Leopardi2006}, and i.i.d.\ uniform~\cite{Muller1959,BrauchartReznikovSaff2018} point sets become uniformly distributed as $N \to \infty$, as minimizers do~\cite{BorodachovHardinSaff2019}, so they reproduce the leading $N^2$ term of the minimal energy. They fall short at lower order: they have no local triangular order (Figure~\ref{fig:families}(c)), and i.i.d.\ points also cluster, so that their excess grows as $\frac{1}{4}N\ln N$. Random points that repel each other still pay a linear excess: exactly $0.1244\,N$ in expectation for the zeros of a random elliptic polynomial~\cite{ArmentanoBeltranShub2011}, and $0.2301\,N$ for the spherical ensemble~\cite{Krishnapur2009}, a determinantal process~\cite{AlishahiZamani2015,BeltranMarzoOrtega2016}.
\end{enumerate}

\subsection{The Three Families on the Sphere}
\label{app:three-families}
\begin{figure}[!htbp]
  \centering
  \includegraphics[width=0.78\textwidth]{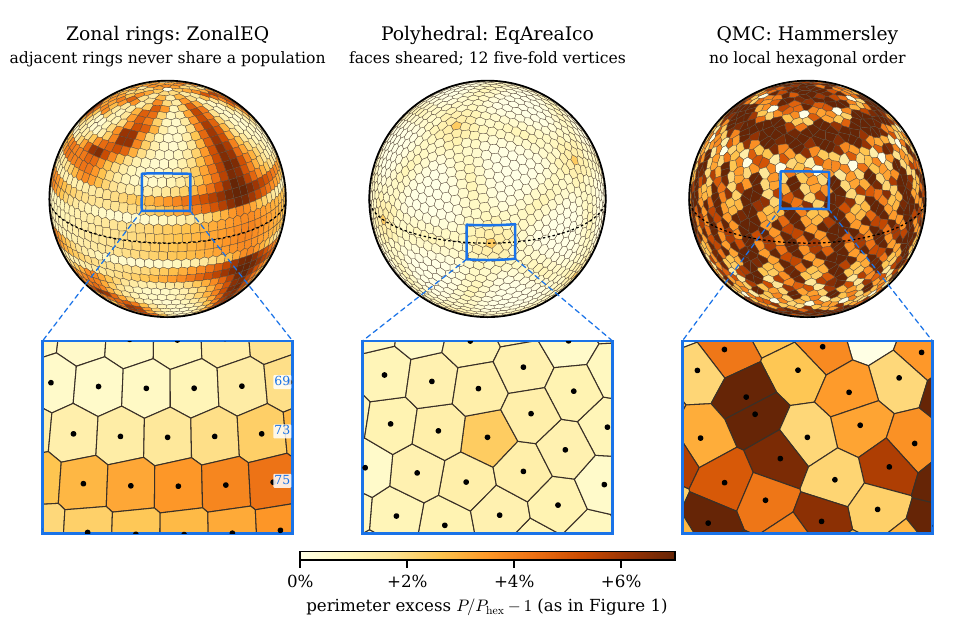}\\[-1pt]
  {\small\textbf{(a)}~Perimeter excess $P/P_{\mathrm{hex}} - 1$}\\[10pt]
  \includegraphics[width=0.78\textwidth]{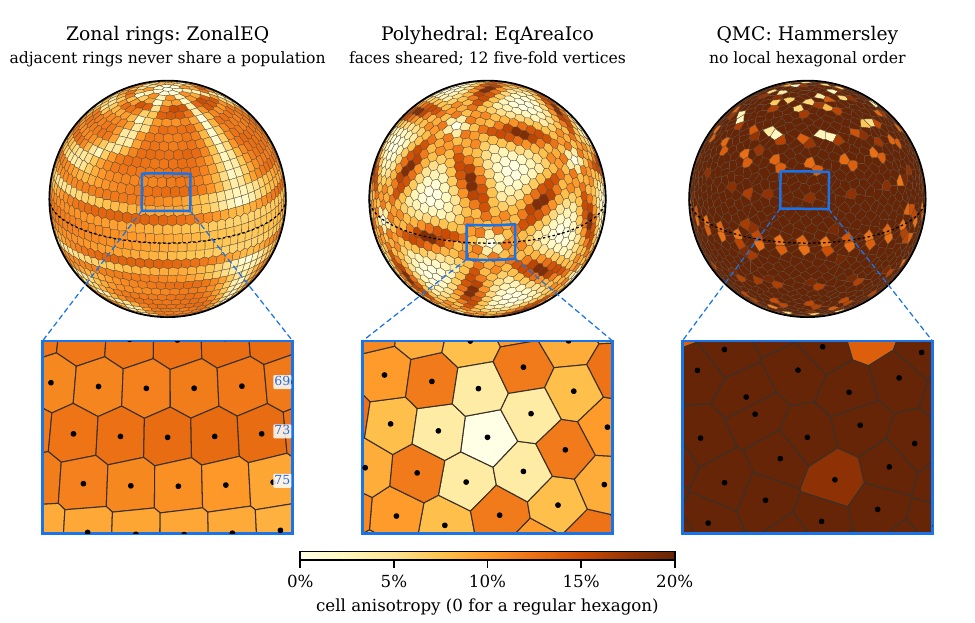}\\[-1pt]
  {\small\textbf{(b)}~Anisotropy $\sqrt{\lambda_{\max}/\lambda_{\min}} - 1$}
  \caption{One construction from each family of Section~\ref{sec:existing} at $N = 1{,}992$, from left to right \texttt{ZonalEQ}~\cite{Leopardi2006}, \texttt{EqAreaIco}~\cite{HardinMichaelsSaff2016}, and \texttt{Hammersley}~\cite{CuiFreeden1997,WongLukHeng1997}, with each spherical Voronoi cell shaded \textbf{(a)}~by its perimeter excess, as in Figure~\ref{fig:spiral-vs-terraced}, and \textbf{(b)}~by its anisotropy (both $0$ for a regular hexagon). Below each sphere, an equal-area zoom of the boxed region. \texttt{ZonalEQ}: the rings are spaced for square rather than hexagonal cells, and the bands mark where neighboring rings, whose populations differ (labels in the zoom), drift in and out of step. \texttt{EqAreaIco}: in (a) the cells are nearly perfect except at the $12$ five-fold corners (the zoom is centered on one); the shear of the area-preserving map is small in each cell, so (a) looks pale, but it is spread over every face and concentrates along the face edges in (b); its energy cost is the excess of $0.00198\,N$ in Table~\ref{tab:prior}. \texttt{Hammersley}: no cell is close to a regular hexagon.}
  \Description{Two rows of three spheres covered by Voronoi cells shaded from pale (regular hexagon) to dark brown, each sphere with a rectangular zoom below; the top row is shaded by perimeter excess (0 to 7 percent), the bottom row by anisotropy (0 to 20 percent). Left column, ZonalEQ: curved pale and dark bands in both rows, and in the zoom three rows of cells labeled with ring populations 69, 73, and 75. Middle column, EqAreaIco: in the top row almost uniformly pale with a few faint spots at the icosahedron's corners; in the bottom row pale patches separated by darker lines that trace the edges of an icosahedron; the zoom shows a pentagonal cell. Right column, Hammersley: dark, irregular cells almost everywhere in both rows.}
  \label{fig:three-families}
\end{figure}

\paragraph{Measuring departures from the triangular lattice.} The Voronoi cells of the triangular ($A_2$) lattice are regular hexagons, and among all partitions of the plane into cells of equal area, the regular hexagonal tiling has the least perimeter~\cite{Hales2001}. The triangular lattice is also the lattice that minimizes the energies behind the Brauchart--Hardin--Saff constant: the Epstein zeta function~\cite{Rankin1953,Cassels1959}, theta functions~\cite{Montgomery1988}, and, among lattices, the renormalized logarithmic energy~\cite{SandierSerfaty2012,BeterminSandier2018}, whose value at this lattice gives $\CBHS$. We therefore measure how far each cell is from the triangular lattice by its \emph{perimeter excess} $P/P_{\mathrm{hex}} - 1$, where $P$ is its geodesic perimeter and $P_{\mathrm{hex}}$ is the perimeter of a regular spherical hexagon of the same area. It vanishes for a perfect cell, is proportional to the strain energy for small strains (see below), and is at least $2.4\%$ for every pentagon. No theorem bounds the logarithmic energy by the perimeter excess, so we use it only as a geometric diagnostic. Empirically, at $N \approx 4000$ and $10^4$, the mean perimeter excess of the five constructions of Figures~\ref{fig:spiral-vs-terraced} and~\ref{fig:three-families} tracks $(E - \EBHS)/N$ with linear correlation $0.99$ (\texttt{Hammersley}, an outlier, excluded). Because the energy is a sum over points, we report the mean perimeter excess together with the share of cells more than $1\%$ above $P_{\mathrm{hex}}$, which shows how concentrated the distortion is; the median would describe only the typical cell and hide the seams where \texttt{Terraced}\-\texttt{Zonal} pays its energy.

Figure~\ref{fig:three-families} shows one published construction from each family of Section~\ref{sec:existing} at the same $N$, shaded (a) by perimeter excess and (b) by the \emph{anisotropy} $\sqrt{\lambda_{\max}/\lambda_{\min}} - 1$, where $\lambda_{\max} \ge \lambda_{\min}$ are the eigenvalues of the cell's second-moment tensor in the tangent plane, which measures how elongated the cell is. The two views complement each other: the anisotropy grows linearly with a small shear, so it makes visible the weak shear of the polyhedral family, which the perimeter excess, quadratic in the shear, shows only faintly; but it scores nearly regular pentagons and heptagons close to $0$.

\paragraph{Relation to the strain energy.} For small strains both measures can be compared with the energy. The energy per point of a lattice with unit area per point depends only on the lattice's shape~\cite{SandierSerfaty2012,BeterminSandier2018}, and Lemma~\ref{lem:kronecker} evaluates it for one family of shapes: a ring with $m_k = e^{\gamma} y_k$ points is the triangular lattice under a pure shear, compressed by $e^{-\gamma}$ along the ring and stretched by $e^{\gamma}$ across it, and by~\eqref{eq:phi-strain-taylor} this costs $\Phi_{\mathrm{strain}} = \frac{1}{4}\gamma^2 + \mathcal{O}(\gamma^3)$ per point. Because the lattice has hexagonal symmetry, its elastic energy is isotropic~\cite[\S10]{LandauLifshitz1986}: a shear of the same size costs the same in every direction. The perimeter excess is also quadratic in the shear, since the isoperimetric deficit of a polygon is controlled by the variance of its side lengths and radii~\cite{IndreiNurbekyan2015} (for whole tilings, see~\cite{CarocciaMaggi2016}); for the Voronoi cell of the sheared lattice we find numerically $0.750\,\gamma^2$ in every direction. The anisotropy, the usual eigenvalue-ratio index of a cell's second-moment tensor~\cite{SchroderTurk2011}, is instead linear, $1.20\,\gamma$. To leading order, therefore,
\[
  \Phi_{\mathrm{strain}} \;\approx\; \tfrac{1}{3}\bigl(P/P_{\mathrm{hex}} - 1\bigr) \;\approx\; \tfrac{1}{5.8}\bigl(\sqrt{\lambda_{\max}/\lambda_{\min}} - 1\bigr)^2.
\]
The perimeter excess is thus a direct proxy for the strain energy, whereas the anisotropy has to be squared; its mean overweights a weak shear spread over many cells. For \texttt{EqAreaIco}, whose cells are all slightly sheared, one third of the mean perimeter excess at $N = 9{,}992$ is $0.00194$, against the exact $(E - \EBHS)/N = 0.00186$. The relation holds only for small strains away from defects. Pentagons, heptagons, and dislocations carry their own energy, with long-range elastic interactions~\cite{Bowick2000}. Large strains break it too (the square lattice has perimeter excess $7.5\%$ and anisotropy $0$, yet costs only $0.0053$ per point more than the triangular lattice), so the correlation reported above is a ranking, not an identity.

\section{Algorithm Details}
\label{app:algorithm}
Algorithms~\ref{alg:stage1} and~\ref{alg:stage23} are not separate algorithms: they are Algorithm~\ref{alg:terraced-zonal} written out in full, stage by stage. Algorithm~\ref{alg:stage1} expands lines~1--3 (Stage~1) and Algorithm~\ref{alg:stage23} expands lines~4--9 (Stages~2--3); their line numbers are local. The subsections below explain the individual steps.

\subsection{Ring Discretization}
\label{app:alg-discretization}
This is step~1 of Stage~1 (Section~\ref{sec:stage1}).
Following the polar-cap and collar geometry of Rakhmanov--Saff--Zhou~\cite{RSZ1994} and Leopardi~\cite{Leopardi2006}, but using the equilateral triangular ($A_2$) meridional ring spacing $h_0 = \sqrt{2\pi\sqrt{3}/N}$ rather than the square-cell spacing $\sqrt{4\pi/N}$, set the polar cap colatitude $\theta_{\mathrm{cap}}$ (so each polar cap has area $2\pi(1-\cos\theta_{\mathrm{cap}}) = 4\pi/N$) and ring count $K$:
\begin{equation}
  \label{eq:discretization}
  \begin{gathered}
  \theta_{\mathrm{cap}} = \arccos\!\left(1 - \frac{2}{N}\right), \qquad
  h_0 = \sqrt{\frac{2\pi\sqrt{3}}{N}}, \\
  K = \max\!\left(1,\; \operatorname{round}\!\left(\frac{\pi - 2\theta_{\mathrm{cap}}}{h_0}\right)\right).
  \end{gathered}
\end{equation}
Partition $[\theta_{\mathrm{cap}}, \pi - \theta_{\mathrm{cap}}]$ into $K$ equal-colatitude strips with edges $\vartheta_k = \theta_{\mathrm{cap}} + \frac{k}{K}(\pi - 2\theta_{\mathrm{cap}})$ for $k = 0,\dots,K$ (strip $k$ holds ring $k$), and define the ideal real-valued equal-area ring populations via Archimedes' hat-box theorem (Lambert cylindrical equal-area projection~\cite{Snyder1992,Leopardi2006}):
\begin{equation}
  \label{eq:y-ideal}
  \begin{gathered}
  \Delta z_k^{\mathrm{ideal}} = \cos\vartheta_k - \cos\vartheta_{k+1}, \\
  y_k = (N - 2)\,\frac{\Delta z_k^{\mathrm{ideal}}}{\sum_{j=0}^{K-1} \Delta z_j^{\mathrm{ideal}}}, \qquad k = 0,\dots,K-1,
  \end{gathered}
\end{equation}
together with the rounded baseline $y_k^{\mathrm{base}} = \max(1, \operatorname{round}(y_k)) \in \Zbb_{\ge 1}$.

\subsection{Restoring the Point Count}
\label{app:alg-restore}
This is step~4 of Stage~1 (Section~\ref{sec:stage1}).
Let $\mathrm{diff} = (N - 2) - \sum_{k=0}^{K-1} m_k$ be the residual point count after unconstrained Viterbi backtracking, and decompose $(m_0,\dots,m_{K-1})$ into its terraces $\mathcal{T}_t = \{k_t, \dots, k_{t+1} - 1\}$ for $t = 0,\dots,T-1$ (so $k_0 = 0$ and $k_T = K$), with populations $M_t$ and lengths $L_t = k_{t+1} - k_t$. While $\mathrm{diff} \neq 0$, select the admissible move with minimum marginal stretching cost per point $\Delta\Psi / |\Delta S|$ among:
\begin{itemize}
  \item \textbf{Right boundary shift:} For any $t \in \{0,\dots,T-2\}$ with $L_{t+1} > 1$, moving the first ring of terrace $t+1$ into terrace $t$ ($k_{t+1} \leftarrow k_{t+1} + 1$) changes its population from $M_{t+1}$ to $M_t$ ($\Delta S = M_t - M_{t+1}$, zero new seams), admissible if $0 < \Delta S \cdot \operatorname{sgn}(\mathrm{diff}) \le |\mathrm{diff}|$.
  \item \textbf{Left boundary shift:} For any $t \in \{0,\dots,T-2\}$ with $L_t > 1$, moving the last ring of terrace $t$ into terrace $t+1$ ($k_{t+1} \leftarrow k_{t+1} - 1$) changes its population from $M_t$ to $M_{t+1}$ ($\Delta S = M_{t+1} - M_t$, zero new seams), admissible if $0 < \Delta S \cdot \operatorname{sgn}(\mathrm{diff}) \le |\mathrm{diff}|$.
  \item \textbf{Whole-terrace increment/decrement:} For any $t \in \{0,\dots,T-1\}$ of length $L_t \le |\mathrm{diff}|$, changing $M_t \leftarrow M_t + \operatorname{sgn}(\mathrm{diff}) \ge 1$ changes the total sum by $\Delta S = \operatorname{sgn}(\mathrm{diff})\,L_t$ with zero new seams.
\end{itemize}
If none of these moves is admissible, an \textbf{endpoint peel} changes the first ring of some terrace of length at least $2$ by $\operatorname{sgn}(\mathrm{diff})$, at the cost of at most one new seam ($\Delta S = \operatorname{sgn}(\mathrm{diff})$). Every admissible move satisfies $0 < \Delta S \cdot \operatorname{sgn}(\mathrm{diff}) \le |\mathrm{diff}|$, so $|\mathrm{diff}|$ strictly decreases, and an admissible move always exists; the loop therefore terminates with $\sum_k m_k = N - 2$ after at most $|\mathrm{diff}_0|$ iterations, where $\mathrm{diff}_0$ is the initial residual.

\begin{figure}[H]
  \centering
  \begin{minipage}{0.85\textwidth}
  \renewcommand{\thealgorithm}{1a}\renewcommand{\theHalgorithm}{1a}%
  \begin{algorithm}[H]
  \small
  \caption{Stage~1 of Algorithm~\ref{alg:terraced-zonal} (\texttt{Terraced}\-\texttt{Zonal}) in full: ring populations. Expands lines~1--3 of Algorithm~\ref{alg:terraced-zonal}.}
  \label{alg:stage1}
  \begin{algorithmic}[1]
    \Require Point count $N \ge 3$
    \Statex \textbf{Constants}$^\dagger$\textbf{:} stretch stiffness $c_{\mathrm{stretch}} = \frac{1}{4}$~\eqref{eq:stretch-cost}
    \Statex \phantom{\textbf{Constants}$^\dagger$\textbf{:} }seam cost $c_{\mathrm{step}} = \log(1 + e^{-\pi\sqrt{3}}) \approx 0.004324$~\eqref{eq:c-step}
    \Statex \textbf{Design parameter:} window half-width $w = \max(20, \operatorname{round}(0.65\,N^{1/3}))$
    \Ensure Ring count $K$
    \Statex \hspace*{-\algorithmicindent}\phantom{\textbf{Ensure:} }Ideal populations $y_0, \dots, y_{K-1}$
    \Statex \hspace*{-\algorithmicindent}\phantom{\textbf{Ensure:} }Integer populations $m_0, \dots, m_{K-1} \ge 1$ with $\sum_k m_k = N - 2$
    \State $\theta_{\mathrm{cap}} \gets \arccos(1 - 2/N)$ \Comment{polar caps of area $4\pi/N$}
    \State $h_0 \gets \sqrt{2\pi\sqrt{3}/N}$ \Comment{ring spacing}
    \State $K \gets \max\bigl(1, \operatorname{round}((\pi - 2\theta_{\mathrm{cap}})/h_0)\bigr)$ \Comment{\eqref{eq:discretization}}
    \State Compute $y_k$ and $y_k^{\mathrm{base}} \gets \max(1, \operatorname{round}(y_k))$ for $0 \le k < K$ \Comment{\eqref{eq:discretization}--\eqref{eq:y-ideal}}
    \State $W \gets 2w + 1$ \Comment{candidates per ring}
    \State $m_{k,i} \gets y_k^{\mathrm{base}} + i - w$ for $0 \le i < W$ \Comment{\eqref{eq:window}}
    \State $D[0,i] \gets \psi_{\mathrm{stretch}}(m_{0,i}, y_0)$ for $0 \le i < W$ \Comment{\eqref{eq:stretch-cost}; $+\infty$ if $m_{0,i} < 1$}
    \For{$k = 1, \dots, K-1$} \Comment{Viterbi--Potts DP}
      \State $j^* \gets \arg\min_{j} D[k-1, j]$, \ $s_k \gets y_k^{\mathrm{base}} - y_{k-1}^{\mathrm{base}}$
      \For{$i = 0, \dots, W-1$}
        \State $D[k,i] \gets D[k-1, j^*] + c_{\mathrm{step}}\,(y_{k-1} + y_k)/2$, \ $\mathrm{back}[k,i] \gets j^*$ \Comment{seam$^\dagger$}
        \If{$0 \le i + s_k < W$ \textbf{and} $D[k-1, i+s_k] < D[k,i]$} \Comment{stay}
          \State $D[k,i] \gets D[k-1, i+s_k]$, \ $\mathrm{back}[k,i] \gets i + s_k$
        \EndIf
        \State $D[k,i] \gets D[k,i] + \psi_{\mathrm{stretch}}(m_{k,i}, y_k)$ \Comment{$+\infty$ if $m_{k,i} < 1$}
      \EndFor
    \EndFor
    \State $(m_0, \dots, m_{K-1}) \gets$ backtrack through $\mathrm{back}$ from $\arg\min_i D[K-1, i]$
    \State Split $(m_0, \dots, m_{K-1})$ into maximal constant runs (terraces) $\mathcal{T}_0, \dots, \mathcal{T}_{T-1}$
    \While{$\sum_k m_k \neq N - 2$} \Comment{mass conservation}
      \State Apply the seam-free boundary shift or terrace $\pm 1$ minimizing $\Delta\Psi / |\Delta S|$;
      \Statex \hspace{\algorithmicindent} if none is admissible, peel one end ring of a terrace by $\pm 1$
    \EndWhile
  \end{algorithmic}
  \vspace{2pt}\noindent $^\dagger$\,Specific to the logarithmic energy; all other steps depend only on the geometry.
\end{algorithm}
  \end{minipage}
  \Description{Numbered pseudocode for Stage 1: compute the ring count and ideal ring populations, run a dynamic program over a window of candidate populations per ring with a seam penalty for population changes, backtrack, and adjust terraces until the populations sum to N minus 2.}
\end{figure}

\subsection{Multipole Latitude Relaxation}
\label{app:alg-relax}
This is step~3 of Stage~2 (Section~\ref{sec:stage2}).
Initialize $z \leftarrow z^{\mathrm{zonal}}$. For $4$ iterations, compute the log-radii $\xi_k = \log\zeta_k = \frac{1}{2}\log\frac{1 - z_k}{1 + z_k}$ of Lemma~\ref{lem:cyclotomic} and the truncated multipole gradient
\begin{equation}
  \label{eq:multipole-grad}
  (\nabla_\xi \Emult)_k \;:=\; \sum_{d=1}^{6} \bigl(F_{k-d,k} - F_{k,k+d}\bigr),
  \qquad
  F_{ab} \;:=\; \frac{m_a m_b\,\rho_{ab}\,(\rho_{ab} - \cos\chi_{ab})}{1 - 2\rho_{ab}\cos\chi_{ab} + \rho_{ab}^2} \quad (a < b),
\end{equation}
with $\rho_{ab}$ and $\chi_{ab}$ as in Lemma~\ref{lem:cyclotomic} and terms whose indices fall outside $\{0, \dots, K-1\}$ omitted. Since $\partial\rho_{ab}/\partial\xi_b = -\Lambda_{ab}\rho_{ab}$ and $\gcd(m_a, m_b)\operatorname{lcm}(m_a, m_b) = m_a m_b$, $F_{ab}$ is the derivative of the $(a, b)$ term of~\eqref{eq:e-mult-exact} with respect to $\xi_b$, and minus its derivative with respect to $\xi_a$. Hence~\eqref{eq:multipole-grad} is $\partial\Emult/\partial\xi_k$ restricted to rings at most $6$ apart. Pairs with $\Lambda_{ab}(\xi_a - \xi_b) \le -36$, whose terms are below double precision, are set to zero. Update
\begin{equation}
  \label{eq:z-fixed-point}
  z_k^{\mathrm{new}} = z_k^{\mathrm{zonal}} + \frac{2\,(\nabla_\xi \Emult)_k}{m_k (N - 1)},
\end{equation}
accepting $z \leftarrow z^{\mathrm{new}}$ whenever $1 > z_0^{\mathrm{new}} > z_1^{\mathrm{new}} > \dots > z_{K-1}^{\mathrm{new}} > -1$.

\begin{figure}[!htbp]
  \centering
  \includegraphics[width=\textwidth]{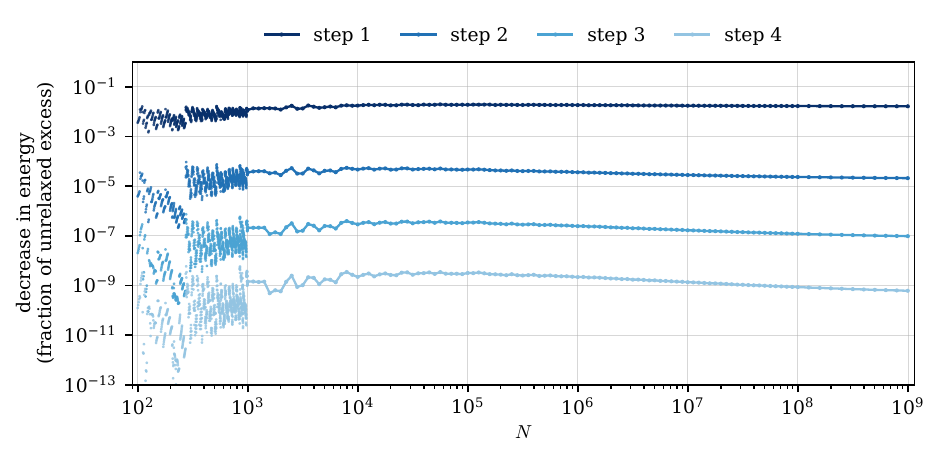}
  \caption{\textbf{Effect of each relaxation step on the energy.} Decrease in the exact energy produced by each of the four relaxation steps of Stage~2, relative to the excess $E(X_N) - \EBHS(N)$ before relaxation, for every $N$ from $100$ to $1{,}000$ (dots) and for $111$ log-spaced values of $N$ from $10^3$ to $10^9$ (lines). Each step decreases the energy at every $N$ shown. For $N \ge 10^3$, the first step reduces the excess by $1.2$--$1.9\%$, and each subsequent step contributes at least two orders of magnitude less than its predecessor.}
  \Description{Log-log plot of the energy decrease per relaxation step against N from 100 to 10 to the 9. Step 1 sits near 0.01 to 0.02; step 2 near 2 times 10 to the minus 5; step 3 near 2 times 10 to the minus 7; step 4 near 10 to the minus 9. The curves are flat for N above 1000 and scattered between 100 and 1000.}
  \label{fig:relaxation-steps}
\end{figure}

Few relaxation steps are needed, if any (Figure~\ref{fig:relaxation-steps}). The relaxation lowers the excess $E(X_N) - \EBHS(N)$ by only $1.2$--$1.9\%$ for $N \ge 10^3$, and almost all of this comes from the first step: steps~$2$--$4$ together change the excess by less than $10^{-4}$ of its value. A single step would therefore give essentially the same configurations, and omitting the relaxation altogether would raise the excess by less than $2\%$. All results in this paper use four steps; Theorem~\ref{thm:energy} applies to the configurations before relaxation.

\subsection{Stage 3: Coordinate Synthesis}
\label{sec:stage3}

Place $\vx_0 = (0,0,1)$ and $\vx_{N-1} = (0,0,-1)$. For each ring $k \in \{0,\dots,K-1\}$ with radius $r_k = \sqrt{\max(0, 1 - z_k^2)}$, generate its $m_k$ points for $j = 0,\dots,m_k - 1$:
\begin{equation}
  \label{eq:coord-synthesis}
  \phi_{k,j} = 2\pi\left(\frac{j}{m_k} + \alpha_k\right), \qquad
  \vx_{k,j} = \bigl(r_k \cos\phi_{k,j},\; r_k \sin\phi_{k,j},\; z_k\bigr).
\end{equation}

\begin{figure}[H]
  \centering
  \begin{minipage}{0.85\textwidth}
  \renewcommand{\thealgorithm}{1b}\renewcommand{\theHalgorithm}{1b}%
  \begin{algorithm}[H]
  \small
  \caption{Stages~2--3 of Algorithm~\ref{alg:terraced-zonal} (\TZ{Q}) in full: heights, phases, and coordinates. Expands lines~4--9 of Algorithm~\ref{alg:terraced-zonal}; continues from Algorithm~\ref{alg:stage1}.}
  \label{alg:stage23}
  \begin{algorithmic}[1]
    \Require Ring count $K$ and integer populations $m_0,\dots,m_{K-1}$ from Stage~1 (Algorithm~\ref{alg:stage1})
    \Statex \hspace*{-\algorithmicindent}\phantom{\textbf{Require:} }Matched height moments $Q \ge 0$
    \Statex \textbf{Design parameters:} $n_{\mathrm{relax}} = 4$ relaxation steps; multipole band $d_{\max} = 6$
    \Ensure Configuration $X_N = \{\vx_0,\dots,\vx_{N-1}\} \subset \Sbb$
    \Statex \textbf{Stage 2: latitudes and phases} (Section~\ref{sec:stage2}, Appendix~\ref{app:alg-relax}) \hfill $\triangleright$ $\mathcal{O}(N^{1/2}\log N)$ time
    \State $z_k^{\mathrm{zonal}} \gets (m_{\mathrm{below},k} - m_{\mathrm{above},k})/(N - 1)$ for $0 \le k < K$ \Comment{\eqref{eq:z-zonal}$^\dagger$}
    \State $z_k \gets z_k^{\mathrm{zonal}}$ for $0 \le k < K$ \Comment{start of the relaxation}
    \State $\alpha_0 \gets 0$
    \For{$k = 1, \dots, K-1$}
      \State $\alpha_k \gets \bigl(\alpha_{k-1} + 1/(2\operatorname{lcm}(m_{k-1}, m_k))\bigr) \bmod 1$ \Comment{\eqref{eq:phase-locking}}
    \EndFor
    \For{$n_{\mathrm{relax}}$ iterations} \Comment{stereographic multipole relaxation$^\dagger$}
      \State $\xi_k \gets \frac{1}{2}\log\frac{1 - z_k}{1 + z_k}$ for $0 \le k < K$
      \State $z_k^{\mathrm{new}} \gets z_k^{\mathrm{zonal}} + 2(\nabla_\xi \Emult)_k / (m_k(N - 1))$ \Comment{\eqref{eq:multipole-grad}--\eqref{eq:z-fixed-point}, band $d_{\max}$}
      \If{$1 > z_0^{\mathrm{new}} > \dots > z_{K-1}^{\mathrm{new}} > -1$}
        \State $z \gets z^{\mathrm{new}}$
      \EndIf
    \EndFor
    \If{$Q > 0$} \Comment{integration only\intmark}
      \State project $z$ onto $r_1(z) = \dots = r_{\min(K-1, Q)}(z) = 0$ \Comment{\eqref{eq:zonal-moment-constraint}, Appendix~\ref{app:qmc}}
    \EndIf
    \Statex \textbf{Stage 3: coordinate synthesis} (Appendix~\ref{sec:stage3}) \hfill $\triangleright$ $\Theta(N)$ time
    \State $\vx_0 \gets (0,0,1)$, \ $\vx_{N-1} \gets (0,0,-1)$
    \For{$k = 0, \dots, K-1$}
      \State $r_k \gets \sqrt{1 - z_k^2}$
      \For{$j = 0, \dots, m_k - 1$}
        \State $\phi_{k,j} \gets 2\pi(j/m_k + \alpha_k)$
        \State $\vx_{k,j} \gets (r_k\cos\phi_{k,j},\, r_k\sin\phi_{k,j},\, z_k)$ \Comment{\eqref{eq:coord-synthesis}}
      \EndFor
    \EndFor
    \State \Return $X_N$
  \end{algorithmic}
  \vspace{2pt}\noindent $^\dagger$\,Specific to the logarithmic energy; \intmark\,specific to integration; all other steps depend only on the geometry.
\end{algorithm}
  \end{minipage}
  \Description{Numbered pseudocode for Stages 2 and 3: compute closed-form ring heights and phase offsets, apply four multipole relaxation steps to the heights, each accepted only if the heights stay ordered, then place the two poles and the points of each ring.}
\end{figure}

\subsection{Constants and Design Parameters}
\label{app:constants}
Table~\ref{tab:constants} lists every numerical constant used by Algorithm~\ref{alg:terraced-zonal}. Apart from the parameter $Q$, the constants fall into three groups: constants derived from the geometry or the energy, which leave no freedom; design parameters of Stage~1 and Stage~2, which we fixed with a wide margin and to which the output is insensitive; and numerical safeguards of the integration step. The sensitivity checks quoted in the table use the reference implementation at the stated values of $N$.

\begin{table}[!htbp]
  \centering
  \caption{Constants of Algorithm~\ref{alg:terraced-zonal}, where they come from, and how much the output depends on them.}
  \label{tab:constants}
  \small
  \renewcommand{\arraystretch}{1.12}
  \begin{tabular}{@{}>{\raggedright\arraybackslash}p{0.19\textwidth}>{\raggedright\arraybackslash}p{0.375\textwidth}>{\raggedright\arraybackslash}p{0.375\textwidth}@{}}
    \toprule
    Constant & Origin & Role and sensitivity \\
    \midrule
    \multicolumn{3}{@{}l}{\emph{Derived from the geometry or the energy (no freedom)}} \\
    $h_0 = \sqrt{2\pi\sqrt{3}/N}$ & row spacing of a triangular lattice in which each point has area $4\pi/N$ (Section~\ref{sec:terraces}) & sets $K \approx \pi/h_0 \approx 0.952\,N^{1/2}$ rings \\
    polar cap $\theta_{\mathrm{cap}}$ & $\theta_{\mathrm{cap}} = \arccos(1 - 2/N)$, so each polar cap has area $4\pi/N$, one point's share~\eqref{eq:discretization} & reserves one point for each pole \\
    $c_{\mathrm{stretch}} = \frac{1}{4}$ & stiffness of the triangular-lattice energy under shear (Lemma~\ref{lem:kronecker}) & weight of stretching in Stage~1~\eqref{eq:stretch-cost} \\
    exponents $\frac{1}{3}, \frac{2}{3}$ in $y^{1/3}m^{2/3}$ & the only split for which a terrace's best population is the mean of its ideal populations to second order (Lemma~\ref{lem:unbiased}) & removes a systematic bias in the terrace populations \\
    $c_{\mathrm{step}} \approx 0.004324$ & $c_{\mathrm{step}}$~\eqref{eq:c-step}, the binding energy per point that two equal, staggered adjacent rings lose when their populations differ & seam price in Stage~1~\eqref{eq:seam-cost}; sets the terrace length $L_* = \Theta(N^{1/3})$. The all-pairs value $c_{\mathrm{seam}} \approx 0.0042867$~\eqref{eq:c-seam} enters only the analysis \\
    stagger $\alpha_k - \alpha_{k-1}$ & $1/(2\operatorname{lcm}(m_{k-1}, m_k))$, a half spacing; inside a terrace $1/(2m)$, which interlocks equal rings into a triangular lattice~\eqref{eq:phase-locking} & creates the triangular lattice inside each terrace (Figure~\ref{fig:concept}) \\
    \addlinespace
    \multicolumn{3}{@{}l}{\emph{Design parameters (fixed with margin)}} \\
    window $w$ & $w = \max(20, \operatorname{round}(0.65\,N^{1/3}))$~\eqref{eq:window}: optimal terraces need population excursions of at most $0.312\,N^{1/3}$ (Appendix~\ref{app:proof-energy-thm}), so $0.65$ leaves a factor of about $2$; the floor $20$ covers small $N$ & sets $W = \Theta(N^{1/3})$ and the $\Theta(N^{5/6})$ cost of the Stage~1 dynamic program. At $49$ values of $N \in [300, 10^6]$, doubling $w$ never changed $m$; the largest offset used was $0.67\,w$, and halving $w$ changed $m$ at $41$ of them \\
    $4$ relaxation steps & chosen in advance; not covered by Theorem~\ref{thm:energy} & lowers the excess by $1.2$--$1.9\%$ for $N \ge 10^3$; the first step gives over $99.6\%$ of this (Figure~\ref{fig:relaxation-steps}) \\
    band $d_{\max} = 6$; mask at $-36$ & a ring pair with $\Lambda_{ab}(\xi_a - \xi_b) \le -36$ contributes below $e^{-36} \approx 2 \times 10^{-16}$, under double precision; inside a terrace this holds for $d \ge 7$, since $\pi\sqrt{3} \cdot 7 > 36$ (Remark~\ref{rem:cost}) & no effect beyond rounding: bands $6$ and $10$ give identical heights at $N = 1{,}200$ to $10^6$ \\
    \addlinespace
    \multicolumn{3}{@{}l}{\emph{Integration only ($Q > 0$; Appendix~\ref{app:qmc})}} \\
    $Q = 96$ & smooth integration error saturates by $Q \approx 24$, where 1D Legendre truncation reaches $10^{-15}$ (Figure~\ref{fig:qmc-sweep}); $96$ leaves a factor of $4$ & number of matched height moments, $Q' = \min(K - 1, Q)$. At $N = 10{,}242$, any $Q \in [24, 92]$ gives the same error ($4.7 \times 10^{-11}$) and changes $(E - \EBHS)/N^{2/3}$ by $< 1.1\%$ \\
    $6$ Gauss--Newton steps; step cap $0.2/K$; ridge and stopping tolerance $10^{-14}$; clip $10^{-9}$; guard $10^{-12}$ & numerical safeguards: the cap keeps each step below a tenth of the mean ring spacing in height ($2/K$), so the rings stay ordered; the tolerances sit at double precision & residual moments end below $4 \times 10^{-15}$ (Table~\ref{tab:qmc-energy}); no other effect \\
    \bottomrule
  \end{tabular}
\end{table}

\section{Proof for the Complexity Analysis}
\label{app:proof-complexity}

\subsection{The Unbiased Stretching Cost}
\label{app:proof-unbiased}

Lemma~\ref{lem:kronecker} gives the strain cost of ring $k$ to leading order as $c_{\mathrm{stretch}}(m_k - y_k)^2/y_k$. At that order the denominator could equally be $m_k$, or any $y_k^{1-\alpha} m_k^{\alpha}$. Stage~1 uses $y^{1/3} m^{2/3}$~\eqref{eq:stretch-cost} because it is the unique choice for which the best constant population on a terrace is the mean ideal population to second order, which prevents systematic mass drift in the unconstrained Viterbi DP:

\begin{lemma}[Unbiased Stationary Point of $\psi_{\mathrm{stretch}}$]
\label{lem:unbiased}
Let $\mathcal{T}$ be a terrace with ideal populations $y_k = \bar{y} + \delta_k$, $\sum_{k \in \mathcal{T}} \delta_k = 0$, and for $\alpha \in \Rbb$ consider the homogeneous cost $F_\alpha(m) = \sum_{k \in \mathcal{T}} (m - y_k)^2 / (y_k^{1-\alpha} m^{\alpha})$. Then
\[
  F_\alpha'(\bar{y}) \;=\; (2 - 3\alpha)\sum_{k \in \mathcal{T}}\frac{\delta_k^2}{\bar{y}^2} \;+\; \mathcal{O}\!\left(\sum_{k \in \mathcal{T}}\frac{|\delta_k|^3}{\bar{y}^3}\right).
\]
In particular, the minimizer of $F_\alpha$ equals the arithmetic mean $\bar{y} = \frac{1}{|\mathcal{T}|}\sum_{k \in \mathcal{T}} y_k$ through second order in $\delta_k/\bar{y}$ if and only if $\alpha = 2/3$.
\end{lemma}

\begin{proof}
For a single ring, differentiating at $m = \bar{y}$ gives
\[
  \left.\frac{\partial}{\partial m}\frac{(m - y_k)^2}{y_k^{1-\alpha} m^\alpha}\right|_{m=\bar{y}}
  \;=\; \frac{-2\delta_k}{\bar{y}(1 + \delta_k/\bar{y})^{1-\alpha}} - \frac{\alpha\delta_k^2}{\bar{y}^2(1 + \delta_k/\bar{y})^{1-\alpha}}.
\]
Using the first-order binomial expansion $(1 + \delta_k/\bar{y})^{-(1-\alpha)} = 1 - (1-\alpha)\frac{\delta_k}{\bar{y}} + \mathcal{O}(\delta_k^2/\bar{y}^2)$, we obtain
\begin{equation}
  \label{eq:exact-unbiased}
  \left.\frac{\partial}{\partial m}\frac{(m - y_k)^2}{y_k^{1-\alpha} m^\alpha}\right|_{m=\bar{y}}
  \;=\; -\frac{2\delta_k}{\bar{y}} \;+\; (2 - 3\alpha)\frac{\delta_k^2}{\bar{y}^2} \;+\; \mathcal{O}\!\left(\frac{\delta_k^3}{\bar{y}^3}\right).
\end{equation}
Summing over $k \in \mathcal{T}$ cancels the linear term since $\sum_{k \in \mathcal{T}} \delta_k = 0$, leaving the stated expression. The quadratic bias coefficient $2 - 3\alpha$ vanishes if and only if $\alpha = 2/3$.
\end{proof}

\subsection{Stage-1 Terraces Are Nearly Unbiased}
\label{app:stage1-unbiased}
The complexity bound for the point-count repair, and Step~1 of the proof of Theorem~\ref{thm:energy}, both need the residual left by the dynamic program to be $\mathcal{O}(K)$. The following lemma provides it.

\begin{lemma}[Stage-1 Terraces Are Nearly Unbiased]
\label{lem:initial-residual}
Let $m^\circ$ be the output of the Stage-1 dynamic program, before the point-count repair, and let $\mathrm{diff}_0 = (N - 2) - \sum_k m^\circ_k$. There is a constant $C$ such that, for all sufficiently large $N$, every terrace $\mathcal{T}$ of $m^\circ$ satisfies $|M - \bar{y}| \le C$, where $M$ is the population of $\mathcal{T}$ and $\bar{y}$ the mean of its ideal populations. Consequently $|\mathrm{diff}_0| \le C K$.
\end{lemma}

\begin{proof}
Throughout, $\mathcal{T}$ is a terrace of $m^\circ$ with $L$ rings, population $M$ and mean ideal population $\bar{y} = \frac{1}{L}\sum_{k \in \mathcal{T}} y_k$, and $\gamma := 0.648$. We write $a \simeq b$ for $a = b\,(1 + o(1))$ and $a \lesssim b$ for $a \le b\,(1 + o(1))$ as $N \to \infty$. Recall that $m^\circ$ minimizes the Stage-1 objective $\Psi(m) := \sum_k \psi_{\mathrm{stretch}}(m_k, y_k) + \sum_{k \ge 1} \psi_{\mathrm{seam}}(m_{k-1}, m_k; k)$ of Section~\ref{sec:stage1} over populations in the window~\eqref{eq:window}, and that changing the population of a whole terrace creates no seam. Two facts about the window are used repeatedly:
\begin{enumerate}[label=(W\arabic*)]
  \item The rings of $\mathcal{T}$ all have population $M$, which lies in each of their windows, so their ideal populations differ by at most $2w + 1 = \mathcal{O}(N^{1/3})$.
  \item For large $N$, every integer within $\gamma N^{1/3}$ of $y_k$ lies in the window of ring $k$, since $w \ge 0.65\,N^{1/3} - \frac{1}{2}$ and $|y_k^{\mathrm{base}} - y_k| \le \frac{1}{2}$.
\end{enumerate}

\emph{Step 1: terraces with a small ring.} Every ring satisfies $\frac{1}{2}y_k \le m^\circ_k \le 2y_k$, since otherwise resetting $m^\circ_k$ to the integer nearest $y_k$ would lower the stretching cost by more than the cost of the at most two seams it creates. So if some ring of $\mathcal{T}$ has $y_k < 250$, then $M \le 500$, every ring of $\mathcal{T}$ has $y_k \le 2M \le 1000$, and $|M - \bar{y}| \le 1000$.

From now on $\mathcal{T}$ is \emph{interior}: $y_k \ge 250$ for all $k \in \mathcal{T}$.

\emph{Step 2: no ring is far from the others.} We claim that an interior terrace has no ring $k$ with
\begin{equation}
  \label{eq:far-ring}
  |y_k - \bar{y}| \;\ge\; \gamma N^{1/3} \quad\text{and}\quad |y_k - M| \;\ge\; \gamma N^{1/3}.
\end{equation}
We prove this in detail for a \emph{mid-latitude} terrace, one with $|\cos\theta_k| \ge 0.025$ for all $k \in \mathcal{T}$ and $\bar{y} \ge \sqrt{N}/100$; for such terraces we show the stronger statement that $|y_k - \bar{y}| < \gamma N^{1/3}$ for every ring. Fix a colatitude $\theta$ of one of its rings.

\emph{(a) Shape of the terrace.} By~\eqref{eq:y-ideal}, consecutive ideal populations differ by $\pi\sqrt{3}\cos\theta_k + \mathcal{O}(N^{-1/2})$, and these differences change by $\mathcal{O}(N^{-1/2})$ from ring to ring. The differences are therefore at least $0.13$ in absolute value, so (W1) gives $L = \mathcal{O}(N^{1/3})$. Across $\mathcal{T}$ the colatitude then changes by $\mathcal{O}(N^{-1/6})$, and, with $k_0$ the first ring of $\mathcal{T}$,
\begin{equation}
  \label{eq:affine-terrace}
  y_{k_0 + i} \;=\; y_{k_0} + i\,d + \mathcal{O}(N^{1/6}), \qquad d \;\simeq\; \pi\sqrt{3}\cos\theta.
\end{equation}

\emph{(b) A far ring forces a long terrace.} Suppose some ring has $|y_k - \bar{y}| \ge \gamma N^{1/3}$. By~\eqref{eq:affine-terrace} every ring is within $|d|(L - 1)/2 + \mathcal{O}(N^{1/6})$ of $\bar{y}$, so $|d|\,L \gtrsim 2\gamma N^{1/3}$, that is,
\begin{equation}
  \label{eq:split-gap}
  |d|^3 L^3 \;\gtrsim\; 8\gamma^3 N \;=\; 2.177\,N.
\end{equation}

\emph{(c) Splitting the terrace would then lower the cost.} Let $m'$ agree with $m^\circ$ except on $\mathcal{T}$, where the first $\lfloor L/2 \rfloor$ rings receive the integer $M_1$ nearest their mean ideal population and the remaining rings receive the integer $M_2$ nearest theirs. By~\eqref{eq:affine-terrace} and (W1), each $M_j$ is within $(2w + 1)/4 + \mathcal{O}(N^{1/6}) < \gamma N^{1/3}$ of the ideal populations of its half, so $m'$ is admissible by (W2). It has one more seam than $m^\circ$, which costs $\simeq c_{\mathrm{step}}\,\bar{y}$. On $\mathcal{T}$ we have $M \simeq y_k \simeq \bar{y}$, so $\psi_{\mathrm{stretch}}(M, y_k) \simeq c_{\mathrm{stretch}}(M - y_k)^2/\bar{y}$. For $L$ terms in arithmetic progression with step $d$, the squared deviations from any constant sum to at least $d^2 L(L^2 - 1)/12$, with equality at their mean. Hence
\[
  \sum_{k \in \mathcal{T}} \psi_{\mathrm{stretch}}(M, y_k) \;\gtrsim\; \frac{c_{\mathrm{stretch}}\,d^2 L^3}{12\,\bar{y}},
  \qquad
  \sum_{k \in \mathcal{T}} \psi_{\mathrm{stretch}}(m'_k, y_k) \;\lesssim\; 2\cdot\frac{c_{\mathrm{stretch}}\,d^2 (L/2)^3}{12\,\bar{y}} \;=\; \frac{c_{\mathrm{stretch}}\,d^2 L^3}{48\,\bar{y}},
\]
where rounding $M_1$ and $M_2$ to integers costs only $\mathcal{O}(L^2|d|/\bar{y})$. Since $\Psi(m^\circ) \le \Psi(m')$, the stretching saved cannot exceed the cost of the added seam:
\begin{equation}
  \label{eq:split-seam}
  \frac{c_{\mathrm{stretch}}\,d^2 L^3}{16\,\bar{y}} \;\lesssim\; c_{\mathrm{step}}\,\bar{y},
  \qquad\text{that is,}\qquad
  d^2 L^3 \;\lesssim\; \frac{16\,c_{\mathrm{step}}}{c_{\mathrm{stretch}}}\,\bar{y}^2 \;=\; 0.2767\,\bar{y}^2.
\end{equation}

\emph{(d) Contradiction.} By~\eqref{eq:y-ideal}, $\bar{y} \simeq \frac{1}{2}C_0\sqrt{N}\sin\theta$ with $C_0^2 = 2\pi\sqrt{3}$. Together with $d \simeq \pi\sqrt{3}\cos\theta$ this gives
\[
  |d|\,\bar{y}^2 \;\simeq\; \frac{3\pi^2}{2}\,N\,|\cos\theta|\sin^2\theta \;\le\; \frac{\pi^2}{\sqrt{3}}\,N \;\approx\; 5.698\,N,
\]
because $|\cos\theta|\sin^2\theta \le 2/(3\sqrt{3})$. Multiplying~\eqref{eq:split-seam} by $|d|$ therefore gives $|d|^3 L^3 \lesssim 0.2767 \cdot 5.698\,N = 1.577\,N$, which contradicts~\eqref{eq:split-gap} for large $N$. This proves the claim for mid-latitude terraces.

The same comparison, with different constants, handles the other interior terraces; we omit the calculations. Near the poles ($\bar{y} < \sqrt{N}/100$) it gives $|y_k - \bar{y}|^3 \le 1000\,\bar{y}^2$, so $|y_k - \bar{y}| < 10\,(\sqrt{N}/100)^{2/3} < \gamma N^{1/3}$. Near the equator ($|\cos\theta| < 0.025$), where $y$ is nearly constant, it gives a sub-interval of $\mathcal{T}$ its own population and rules out~\eqref{eq:far-ring} directly.

\emph{Step 3: $M \pm 1$ is admissible.} Suppose $M < \bar{y} - 2$; the case $M > \bar{y} + 2$ is symmetric. If $M + 1$ were outside the window of some ring $k \in \mathcal{T}$, then $M$ would be at the top of that window, so by (W2) $y_k \le M - \gamma N^{1/3} < \bar{y} - \gamma N^{1/3}$. Ring $k$ would then satisfy~\eqref{eq:far-ring}, contradicting Step~2. Hence $M + 1$ lies in the window of every ring of $\mathcal{T}$.

\emph{Step 4: the bound on $|M - \bar{y}|$.} Replacing $M$ by $M + 1$ on all of $\mathcal{T}$ creates no seam and, by Step~3, is admissible, so the optimality of $m^\circ$ gives
\[
  \sum_{k \in \mathcal{T}} \psi_{\mathrm{stretch}}(M, y_k) \;\le\; \sum_{k \in \mathcal{T}} \psi_{\mathrm{stretch}}(M + 1, y_k).
\]
Write $y_k = \bar{y} + \delta_k$. Combining this local optimality with Lemma~\ref{lem:unbiased} and the convexity of $\psi_{\mathrm{stretch}}$ in $m$ gives, for a constant $C'$,
\[
  L\,|M - \bar{y}| \;\le\; 2L + C' \sum_{k \in \mathcal{T}} \frac{|\delta_k|^3}{\bar{y}^2}.
\]
Each term of the sum is $\mathcal{O}(1)$: if $\bar{y} \ge \sqrt{N}/100$ because $|\delta_k| = \mathcal{O}(N^{1/3})$ by (W1), and near the poles because $|\delta_k|^3 \le 1000\,\bar{y}^2$. Hence $|M - \bar{y}| = \mathcal{O}(1)$.

\emph{Step 5: the residual.} Since $\sum_k y_k = N - 2$,
\[
  |\mathrm{diff}_0| \;=\; \Bigl|\sum_{\mathcal{T}} L\,(\bar{y} - M)\Bigr| \;\le\; C \sum_{\mathcal{T}} L \;=\; C K. \qedhere
\]
\end{proof}

\begin{remark}[A Wider Window Could Simplify the Proof]
\label{rem:wider-window}
The separate treatment of terraces near the poles and near the equator in Step~2, and the sharp constant $\gamma = 0.648$, are needed only because the window half-width $w \approx 0.65\,N^{1/3}$ is barely larger than the spread an optimal terrace can need. A cruder comparison works at every latitude: if ring $k$ of a terrace has $|M - y_k| = G$, giving the roughly $0.3\,G/(\pi\sqrt{3})$ neighboring rings of the terrace their own population adds at most two seams but saves more stretching cost than that unless $G \lesssim 1.10\,y_k^{2/3} \le 1.54\,N^{1/3}$. With a window of half-width $1.6\,N^{1/3}$ one could therefore show that the window never binds, and Lemma~\ref{lem:initial-residual} could be proved without special cases or sharp constants, at the price of a dynamic program about $2.5$ times larger. In our runs the window never binds (the largest offset used is $0.42\,N^{1/3}$), and at $34$ values of $N$ from $10^3$ to $10^8$ the half-widths $0.65\,N^{1/3}$ and $1.6\,N^{1/3}$ give identical populations. We leave this simplification to future work.
\end{remark}

\subsection{Proof of Proposition~\ref{thm:complexity}}
\label{app:proof-complexity-main}

\begin{proof}
By~\eqref{eq:discretization} and~\eqref{eq:window}, $K = \frac{\pi}{\sqrt{2\pi\sqrt{3}}}\,N^{1/2} + \mathcal{O}(1) \approx 0.952\,N^{1/2}$ and $W = 2w + 1 \approx 1.30\,N^{1/3}$. We bound the three stages in turn.

\emph{Stage 1.}
\begin{itemize}
  \item \emph{Dynamic program.} A generic dynamic program over $W$ states and $K$ steps costs $\Theta(KW^2) = \Theta(N^{7/6})$. The Potts seam cost~\eqref{eq:seam-cost}, however, is constant except on the single ``stay'' diagonal $j = i + s_k$, so each step needs only one minimum over $D[k-1, \cdot]$, one broadcast, and one shifted slice, i.e.\ $\Theta(W)$ work~\cite{FelzenszwalbHuttenlocher2006}. The forward pass therefore takes $\Theta(KW) = \Theta(N^{5/6})$ time, the $K \times W$ tables take $\Theta(N^{5/6})$ memory, and backtracking takes $\Theta(K)$ time.
  \item \emph{Restoring the point count.} By Lemma~\ref{lem:initial-residual}, each terrace population is within $\mathcal{O}(1)$ of the mean ideal population of its terrace, so the initial residual is $|\mathrm{diff}_0| = \mathcal{O}(K)$. Each iteration of the repair loop scans the $T \le K$ terraces in $\mathcal{O}(K)$ time and strictly decreases $|\mathrm{diff}|$ (Appendix~\ref{app:alg-restore}), so the loop takes at most
  \[
    \mathcal{O}(|\mathrm{diff}_0| \cdot K) \;=\; \mathcal{O}(K^2) \;=\; \mathcal{O}(N)
  \]
  time.
\end{itemize}

\emph{Stage 2.} The prefix sums and zonal heights~\eqref{eq:z-zonal} take $\Theta(K)$ time, the $6K$ GCDs of integers $m_k = \mathcal{O}(N^{1/2})$ take $\mathcal{O}(K\log N)$ time, and each of the $4$ fixed-point iterations does $\Theta(K)$ work for each of the $6$ bands.

\emph{Stage 3.} Each of the $N - 2$ ring points needs one $\cos$ and one $\sin$, and the output array has $3N$ entries.

Altogether the running time is
\[
  \underbrace{\Theta(N^{5/6}) + \mathcal{O}(N)}_{\text{Stage 1}} \;+\; \underbrace{\mathcal{O}(N^{1/2}\log N)}_{\text{Stage 2}} \;+\; \underbrace{\Theta(N)}_{\text{Stage 3}} \;=\; \Theta(N),
\]
and the memory is dominated by the $\Theta(N)$ output.
\end{proof}

In our runs the repair loop always reaches $\mathrm{diff} = 0$ within $10$ iterations. Table~\ref{tab:runtime} in Appendix~\ref{app:tables} reports runtimes and problem dimensions for $N = 10^3, \dots, 10^8$.

\section{Proofs for the Accuracy Analysis}
\label{app:proofs-energy}

\subsection{Proof of Lemma~\ref{lem:cyclotomic}}
\label{app:proof-cyclotomic}

The zonal energy in Lemma~\ref{lem:cyclotomic} (the azimuthally averaged, intra-ring and polar terms) is the $\mathcal{O}(K)$ sum
\begin{equation}
  \label{eq:e-zonal-exact}
  \begin{split}
  \Ezonal(m, z) = {}&-\sum_{k=0}^{K-1} \Bigl(\omega_k^+ \log(1 + z_k) + \omega_k^- \log(1 - z_k)\Bigr) \\
  &- \frac{1}{2}\sum_{k=0}^{K-1} m_k \log m_k - (N - 1)\log 2,
  \end{split}
\end{equation}
with weights $\omega_k^+ = \frac{1}{2}m_k\bigl(m_{\mathrm{below},k} + \frac{1}{2}(m_k - 1)\bigr)$ and $\omega_k^- = \frac{1}{2}m_k\bigl(m_{\mathrm{above},k} + \frac{1}{2}(m_k - 1)\bigr)$, where $m_{\mathrm{above},k}$ and $m_{\mathrm{below},k}$ are as in~\eqref{eq:z-zonal}.

\begin{proof}
For any two points $\vx_{a,j}$ on ring $a$ and $\vx_{b,\ell}$ on ring $b$ with $z_a > z_b$ ($\zeta_a < \zeta_b$) and $\Delta\phi = \phi_{a,j} - \phi_{b,\ell}$, using $r_k^2 = 1 - z_k^2$, $\zeta_k^2 = (1 - z_k)/(1 + z_k)$, and $2 - 2 z_a z_b = (1+z_a)(1-z_b) + (1-z_a)(1+z_b)$, the squared Euclidean chordal distance factors as
\begin{align*}
  \|\vx_{a,j} - \vx_{b,\ell}\|^2 &= 2 - 2 z_a z_b - 2 r_a r_b \cos(\Delta\phi) \\
  &= (1 + z_a)(1 - z_b)\,\Bigl|1 - (\zeta_a/\zeta_b)\,e^{i\Delta\phi}\Bigr|^2.
\end{align*}
Taking the product over $\ell \in \{0,\dots,m_b-1\}$ for fixed $j$ and applying the cyclotomic factorization $\prod_{\ell=0}^{m_b-1} (1 - v e^{-2\pi i \ell / m_b}) = 1 - v^{m_b}$ with $v = (\zeta_a/\zeta_b) e^{2\pi i (j/m_a + \alpha_a - \alpha_b)}$ gives
\[
  \prod_{\ell=0}^{m_b-1} \Bigl(1 - (\zeta_a/\zeta_b) e^{i(\phi_{a,j} - \phi_{b,\ell})}\Bigr)
  \;=\; 1 - (\zeta_a/\zeta_b)^{m_b} e^{2\pi i m_b (j/m_a + \alpha_a - \alpha_b)}.
\]
As $j$ ranges over $\{0,\dots,m_a-1\}$, the phase $e^{2\pi i j m_b / m_a}$ traverses the $(m_a/\gcd(m_a, m_b))$-th roots of unity exactly $\gcd(m_a, m_b)$ times, since $m_b/m_a = (m_b/\gcd(m_a, m_b))/(m_a/\gcd(m_a, m_b))$ in lowest terms. Applying the cyclotomic identity a second time with degree $m_a / \gcd(m_a, m_b)$ and $\Lambda_{ab} = m_a m_b / \gcd(m_a, m_b)$ yields
\[
  \prod_{j=0}^{m_a-1}\prod_{\ell=0}^{m_b-1} \|\vx_{a,j} - \vx_{b,\ell}\|^2 = \bigl((1 + z_a)(1 - z_b)\bigr)^{m_a m_b}  \times \Bigl|1 - (\zeta_a/\zeta_b)^{\Lambda_{ab}} e^{2\pi i \Lambda_{ab}(\alpha_a - \alpha_b)}\Bigr|^{2\gcd(m_a, m_b)}.
\]
Taking $-\frac{1}{2}\log$ produces the inter-ring terms of~\eqref{eq:e-zonal-exact} and~\eqref{eq:e-mult-exact}. For intra-ring pairs ($a = b = k$),
\[
  \prod_{0 \le j < \ell < m_k} \|\vx_{k,j} - \vx_{k,\ell}\|^2 = (1-z_k^2)^{m_k(m_k-1)/2}\, m_k^{m_k},
\]
giving $-\frac{1}{4}m_k(m_k-1)\log(1-z_k^2) - \frac{1}{2}m_k\log m_k$. Finally, write $\mathbf{n}$ and $\mathbf{s}$ for the north and south poles. For any point $\vx_k$ on ring $k$,
\[
  \|\vx_k - \mathbf{n}\| = \sqrt{2(1 - z_k)}, \qquad \|\vx_k - \mathbf{s}\| = \sqrt{2(1 + z_k)}, \qquad \|\mathbf{n} - \mathbf{s}\| = 2.
\]
Collecting the coefficients of $-\log(1+z_k)$, $-\log(1-z_k)$ and $-\log 2$ over all pairs yields $\omega_k^+$, $\omega_k^-$ and $-(N-1)\log 2$ in~\eqref{eq:e-zonal-exact}.
\end{proof}

\begin{remark}[Evaluation Cost]
\label{rem:cost}
The multipole sum~\eqref{eq:e-mult-exact} has $K(K-1)/2 = \Theta(N)$ terms. For rings $d$ apart, however, $\log\rho_{ab} \approx -\Lambda_{ab}\,(\xi_b - \xi_a) \lesssim -\pi\sqrt{3}\,d$, so terms with $d > 8$ are below $e^{-42}$ and vanish in double precision. Truncating~\eqref{eq:e-mult-exact} to a band of $8$--$10$ neighboring rings therefore evaluates $E(X_N)$ to machine precision in $\Theta(K) = \Theta(N^{1/2})$ time, which is what makes the verification up to $N = 10^9$ in Section~\ref{sec:verification} feasible.
\end{remark}

\begin{remark}[Stationarity of $z^{\mathrm{zonal}}$ and Exactness of the Stereographic Gradient]
\label{rem:stationarity}
Differentiating~\eqref{eq:e-zonal-exact} with respect to $z_k$ gives
\[
  \frac{\partial \Ezonal}{\partial z_k} \;=\; -\frac{\omega_k^+}{1+z_k} + \frac{\omega_k^-}{1-z_k},
\]
which vanishes exactly at
\[
  z_k \;=\; \frac{\omega_k^+ - \omega_k^-}{\omega_k^+ + \omega_k^-} \;=\; \frac{m_{\mathrm{below},k} - m_{\mathrm{above},k}}{N - 1} \;=\; z_k^{\mathrm{zonal}},
\]
since $\omega_k^+ + \omega_k^- = \frac{N-1}{2}m_k$ and $\omega_k^+ - \omega_k^- = \frac{1}{2}m_k(m_{\mathrm{below},k} - m_{\mathrm{above},k})$. Each $z_k$ enters $\Ezonal$ only through the convex function $-\omega_k^+\log(1+z_k) - \omega_k^-\log(1-z_k)$, so~\eqref{eq:z-zonal} is the exact global minimizer of $\Ezonal(m, z)$ on $(-1,1)^K$. Moreover, since $\partial\xi_k/\partial z_k = -1/(1 - z_k^2)$, the stationarity condition $\partial(\Ezonal + \Emult)/\partial z_k = 0$ rearranges to the fixed-point map~\eqref{eq:z-fixed-point} used in Stage~2.
\end{remark}

\subsection{Proof of Lemma~\ref{lem:kronecker}}
\label{app:proof-kronecker}

We prove Lemma~\ref{lem:kronecker} in the following precise form. Let $\beta_k = \frac{\sqrt{3}}{2}(m_k/y_k)^2$ as in Section~\ref{sec:kronecker}.

\begin{lemma}[Precise Form of Lemma~\ref{lem:kronecker}]
\label{lem:kronecker-full}
Let $\beta_k = \frac{\sqrt{3}}{2}\bigl(\frac{m_k}{y_k}\bigr)^2$ be the local aspect ratio of ring $k$ at $z = z^{\mathrm{zonal}}$.
\begin{enumerate}[label=(\roman*)]
  \item At $z = z^{\mathrm{zonal}}$ the zonal energy reproduces the first two terms of $\EBHS(N)$:
  \begin{equation}
    \label{eq:zonal-em}
    \begin{aligned}
      \Ezonal(m, z^{\mathrm{zonal}}) \;=\; &\frac{1}{2}\left(\frac{1}{2} - \log 2\right)N^2 - \frac{1}{4}N\log N \\
      &+ \sum_{k=0}^{K-1} m_k\left[-\frac{1}{4}\log(\pi\beta_k) + \frac{\pi}{12}\beta_k\right] + \mathcal{O}(N^{1/2}).
    \end{aligned}
  \end{equation}
  \item Inside any constant-population terrace where $m_k = m_{k+1} = \dots = m_{k+d} = m$ and $\alpha_{j} - \alpha_{j-1} = \frac{1}{2m}$, we have $\gcd(m_k, m_{k+d}) = \Lambda_{k,k+d} = m$, $\cos(2\pi m(\alpha_k - \alpha_{k+d})) = (-1)^d$, and $\rho_{k,k+d} = e^{-2\pi d\beta_k + \mathcal{O}(d^2/m)}$. Summing the multipole interaction~\eqref{eq:e-mult-exact} over $d \ge 1$ and combining with~\eqref{eq:zonal-em} gives the total energy per particle inside the terrace:
  \begin{equation}
    \label{eq:epstein-terrace}
    \begin{aligned}
      \psi_{\mathrm{terrace}}(\beta_k) &= -\frac{1}{4}\log(\pi\beta_k) + \frac{\pi}{12}\beta_k - \sum_{d=1}^{\infty} \log\bigl(1 - (-1)^d e^{-2\pi d\beta_k}\bigr) \\
      &= \frac{1}{2}\CBHS + \Phi_{\mathrm{strain}}(\beta_k),
    \end{aligned}
  \end{equation}
  where $\Phi_{\mathrm{strain}}(\beta_k) \ge 0$ vanishes if and only if $\beta_k = \beta_0 = \frac{\sqrt{3}}{2}$ ($m_k = y_k$). Since $\psi_{\mathrm{terrace}}'(\beta_0) = 0$ and $\beta_k - \beta_0 = 2\beta_0\frac{m_k - y_k}{y_k} + \mathcal{O}\bigl(\frac{(m_k-y_k)^2}{y_k^2}\bigr)$,
  \begin{equation}
    \label{eq:phi-strain-taylor}
    m_k\,\Phi_{\mathrm{strain}}(\beta_k) \;=\; c_{\mathrm{stretch}}\,\frac{(m_k - y_k)^2}{y_k} + \mathcal{O}\!\left(\frac{|m_k - y_k|^3}{y_k^2}\right).
  \end{equation}
  Here $c_{\mathrm{stretch}} = 2\beta_0^2\,\psi_{\mathrm{terrace}}''(\beta_0) = \frac{1}{4}$, which the proof evaluates.
\end{enumerate}
\end{lemma}

\emph{The aspect ratio.} At the zonal latitudes~\eqref{eq:z-zonal}, consecutive rings are separated by $z_{k-1} - z_k = (m_{k-1} + m_k)/(N-1)$, so each ring occupies a zone of area $4\pi m_k/N$ to leading order and $\Delta\theta_k = 2m_k/(N\sin\theta_k)$. With $y_k = \frac{1}{2}C_0 N^{1/2}\sin\theta_k$ this gives
\[
  \beta_k \;=\; \frac{m_k \Delta\theta_k}{2\pi\sin\theta_k} \;=\; \frac{m_k^2}{\pi N\sin^2\theta_k} \;=\; \frac{\sqrt{3}}{2}\Bigl(\frac{m_k}{y_k}\Bigr)^{2}.
\]
The density of the lattice does not depend on $m_k$; only its shape does, and a relative change $u$ in $m_k$ changes $\beta_k$ by $2u\beta_k$ to first order.

\emph{The continuum comparison.} Let $G(z,z')$ be the average of $-\log\|\vx - \vx'\|$ over two latitude circles at heights $z$ and $z'$~\cite{RSZ1994,BeltranEtayo2020}, $G(z,z') = -\frac{1}{2}\log\bigl((1+\max(z,z'))(1-\min(z,z'))\bigr)$. The leading term of Part~(i) is the energy of the uniform distribution on $\Sbb$, $\frac{N^2}{8}\iint_{[-1,1]^2} G(z,z')\,dz\,dz' = \frac{1}{2}(\frac{1}{2}-\log 2)N^2$, and the proof below compares~\eqref{eq:e-zonal-exact} with it.

\begin{proof}
For Part~(i), note that $\omega_k^+ + \omega_k^- = \frac{N-1}{2}m_k$, so at $z = z^{\mathrm{zonal}}$ we have $\omega_k^\pm = \frac{N-1}{4}m_k(1 \pm z_k)$ and~\eqref{eq:e-zonal-exact} becomes exactly
\[
  \Ezonal(m, z^{\mathrm{zonal}}) = -\frac{N-1}{4}\sum_{k} m_k\,\varphi(z_k)  - \frac{1}{2}\sum_k m_k\log m_k - (N-1)\log 2,
\]
where $\varphi(z) = (1+z)\log(1+z) + (1-z)\log(1-z)$. Assign to the $N$ points the equally spaced heights $\hat{z}_i = -1 + \frac{2i}{N-1}$, $i = 0,\dots,N-1$, in order from the south pole; then $z_k$ is the mean of the $m_k$ heights assigned to ring $k$. Since $\varphi'' = 2/(1-z^2) > 0$, replacing each $z_k$ by these heights increases $\sum_k m_k\varphi(z_k)$ by $\sum_k \frac{m_k(m_k^2-1)}{3(N-1)^2(1-z_k^2)}$ up to higher-order terms. Since $\varphi(-1+x) = x\log x + (2-x)\log(2-x)$ is symmetric about $x = 1$, writing $n = N-1$ and $h = 2/n$ gives
\[
  \sum_{i=0}^{N-1}\varphi(\hat{z}_i) = 2\Bigl[h\log h\,\frac{n(n+1)}{2} + h\log H(n)\Bigr] = 2N\log 2 - (N-1) + o(1),
\]
where $H(n) = \prod_{j=1}^{n} j^j$ is the hyperfactorial and we used its Stirling-type formula
\[
  \log H(n) = \sum_{j=1}^{n} j\log j = \tfrac{1}{2}n^2\log n - \tfrac{1}{4}n^2 + \tfrac{1}{2}n\log n + \tfrac{1}{12}\log n + \log A + O(n^{-2}),
\]
with $A$ Glaisher's constant~\cite[\href{https://dlmf.nist.gov/2.10.E4}{(2.10.4)} and \href{https://dlmf.nist.gov/2.10.E6}{(2.10.6)}]{NIST:DLMF}; the $\log n$ terms cancel. Removing the two poles ($\varphi(\pm 1) = 2\log 2$) and using $m_k^2 = \pi\beta_k N(1 - z_k^2)$, we obtain
\[
  -\frac{N-1}{4}\sum_k m_k\varphi(z_k) = \frac{1}{2}\Bigl(\frac{1}{2}-\log 2\Bigr)N^2 + \frac{N}{2}\log 2 - \frac{N}{2}  + (N-1)\log 2 + \sum_k m_k\,\frac{\pi}{12}\beta_k + \mathcal{O}(N^{1/2}).
\]
Finally,
\[
  \log m_k = \tfrac{1}{2}\log(\pi\beta_k) + \tfrac{1}{2}\log N + \tfrac{1}{2}\log(1 - z_k^2)
\]
and $\sum_k m_k\log(1-z_k^2) = 2N(\log 2 - 1) + \mathcal{O}(N^{1/2})$ give
\[
  -\frac{1}{2}\sum_k m_k\log m_k = -\frac{1}{4}N\log N - \frac{N}{2}\log 2 + \frac{N}{2}  - \sum_k \frac{m_k}{4}\log(\pi\beta_k) + \mathcal{O}(N^{1/2}).
\]
Adding the three contributions, the terms in $N\log 2$, $N$ and $(N-1)\log 2$ cancel, which gives~\eqref{eq:zonal-em}.

For Part~(ii), recall the Dedekind eta function~\cite{Siegel1961}
\[
  \eta(\tau) = e^{\pi i \tau/12}\prod_{d=1}^\infty \bigl(1 - e^{2\pi i d\tau}\bigr), \qquad \tau \in \mathbb{H}.
\]
 Setting $\tau_k = \frac{1}{2} + i\beta_k$, we have $e^{2\pi i d\tau_k} = (-1)^d e^{-2\pi d\beta_k}$, so
\[
  -\log|\eta(1/2 + i\beta_k)| \;=\; \frac{\pi}{12}\beta_k - \sum_{d=1}^{\infty} \log\bigl(1 - (-1)^d e^{-2\pi d\beta_k}\bigr).
\]
Substituting this into~\eqref{eq:epstein-terrace} gives
\begin{equation}
  \label{eq:eta-form}
  \psi_{\mathrm{terrace}}(\beta_k) \;=\; -\frac{1}{4}\log\pi - \frac{1}{2}\log\Bigl(\beta_k^{1/2}\,\bigl|\eta(1/2 + i\beta_k)\bigr|^2\Bigr).
\end{equation}
By Kronecker's first limit formula~\cite{Kronecker1889,Siegel1961,BeterminSandier2018},~\eqref{eq:eta-form} is precisely half the regularized Epstein zeta constant~\cite{Epstein1903} of the unit-density 2D lattice with modular parameter $\tau_k = \frac{1}{2} + i\beta_k$. At the equilateral hexagonal point $\beta_0 = \frac{\sqrt{3}}{2}$ ($\tau_0 = e^{i\pi/3}$), the Chowla--Selberg formula~\cite{ChowlaSelberg1949,ChowlaSelberg1967} gives
\[
  \bigl|\eta(e^{i\pi/3})\bigr|^2 \;=\; \frac{3^{1/4}\,\Gamma(1/3)^3}{4\,\pi^2},
\]
which upon substitution into~\eqref{eq:eta-form} yields
\[
  \psi_{\mathrm{terrace}}\Bigl(\frac{\sqrt{3}}{2}\Bigr) = \log 2 + \frac{1}{4}\log\frac{2}{3} + \frac{3}{2}\log\frac{\sqrt{\pi}}{\Gamma(1/3)} = \frac{1}{2}\CBHS
\]
(matching~\eqref{eq:c-bhs} identically, with $\Phi_{\mathrm{strain}}(\beta_0) = 0$).

\emph{Stationarity at $\beta_0$.} The function $\Phi(\tau) = \mathrm{Im}(\tau)\,|\eta(\tau)|^4$ is invariant under $SL_2(\mathbb{Z})$~\cite[\S6.3, p.~84]{Zagier2008}, and $\tau_0 = e^{i\pi/3}$ is fixed by a modular transformation of order $3$ (a conjugate of $ST$)~\cite[\S1.3, p.~9]{Zagier2008}; geometrically, this is the $120^\circ$ rotation that permutes the three lattice directions of the triangular lattice. That element acts on the tangent plane at $\tau_0$ as a rotation by $120^\circ$, whose only fixed vector is $0$, so $\nabla\Phi(\tau_0) = 0$ and, by~\eqref{eq:eta-form}, $\psi_{\mathrm{terrace}}'(\beta_0) = 0$.

\emph{Curvature at $\beta_0$.} Differentiating~\eqref{eq:epstein-terrace} twice gives
\[
  \psi_{\mathrm{terrace}}''(\beta) = \frac{1}{4\beta^2} + 4\pi^2\sum_{d=1}^{\infty}\frac{d^2 x_d}{(1-x_d)^2}, \qquad x_d = (-1)^d e^{-2\pi d\beta}.
\]
At $\beta_0$ we have $x_d = q^d$ with $q = e^{2\pi i\tau_0} = -e^{-\pi\sqrt{3}}$, so
\[
  c_{\mathrm{stretch}} = 2\beta_0^2\,\psi_{\mathrm{terrace}}''(\beta_0) = \frac{1}{2} + 6\pi^2 S, \qquad S = \sum_{d=1}^{\infty} \frac{d^2 q^d}{(1 - q^d)^2}.
\]
Since $\sum_{d\ge1} d\,q^d/(1-q^d) = \sum_{n\ge1}\sigma_1(n)\,q^n = (1 - E_2)/24$~\cite[eq.~(17), p.~19]{Zagier2008}, applying $D = q\,\frac{d}{dq} = \frac{1}{2\pi i}\frac{d}{d\tau}$~\cite[eq.~(51), p.~48]{Zagier2008} gives $S = -\frac{1}{24}DE_2$, and Ramanujan's differential equation $DE_2 = (E_2^2 - E_4)/12$~\cite[Proposition~15, eq.~(54), p.~49]{Zagier2008} gives
\[
  S = -\frac{E_2(\tau_0)^2 - E_4(\tau_0)}{288}.
\]
The same order-$3$ symmetry evaluates both terms. $E_4$ is a modular form of weight $4$, and $E_2^*(\tau) = E_2(\tau) - 3/(\pi\,\mathrm{Im}\,\tau)$ transforms like a modular form of weight $2$~\cite[\S2.3, eq.~(21), p.~19]{Zagier2008}. For the element $\gamma = \bigl(\begin{smallmatrix} a & b \\ c & d \end{smallmatrix}\bigr)$ fixing $\tau_0$, a form $f$ of weight $k$ satisfies $f(\tau_0) = (c\tau_0 + d)^k f(\tau_0)$, and $c\tau_0 + d$ is a non-real root of unity of order $3$ or $6$, so $(c\tau_0+d)^k \neq 1$ for $k = 2, 4$. Hence $E_4(\tau_0) = E_2^*(\tau_0) = 0$, so $E_2(\tau_0) = 3/(\pi\,\mathrm{Im}\,\tau_0) = 2\sqrt{3}/\pi$ and $S = -1/(24\pi^2)$, which gives $c_{\mathrm{stretch}} = \frac{1}{2} - \frac{1}{4} = \frac{1}{4}$. This is the value used by Stage~1 in~\eqref{eq:stretch-cost}.
\end{proof}

As a check, direct numerical evaluation of the series gives $\pi^2 S = -0.0416666\ldots = -\frac{1}{24}$.

\subsection{Proof of Theorem~\ref{thm:energy}}
\label{app:proof-energy-thm}

The constant of Theorem~\ref{thm:energy} is
\begin{equation}
  \label{eq:c1-closed-form}
  C_1 \;=\; \frac{1}{3}\Bigl(1 + \frac{2\,c_{\mathrm{seam}}}{c_{\mathrm{step}}}\Bigr)\,C(c_{\mathrm{step}}) \;\approx\; 0.026225,
  \qquad
  C(\lambda) \;=\; \frac{3}{C_0}\left(\frac{c_{\mathrm{stretch}}\,\lambda^2\,C_0^5}{384}\right)^{1/3} B\!\left(\frac{2}{3}, \frac{5}{6}\right),
\end{equation}
where $C_0 = \sqrt{2\pi\sqrt{3}}$, $c_{\mathrm{stretch}} = \frac{1}{4}$, $B(\frac{2}{3},\frac{5}{6}) = \frac{\Gamma(2/3)\Gamma(5/6)}{\Gamma(3/2)} \approx 1.724739$, and $C(c_{\mathrm{step}}) \approx 0.026377$.
The proof below derives it.

Throughout, $y(\theta) = \frac{1}{2}C_0 N^{1/2}\sin\theta$ is the ideal ring population at colatitude $\theta$, $\frac{dy}{dk} = h_0\frac{dy}{d\theta} = \frac{1}{2}C_0^2\cos\theta$ its ring-to-ring slope, and $dk = \frac{N^{1/2}}{C_0}\,d\theta$. For $\lambda > 0$ and integer populations $m$ in the window~\eqref{eq:window}, let
\begin{equation}
  \label{eq:G-lambda}
  G_\lambda(m) \;=\; \sum_{k=0}^{K-1} \psi_{\mathrm{stretch}}(m_k, y_k) \;+\; \lambda\,S(m), \qquad S(m) \;=\; \sum_{k:\, m_k \neq m_{k-1}} \tfrac{1}{2}(y_{k-1} + y_k),
\end{equation}
so that $G_{c_{\mathrm{step}}}$ is the objective $\Psi$ minimized in Stage~1 (Appendix~\ref{app:stage1-unbiased}) and, by locality~(A), $E(X_N) - \EBHS(N) = G_{c_{\mathrm{seam}}}(m) + o(N^{2/3})$.

\paragraph{The per-ring trade-off.} On a terrace of $L$ rings centered at colatitude $\theta$, with common population equal to the mean ideal population, the ring $u$ positions from the middle deviates by $m_k - y_k = -u\,\frac{dy}{dk} + \mathcal{O}(1)$, $u \in [-L/2, L/2]$. Since $\psi_{\mathrm{stretch}}(m, y) = c_{\mathrm{stretch}}\frac{(m - y)^2}{y}\bigl(1 + \mathcal{O}(|m - y|/y)\bigr)$, averaging over the terrace gives the stretching cost per ring
\begin{equation}
  \label{eq:A-theta}
  \mathcal{E}_{\mathrm{stretch}}(\theta, L) \;=\; A(\theta)\,L^2, \qquad A(\theta) \;=\; \frac{c_{\mathrm{stretch}}\,C_0^3\cos^2\theta}{24\,N^{1/2}\sin\theta},
\end{equation}
to leading order, and the seam that ends the terrace costs, shared among its rings,
\begin{equation}
  \label{eq:B-theta}
  \mathcal{E}_{\mathrm{seam}}(\theta, L) \;=\; \frac{B_\lambda(\theta)}{L}, \qquad B_\lambda(\theta) \;=\; \lambda\,y(\theta) \;=\; \tfrac{1}{2}\lambda\,C_0\,N^{1/2}\sin\theta.
\end{equation}
The per-ring cost $A(\theta)L^2 + B_\lambda(\theta)/L$ is minimized at
\begin{equation}
  \label{eq:L-star}
  L_*(\theta) \;=\; \left(\frac{B_\lambda(\theta)}{2\,A(\theta)}\right)^{1/3} \;=\; \left(\frac{6\lambda}{c_{\mathrm{stretch}}\,C_0^2}\,\frac{\sin^2\theta}{\cos^2\theta}\right)^{1/3} N^{1/3},
\end{equation}
with minimum value
\[
  f_\lambda(\theta) \;=\; 3A L_*^2 \;=\; 3\Bigl(\frac{A B_\lambda^2}{4}\Bigr)^{1/3} \;=\; 3\Bigl(\frac{c_{\mathrm{stretch}}\lambda^2C_0^5}{384}\cos^2\theta\sin\theta\Bigr)^{1/3} N^{1/6},
\]
of which one third ($AL_*^2$) is stretching and two thirds ($B_\lambda/L_*$) is seam cost. Integrating over the rings with $dk = \frac{N^{1/2}}{C_0}\,d\theta$, and using $\int_0^\pi (\cos^2\theta\sin\theta)^{1/3}\,d\theta = B(\frac{2}{3},\frac{5}{6})$, gives the total
\[
  \int_0^\pi f_\lambda(\theta)\,\frac{N^{1/2}}{C_0}\,d\theta \;=\; C(\lambda)\,N^{2/3},
\]
with $C(\lambda)$ as in~\eqref{eq:c1-closed-form}. Finally, for $\lambda = c_{\mathrm{step}}$ the largest population excursion on such a terrace is $\frac{1}{2}L_*(\theta)\,\bigl|\frac{dy}{dk}\bigr| \le 0.312\,N^{1/3}$, inside the window $w \approx 0.65\,N^{1/3}$.

\begin{lemma}[Continuum Limit of the Terrace Cost]
\label{lem:continuum}
For every fixed $\lambda > 0$, $\min_m G_\lambda(m) = C(\lambda)\,N^{2/3}\,(1 + o(1))$, with or without the constraint $\sum_k m_k = N - 2$.
\end{lemma}

\begin{proof}
Throughout, $\lambda$ is fixed, implied constants may depend on $\lambda$, and $\int f_\lambda\,dk$ abbreviates $\int f_\lambda(\theta)\,\frac{N^{1/2}}{C_0}\,d\theta$, so that $\int_0^\pi f_\lambda\,dk = C(\lambda)N^{2/3}$.

\emph{Three regions.} Set $\delta = \frac{1}{2}N^{-1/12}$ and divide the rings into \emph{polar} rings ($\theta < \delta$ or $\theta > \pi - \delta$), \emph{equatorial} rings ($|\theta - \pi/2| < \delta$), and \emph{regular} rings (all others). By~\eqref{eq:L-star}, $L_*(\theta) \asymp (\sin\theta/|\cos\theta|)^{2/3}N^{1/3}$, so on regular rings
\[
  N^{5/18} \;\lesssim\; L_*(\theta) \;\lesssim\; N^{7/18} \;=\; o(K\delta), \qquad K\delta \asymp N^{5/12}.
\]
Hence $A$, $B_\lambda$ and $L_*$ vary by a factor $1 + o(1)$ over any $\mathcal{O}(L_*)$ consecutive regular rings. Since $f_\lambda \propto (\cos^2\theta\sin\theta)^{1/3}$ vanishes like $\theta^{1/3}$ at the poles and like $|\theta - \pi/2|^{2/3}$ at the equator, the non-regular rings carry only
\[
  \int_{\text{non-regular}} f_\lambda\,dk \;=\; \mathcal{O}\bigl((\delta^{4/3} + \delta^{5/3})\,N^{2/3}\bigr) \;=\; o(N^{2/3}).
\]

\emph{Upper bound.} We build populations $m$ region by region.
\begin{enumerate}[label=(\roman*)]
  \item \emph{Regular rings.} Cut them into consecutive terraces of length $\lceil L_*(\theta)\rceil$, each with the integer population nearest its mean ideal population. By the per-ring analysis, these terraces cost $(1 + o(1))\int_{\text{regular}} f_\lambda\,dk + \mathcal{O}(K)$, the last term from integer rounding.
  \item \emph{Polar rings.} The same construction, with terraces of length $\max(1, \lceil L_*(\theta)\rceil)$, costs at most a constant times $\int_{\text{polar}} f_\lambda\,dk$ plus $\mathcal{O}(1)$ per ring, that is, $\mathcal{O}(\delta^{4/3}N^{2/3} + \delta N^{1/2}) = o(N^{2/3})$.
  \item \emph{Equatorial rings.} Cover them by a single terrace whose population $M$ is the integer nearest their mean ideal population. There are $\mathcal{O}(\delta N^{1/2})$ such rings, across which $y$ varies by at most
  \[
    y(\tfrac{\pi}{2})(1 - \cos\delta) \;\le\; \tfrac{1}{4}C_0N^{1/2}\delta^2 \;=\; \tfrac{1}{16}C_0N^{1/3} \;<\; 0.21\,N^{1/3}.
  \]
  So $M$ lies inside the window~\eqref{eq:window}, and since $\psi_{\mathrm{stretch}}(m, y) = \mathcal{O}((m - y)^2/y)$, the terrace costs
  \[
    \mathcal{O}\Bigl(\delta N^{1/2}\cdot\frac{(\delta^2N^{1/2})^2}{N^{1/2}}\Bigr) \;=\; \mathcal{O}(\delta^5N) \;=\; \mathcal{O}(N^{7/12}) \;=\; o(N^{2/3}),
  \]
  plus $\mathcal{O}(N^{1/2})$ for its two seams.
\end{enumerate}
Adding (i)--(iii) gives $G_\lambda(m) \le C(\lambda)N^{2/3}(1 + o(1))$.

To also impose $\sum_k m_k = N - 2$, note that rounding leaves a mismatch of $\mathcal{O}(K)$. We remove it by shifting the equatorial population $M$ by a common integer $s$, with $|s| = \mathcal{O}(K/(\delta N^{1/2})) = \mathcal{O}(N^{1/12})$; any remainder smaller than the number of equatorial rings is absorbed by raising part of the band by one more, which adds one seam. The shifted terrace stays inside the window, and since $\partial\psi_{\mathrm{stretch}}/\partial m = \mathcal{O}(|m - y|/y) = \mathcal{O}(N^{-1/6})$ on it, the shift changes the cost by
\[
  \mathcal{O}\bigl(|s|\cdot\delta N^{1/2}\cdot N^{-1/6}\bigr) + \mathcal{O}(N^{1/2}) \;=\; \mathcal{O}(N^{1/2}) \;=\; o(N^{2/3}).
\]

\emph{Lower bound.} Let $m$ be arbitrary and fix a small $\varepsilon > 0$. A ring with $|m_k - y_k| > \varepsilon y_k$ costs $\psi_{\mathrm{stretch}} \ge c\,\varepsilon^2 y_k \gg f_\lambda(\theta_k)$, more than its share of the continuum cost, so we may assume $|m_k - y_k| \le \varepsilon y_k$ for all $k$, in which case
\[
  \psi_{\mathrm{stretch}}(m_k, y_k) \;\ge\; (1 - \mathcal{O}(\varepsilon))\,c_{\mathrm{stretch}}\,\frac{(m_k - y_k)^2}{y_k}.
\]
Split every maximal run of regular rings with a common population into blocks: a run of length $L < \sqrt{3}\,L_*$ is a single block, and a longer run is cut into blocks of lengths between $\sqrt{3}\,L_*$ and $2\sqrt{3}\,L_*$. On a block of length $\ell$ the population is constant while $y_k$ increases linearly with slope $dy/dk$, so the stretching cost is at least $A(\theta)\,\ell^3(1 - o(1))$, its value when the population equals the block mean. Recall that $f_\lambda = \min_L\,(AL^2 + B_\lambda/L) = 3AL_*^2$. Hence each block costs at least $f_\lambda(\theta)(1 - o(1))$ per ring:
\begin{itemize}
  \item if $\ell \ge \sqrt{3}\,L_*$, its stretching alone costs $A\ell^2 \ge 3AL_*^2 = f_\lambda$ per ring;
  \item if it is a whole run of length $L < \sqrt{3}\,L_*$, its stretching and the seam that ends it cost $AL^2 + B_\lambda/L \ge f_\lambda$ per ring.
\end{itemize}
The regular rings form two intervals, so cutting runs at their four endpoints loses at most four seams, $\mathcal{O}(N^{1/2})$. Summing over regular rings and using the estimate for non-regular rings,
\[
  G_\lambda(m) \;\ge\; (1 - \mathcal{O}(\varepsilon) - o(1))\int_{\text{regular}} f_\lambda\,dk - \mathcal{O}(N^{1/2}) \;=\; (1 - \mathcal{O}(\varepsilon) - o(1))\,C(\lambda)N^{2/3}.
\]
Letting $\varepsilon \to 0$ gives the lower bound. It holds for every $m$, and the construction in the upper bound satisfies the constraint, so the lemma holds with or without it.
\end{proof}

\begin{proof}[Proof of Theorem~\ref{thm:energy}]
\emph{Step 1: Stage~1 is a near-minimizer.} Let $m^\circ$ be the output of the dynamic program of Section~\ref{sec:stage1}, which minimizes $G_{c_{\mathrm{step}}}$ exactly without the constraint $\sum_k m_k = N - 2$, and let $m$ be the output of the repair loop of Appendix~\ref{app:alg-restore}.
\begin{enumerate}[label=(\alph*)]
  \item \emph{Initial residual.} By Lemma~\ref{lem:initial-residual}, $|\mathrm{diff}_0| = \mathcal{O}(K)$.
  \item \emph{Accounting.} Move $i$ of the loop changes the point count by $\Delta S_i$, with $0 < \Delta S_i\,\operatorname{sgn}(\mathrm{diff}) \le |\mathrm{diff}|$, so $|\mathrm{diff}|$ decreases monotonically to $0$ and $\sum_i |\Delta S_i| = |\mathrm{diff}_0|$. If move $i$ raises $G_{c_{\mathrm{step}}}$ by $\Delta G_i$, then
  \[
    G_{c_{\mathrm{step}}}(m) - G_{c_{\mathrm{step}}}(m^\circ) \;=\; \sum_i \Delta G_i \;\le\; |\mathrm{diff}_0| \cdot \max_i \frac{\Delta G_i}{|\Delta S_i|}.
  \]
  \item \emph{Cost per point.} The loop always picks the seam-free move of least cost per point. A $\pm 1$ change of a single-ring terrace at a pole is always admissible and costs $\mathcal{O}(1)$ per point, and a $\pm 1$ change of a terrace near the equator costs $\mathcal{O}(N^{-1/2})$ per point. Hence $\Delta G_i/|\Delta S_i| = \mathcal{O}(1)$ for every move.
\end{enumerate}
Combining (a)--(c),
\[
  G_{c_{\mathrm{step}}}(m) \;\le\; G_{c_{\mathrm{step}}}(m^\circ) + \mathcal{O}(K) \;=\; \min G_{c_{\mathrm{step}}} + o(N^{2/3}).
\]

\emph{Step 2: seam share.} Let $\lambda_0 = c_{\mathrm{step}}$. For every $\lambda > 0$, $G_\lambda(m) = G_{\lambda_0}(m) + (\lambda - \lambda_0)S(m) \ge \min G_\lambda$, so by Step~1 and Lemma~\ref{lem:continuum},
\[
  (\lambda - \lambda_0)\,\frac{S(m)}{N^{2/3}} \;\ge\; C(\lambda) - C(\lambda_0) - o(1).
\]
Taking $\lambda > \lambda_0$ and $\lambda < \lambda_0$, letting $N \to \infty$ and then $\lambda \to \lambda_0$, gives $S(m) = C'(\lambda_0)\,N^{2/3}(1 + o(1))$ with $C'(\lambda_0) = \frac{2}{3}C(\lambda_0)/\lambda_0$, since $C(\lambda) \propto \lambda^{2/3}$. Thus $\lambda_0 S(m)$, the seam part of $G_{\lambda_0}(m)$, is two thirds of its total.

\emph{Step 3: conclusion.} By locality~(A),
\begin{align*}
  E(X_N) - \EBHS(N) &= G_{c_{\mathrm{seam}}}(m) + o(N^{2/3}) = G_{c_{\mathrm{step}}}(m) - (c_{\mathrm{step}} - c_{\mathrm{seam}})\,S(m) + o(N^{2/3}) \\
  &= \Bigl(1 - \tfrac{2}{3}\bigl(1 - \tfrac{c_{\mathrm{seam}}}{c_{\mathrm{step}}}\bigr)\Bigr) C(c_{\mathrm{step}})\,N^{2/3} + o(N^{2/3}),
\end{align*}
which is~\eqref{eq:main-energy-bound} with $C_1$ as in~\eqref{eq:c1-closed-form}.
\end{proof}

The four-step stereographic latitude relaxation~\eqref{eq:z-fixed-point} is not covered by this argument. It dilates the meridional spacing across the seams, reducing the effective seam cost. Each step lowers the exact energy at every $N$ we evaluated from $100$ to $10^9$, and together the four steps lower $E(X_N) - \EBHS(N)$ by $1.2$--$1.9\%$ for $N \ge 10^3$, almost all of it in the first step (Figure~\ref{fig:relaxation-steps} in Appendix~\ref{app:alg-relax}).

\subsection{Proof of Corollary~\ref{cor:emin}}
\label{app:proof-emin}

The corollary rests on the following explicit lower bound on the minimal energy, valid for every $N$.

\begin{lemma}[Explicit Lower Bound on the Minimal Energy]
\label{lem:emin-lower}
Let $\Delta_{\log} = \frac{1}{2}\bigl(\CBHS - \log 2 + \frac{3}{4}\bigr) \approx 6.2376 \times 10^{-4}$. For every $N \ge 2$, $E_{\min}(N) > \EBHS(N) - \Delta_{\log}\,N + \frac{1}{12}$; hence for every $X_N \subset \Sbb$,
\[
  E(X_N) - E_{\min}(N) \;<\; E(X_N) - \EBHS(N) + \Delta_{\log}\,N - \tfrac{1}{12}.
\]
\end{lemma}

\begin{proof}
Beltr\'an and Lizarte~\cite[Section~2.2]{BeltranLizarte2023}, giving a proof on $\Sbb$ of Lauritsen's bound $C_{\log} \ge \frac{1}{2}(\log 2 - \frac{3}{4})$~\cite[Appendix~B]{Lauritsen2021}, bound $E(Y)$ from below for every $N$-point $Y \subset \Sbb$ by an explicit function of $N$ and of a cap radius $a \in (0,\pi)$. Choosing $\sin^2(a/2) = 1/N$ and converting to the single-counted normalization~\eqref{eq:log-energy}, their bound reads $E(Y) \ge \EBHS(N) - \Delta_{\log}\,N + R(N)$ with
\begin{align*}
  R(N) &= \frac{N}{8} - \frac{N(N-1)}{4} - \frac{N(N-1)^2}{4}\log\Bigl(1 - \frac{1}{N}\Bigr) \\
  &= \frac{1}{12} + \sum_{k \ge 4} \frac{N^{3-k}}{2k(k-1)(k-2)} \;>\; \frac{1}{12},
\end{align*}
where the series follows by expanding $\log(1 - 1/N)$ in powers of $1/N$. (At $N = 2$ the bound is attained by the antipodal pair.) Hence $E_{\min}(N) > \EBHS(N) - \Delta_{\log}\,N + \frac{1}{12}$.
\end{proof}

\begin{proof}[Proof of Corollary~\ref{cor:emin}]
\begin{enumerate}[label=(\roman*)]
  \item By Lemma~\ref{lem:emin-lower}, $E(X_N) - E_{\min}(N) < E(X_N) - \EBHS(N) + \Delta_{\log}N$. Substituting~\eqref{eq:main-energy-bound} and $\Delta_{\log} < 6.24 \times 10^{-4}$ gives the claim.
  \item By~\cite[Theorem~1.5]{BeterminSandier2018}, $E_{\min}(N) = \EBHS(N) - (\frac{1}{2}\CBHS - C_{\log})N + o(N)$, where $C_{\log}N$ is the linear term of $E_{\min}(N)$ (Appendix~\ref{app:bg-minimum}) and $C_{\log} \le \frac{1}{2}\CBHS$; by Lemma~\ref{lem:emin-lower}, $C_{\log} \ge \frac{1}{2}\CBHS - \Delta_{\log}$. Subtracting this from~\eqref{eq:main-energy-bound} gives
\[
  \lim_{N\to\infty} \frac{E(X_N) - E_{\min}(N)}{N} \;=\; \tfrac{1}{2}\CBHS - C_{\log} \;\in\; [0, \Delta_{\log}],
\]
which vanishes if and only if the Brauchart--Hardin--Saff conjecture $C_{\log} = \frac{1}{2}\CBHS$ holds.
  \item Subtract the assumed expansion $E_{\min}(N) = \EBHS(N) + \mathcal{O}(\log N)$ from~\eqref{eq:main-energy-bound}.
\end{enumerate}
\end{proof}

By Lemma~\ref{lem:emin-lower} and Table~\ref{tab:exact-1e9}, $E(X_N) - E_{\min}(N) < 0.00298\,N$ at $N = 10^3$ and $< 0.00065\,N$ at $N = 10^9$. The two terms of the bound are equal near $N \approx 7 \times 10^4$; beyond that, the bound is limited by the current knowledge of $C_{\log}$ rather than by the construction. In contrast, the upper bound $C_{\log} \le \frac{1}{2}\CBHS$ of B\'etermin and Sandier implies that any construction with $(E(X_N) - \EBHS(N))/N \to \kappa > 0$ satisfies $\liminf_{N\to\infty} (E(X_N) - E_{\min}(N))/N \ge \kappa$; every construction in Table~\ref{tab:prior} has measured $\kappa \ge 0.00198 > \Delta_{\log}$. Corollary~\ref{cor:emin}(iii) shows that \texttt{Terraced}\-\texttt{Zonal} is not within $\mathcal{O}(\log N)$ of the optimum unless $E_{\min}(N)$ itself exceeds $\EBHS(N)$ by $C_1 N^{2/3} + \mathcal{O}(\log N)$. At small $N$, the putative global optima of Lai, Hao, Xiao, and Fu~\cite{Lai2024} give the explicit lower bound $E(X_N) - E_{\min}(N) \ge E(X_N) - E_{\mathrm{Lai}}(N)$, which lies between $0.13$ and $0.71$ for $N = 100, 150, \dots, 500$.

\section{Additional Experiments and Tables}
\label{app:tables}

Lemma~\ref{lem:cyclotomic} with the banded truncation of Remark~\ref{rem:cost} agrees with the brute-force pairwise sum~\eqref{eq:log-energy} to within $4 \times 10^{-13}$ for $N \le 10^4$ (e.g.\ $9.1 \times 10^{-15}$ at $N = 349$ and $3.6 \times 10^{-13}$ at $N = 10^4$), and with the $\mathcal{O}(N\log N)$ Ewald estimator~\cite{Ewald1921} to within $5.9 \times 10^{-6}$ per point.

\subsection{Stress Test at Adversarially Chosen \texorpdfstring{$N$}{N}}
\label{app:stress}
To verify that the $\Theta(N^{2/3})$ scaling holds uniformly across arbitrary integers $N$ without arithmetic or rounding resonances, we evaluated the brute-force $\mathcal{O}(N^2)$ exact pairwise energy at $20$ values $N \in [349, 84{,}410]$, chosen adversarially to target five failure modes:
\begin{enumerate}[label=(\arabic*)]
  \item \emph{Half-integer $K$ (HK)}: $|(\pi - 2\theta_{\mathrm{cap}})/h_0 - (K + 0.5)| \le 1.2\times 10^{-3}$ (maximal global meridional strain);
  \item \emph{Twin primes, even $K$ (TP)}: $(N-2, N)$ twin primes with even $K$ (forced hemisphere asymmetry and coprime ring sums);
  \item \emph{Maximal Viterbi discrepancy (VD)}: largest point-count residual $|\mathrm{diff}_0|$ left by the dynamic program (Section~\ref{sec:stage1}, step~4), up to $53$ points;
  \item \emph{Maximal surrogate strain (SS)}: local maxima of normalized surrogate strain $(\Psi_{\mathrm{stretch}} + \Psi_{\mathrm{seam}})/N^{2/3}$;
  \item \emph{Maximal hemisphere asymmetry (HA)}: worst-case north--south asymmetry $\sum_k |m_k - m_{K-1-k}|$ up to $64$ points.
\end{enumerate}
As shown in Figure~\ref{fig:adversarial} (and itemized target-by-target in Table~\ref{tab:adversarial} in Appendix~\ref{app:tables}), the exact normalized excess $(E(X_N)-\EBHS(N))/N^{2/3}$ remains tightly bounded in $[0.0223, 0.0269]$ across all $20$ adversarial targets ($0.0264 \pm 0.0002$ for $N \ge 10^4$), and \texttt{Terraced}\-\texttt{Zonal} has lower energy than \texttt{ZonalEQ}~\cite{Leopardi2006} and \texttt{GenSpiral}~\cite{RSZ1994,Bauer2000} at all $20$ targets.

\begin{figure}[htbp]
  \centering
  \includegraphics[width=\textwidth]{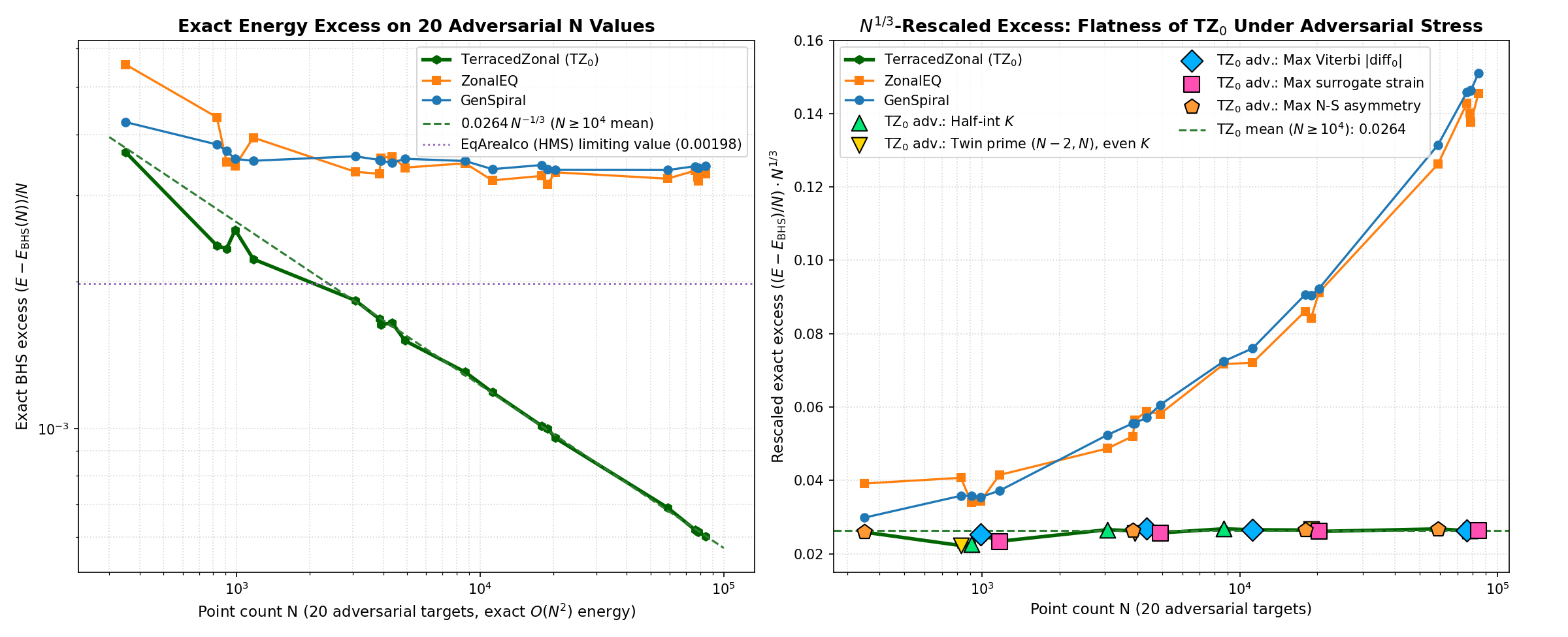}
  \caption{Exact $\mathcal{O}(N^2)$ energy verification at $20$ adversarially chosen values $N \in [349, 84{,}410]$. Left: per-particle BHS excess $(E - \EBHS)/N$. Right: normalized excess $(E - \EBHS)/N^{2/3}$, showing flatness of \TZ{0} around $C_1 \approx 0.026$.}
  \Description{Two panels against N on a log axis from about 300 to 85,000. Left, log-log: the per-point excess of ZonalEQ and GenSpiral stays roughly flat near 0.003 to 0.006, while TerracedZonal falls along a straight dashed line of slope minus one third, crossing below the dotted EqAreaIco reference level at 0.00198 near N equals 2,500. Right: the same excesses multiplied by N to the one third; ZonalEQ and GenSpiral rise from about 0.03 to 0.15, while TerracedZonal stays flat near 0.026, with the five adversarial categories marked by distinct symbols all on the flat line.}
  \label{fig:adversarial}
\end{figure}

\subsection{Test of the Locality Assumption}
\label{app:locality-test}
Locality~(A) neglects three corrections: the drift of $\beta$ across the few rings that interact, the polar caps, where rings are small and terraces short, and the exponentially small interactions across seams. For the unrelaxed configurations of Theorem~\ref{thm:energy} we computed $E(X_N)$ exactly with Lemma~\ref{lem:cyclotomic} and compared $E(X_N) - \EBHS(N)$ with three models: locality~(A) itself; the finer model given by Lemma~\ref{lem:kronecker-full}(i) plus the exact in-terrace lattice sums~\eqref{eq:epstein-terrace} at the local aspect ratios $\beta_k$, with every pair that straddles a seam removed ($c_{\mathrm{seam}}$); and the same model with only the nearest-neighbor pair removed at each seam ($c_{\mathrm{step}}$). Table~\ref{tab:locality} reports the residuals relative to the excess. The residual of locality~(A) halves with each decade of $N$, consistent with $o(N^{2/3})$, and that of the finer $c_{\mathrm{seam}}$ model is $0.04\%$ at $N = 10^8$, while the $c_{\mathrm{step}}$ model drifts to $-0.57\%$, the value $-\frac{2}{3}(1 - c_{\mathrm{seam}}/c_{\mathrm{step}})$ predicted by Theorem~\ref{thm:energy}. The last column compares the terrace lengths chosen by Stage~1 with $L_*(\theta)$ of~\eqref{eq:L-star} (median ratio over terraces away from the poles and the equator).

\begin{table}[!htbp]
  \centering
  \caption{Residuals of three models of $E(X_N) - \EBHS(N)$ for unrelaxed \texttt{Terraced}\-\texttt{Zonal}, as a fraction of the excess, and the median ratio of Stage~1 terrace lengths to $L_*(\theta)$.}
  \label{tab:locality}
  \small
  \begin{tabular}{@{}rrrrrr@{}}
    \toprule
    $N$ & $\frac{E - \EBHS}{N^{2/3}}$ & locality~(A) & lattice, $c_{\mathrm{seam}}$ & lattice, $c_{\mathrm{step}}$ & $L/L_*$ \\
    \midrule
    $10^3$ & $0.024586$ & $17.5\%$ & $8.14\%$ & $8.89\%$ & $1.03$ \\
    $10^4$ & $0.027034$ & $12.1\%$ & $5.23\%$ & $4.79\%$ & $0.99$ \\
    $10^5$ & $0.026970$ & $7.2\%$ & $1.88\%$ & $1.23\%$ & $1.03$ \\
    $10^6$ & $0.026606$ & $3.9\%$ & $0.61\%$ & $-0.05\%$ & $1.00$ \\
    $10^7$ & $0.026339$ & $2.0\%$ & $0.18\%$ & $-0.45\%$ & $1.00$ \\
    $10^8$ & $0.026213$ & $1.0\%$ & $0.04\%$ & $-0.57\%$ & $1.00$ \\
    \bottomrule
  \end{tabular}
\end{table}

\subsection{Tables}

This appendix provides the complete numerical tables referenced in Sections~\ref{sec:intro}, \ref{sec:proof-complexity}, and~\ref{sec:verification}: measured energy excess of prior constructions (Table~\ref{tab:prior}), empirical runtime and problem dimensions from $N = 10^3$ to $N = 10^8$ (Table~\ref{tab:runtime}), exact energy excess computed with Lemma~\ref{lem:cyclotomic} from $N = 10^3$ to $N = 10^9$ (Table~\ref{tab:exact-1e9}), and exact $\mathcal{O}(N^2)$ pairwise energy excess at the $20$ adversarially chosen values $N \in [349, 84{,}410]$ (Table~\ref{tab:adversarial}). In Figure~\ref{fig:all-constructors}, the values of the published constructions for $N \ge 10^3$ are exact sums up to $N \approx 10^5$ and Ewald estimates beyond (for random sets, the analytic expectation). For \texttt{Diamond} and \texttt{Pearl} we report the exact expected energy over the random rotations~\cite[Thm.~2.6]{BeltranEtayo2020}, taking at each $N$ the best of the authors' parameter choices (for \texttt{Diamond}, their published point sets for $N \le 500$ and the ensembles of~\cite{BeltranEtayo2020,BeltranEtayoLopez2023}); for \texttt{EllipticZeros} and \texttt{SphericalEnsemble} we plot the exact expected energies of~\cite{ArmentanoBeltranShub2011} and~\cite{AlishahiZamani2015}. Values for \texttt{Terraced}\-\texttt{Zonal} are exact throughout (Lemma~\ref{lem:cyclotomic}), and those for \texttt{TZ-R} are exact sums. \texttt{TZ-R} starts from terraced configurations produced by Stage~1 of \texttt{Terraced}\-\texttt{Zonal} and simple variants of it (other axial symmetries and seam penalties), and minimizes $E$ numerically with a quasi-Newton method on the sphere (Riemannian L-BFGS) combined with local moves that rearrange lattice defects. We include it only to show how close such an optimization, started from our construction, comes to the putative optima. It is not our contribution: every point interacts with every other, so each evaluation of the gradient of $E$ costs $\Theta(N^2)$ when done directly, an optimizer needs many such evaluations, and it gives no guarantee on the energy it reaches. Our focus is the opposite regime: an explicit configuration, built in a single $\Theta(N)$ pass, whose excess we can analyze. In the reference implementation, the four $K \times W$ Stage~1 arrays (candidate populations, node costs, DP values, and back-pointers) use $28$ bytes per entry, i.e.\ $\approx 34.7\,N^{5/6}$ bytes ($\approx 161$~MB at $N = 10^8$), and the $(N, 3)$ float64 output array uses $24\,N$ bytes.

\begin{table}[!htbp]
  \centering
  \caption{Energy excess per point of previously studied constructions and of \texttt{Terraced}\-\texttt{Zonal}: approximate limiting value of $(E(X_N) - \EBHS(N))/N$. $^{\ast}$Random construction; the value is the expected excess. For \texttt{Diamond} and \texttt{Pearl} the only randomness is an independent rotation of each parallel. For comparison, the excess per point of \texttt{Terraced}\-\texttt{Zonal} decays as $C_1 N^{-1/3}$ (Theorem~\ref{thm:energy}), e.g.\ to $2.6 \times 10^{-4}$ at $N = 10^6$ and $2.6 \times 10^{-5}$ at $N = 10^9$ (Table~\ref{tab:exact-1e9}).}
  \label{tab:prior}
  \small
  \begin{tabular}{@{}llr@{}}
    \toprule
    Family & Construction & $(E - \EBHS)/N$ \\
    \midrule
    Zonal / spiral & \texttt{ZonalEQ}~\cite{Zhou1995,Leopardi2006} & $0.00326$ \\
     & \texttt{Diamond}$^{\ast}$~\cite{BeltranEtayo2020,BeltranEtayoLopez2023} & $0.00315$ \\
     & \texttt{Pearl}$^{\ast}$~\cite{GariboldiGigante2026} & $0.00315$ \\
     & \texttt{GenSpiral}~\cite{RSZ1994,Bauer2000} & $0.00346$ \\
     & \texttt{FibLattice}~\cite{SwinbankPurser2006,Gonzalez2010,Keinert2015} & $0.00548$ \\
    \addlinespace
    Polyhedral & \texttt{EqAreaIco}~\cite{HardinMichaelsSaff2016} & $0.00198$ \\
     & \texttt{HEALPix}~\cite{Gorski2005} & $0.00791$ \\
     & \texttt{Octahedral}~\cite{HolhosRosca2014} & $0.00908$ \\
     & \texttt{RadialIco}~\cite{Teanby2006,HardinMichaelsSaff2016}, \texttt{CubedSphere}~\cite{Ronchi1996,Nair2005} & $\Theta(N)$ \\
    \addlinespace
    QMC / random & \texttt{Hammersley}~\cite{Hammersley1960,CuiFreeden1997,WongLukHeng1997} & $0.0655$ \\
     & \texttt{EllipticZeros}$^{\ast}$~\cite{ArmentanoBeltranShub2011} & $0.1244$ \\
     & \texttt{SphericalEnsemble}$^{\ast}$~\cite{Krishnapur2009,AlishahiZamani2015} & $0.2301$ \\
     & \texttt{RandomJitteredEQ}$^{\ast}$~\cite{Leopardi2006} & $0.2580$ \\
     & \texttt{Random}$^{\ast}$ (i.i.d.\ uniform)~\cite{Muller1959,BrauchartReznikovSaff2018} & $\sim \frac{1}{4}\ln N$ \\
    \midrule
    Zonal (this paper) & \texttt{Terraced}\-\texttt{Zonal} & $\approx 0.026\,N^{-1/3} \to 0$ \\
    \bottomrule
  \end{tabular}
\end{table}

\begin{table}[!htbp]
  \centering
  \caption{Empirical runtime and problem dimensions of \TZ{0}. $K$, $W$: numbers of rings and of candidate populations per ring; $T$: number of terraces; $\mathrm{diff}_0 = (N-2) - \sum_k m_k$: point-count residual left by the dynamic program; MC: iterations of the loop that repairs it (Section~\ref{sec:stage1}, step~4); Stage~1, Stage~2: time for the corresponding stages of Algorithm~\ref{alg:terraced-zonal}; Total: the full construction, including Stage~3. Timings are median wall-clock times on an x86-64 workstation (Python~3.13, NumPy~2.4).}
  \label{tab:runtime}
  \footnotesize
  \setlength{\tabcolsep}{4pt}
  \begin{tabular}{@{}rrrrrrrrrr@{}}
    \toprule
    $N$ & $K$ & $W$ & $K \cdot W$ & $T$ & $\mathrm{diff}_0$ & MC & Stage~1 & Stage~2 & Total \\
    \midrule
    $10^3$ & 29 & 41 & $1.19 \times 10^3$ & 15 & $0$ & 0 & 0.7 ms & 0.9 ms & \textbf{2.1 ms} \\
    $10^4$ & 94 & 41 & $3.85 \times 10^3$ & 27 & $-12$ & 3 & 3.3 ms & 1.0 ms & \textbf{6.2 ms} \\
    $10^5$ & 300 & 61 & $1.83 \times 10^4$ & 47 & $+30$ & 2 & 8.5 ms & 1.4 ms & \textbf{18.1 ms} \\
    $10^6$ & 951 & 131 & $1.25 \times 10^5$ & 75 & $+51$ & 5 & 31.7 ms & 2.1 ms & \textbf{84.2 ms} \\
    $10^7$ & 3,010 & 281 & $8.46 \times 10^5$ & 117 & $+246$ & 8 & 112.4 ms & 4.4 ms & \textbf{0.70 s} \\
    $10^8$ & 9,522 & 605 & $5.76 \times 10^6$ & 179 & $-368$ & 3 & 0.53 s & 11.3 ms & \textbf{8.47 s} \\
    \bottomrule
  \end{tabular}
\end{table}

\begin{table}[!htbp]
  \centering
  \caption{Energy excess of \TZ{0}, computed with the ring-pair formula of Lemma~\ref{lem:cyclotomic} truncated to nearby ring pairs (Remark~\ref{rem:cost}), which takes $\mathcal{O}(N^{1/2})$ time; summing all $N(N-1)/2$ pairs is infeasible at this scale. $K$: number of rings; $T$: number of terraces; mean $L = K/T$: mean terrace length; mean $|\Delta m|$: mean change in population at a seam.}
  \label{tab:exact-1e9}
  \small
  \setlength{\tabcolsep}{5.0pt}
  \begin{tabular}{@{}rrrrrrr@{}}
    \toprule
    $N$ & $K$ & $T$ & Mean $L = K/T$ & Mean $|\Delta m|$ & $(E - \EBHS)/N$ & $(E - \EBHS)/N^{2/3}$ \\
    \midrule
    $10^3$ & 29 & 15 & 1.93 & 6.29 & $0.002429868$ & $0.024299$ \\
    $10^4$ & 94 & 27 & 3.48 & 11.92 & $0.001232969$ & $0.026564$ \\
    $10^5$ & 300 & 47 & 6.38 & 22.00 & $0.000570114$ & $0.026462$ \\
    $10^6$ & 951 & 75 & 12.68 & 43.92 & $0.000261226$ & $0.026123$ \\
    $10^7$ & 3,010 & 117 & 25.73 & 89.21 & $0.000120149$ & $0.025885$ \\
    $10^8$ & 9,522 & 179 & 53.20 & 184.44 & $0.000055538$ & $0.025778$ \\
    $10^9$ & 30,114 & 271 & 111.12 & 385.24 & $\mathbf{0.000025743}$ & $\mathbf{0.025743}$ \\
    \bottomrule
  \end{tabular}
\end{table}

\begin{table}[!htbp]
  \centering
  \caption{Exact $\mathcal{O}(N^2)$ per-particle BHS excess $(E - \EBHS)/N$ and $N^{2/3}$-normalized excess $(E - \EBHS)/N^{2/3}$ at $20$ adversarially chosen values of $N$. Stress categories (Section~\ref{sec:verification}): HA = maximal hemisphere asymmetry; TP = twin primes with even $K$; HK = half-integer $K$; VD = maximal Viterbi discrepancy; SS = maximal surrogate strain. Columns: $K$, $T$ = numbers of rings and terraces; \texttt{EQ} = \texttt{ZonalEQ}, \texttt{GS} = \texttt{GenSpiral}.}
  \label{tab:adversarial}
  \small
  \begin{tabular}{@{}rrlrrrrrr@{}}
    \toprule
    \# & $N$ & Cat. & $K$ & $T$ & $\TZs{0}/N$ & $\TZs{0}/N^{2/3}$ & \texttt{EQ}$/N$ & \texttt{GS}$/N$ \\
    \midrule
    1 & 349 & HA & 17 & 9 & 0.003682 & 0.0259 & 0.005563 & 0.004239 \\
    2 & 829 & TP & 26 & 13 & 0.002369 & 0.0223 & 0.004335 & 0.003814 \\
    3 & 909 & HK & 27 & 15 & 0.002334 & 0.0226 & 0.003512 & 0.003696 \\
    4 & 986 & VD & 29 & 15 & 0.002549 & 0.0254 & 0.003449 & 0.003562 \\
    5 & 1,170 & SS & 31 & 15 & 0.002222 & 0.0234 & 0.003939 & 0.003532 \\
    6 & 3,064 & HK & 52 & 21 & 0.001830 & 0.0266 & 0.003354 & 0.003607 \\
    7 & 3,851 & HA & 58 & 21 & 0.001677 & 0.0263 & 0.003321 & 0.003546 \\
    8 & 3,919 & TP & 58 & 21 & 0.001630 & 0.0257 & 0.003581 & 0.003529 \\
    9 & 4,351 & VD & 62 & 23 & 0.001646 & 0.0269 & 0.003603 & 0.003498 \\
    10 & 4,907 & SS & 65 & 23 & 0.001513 & 0.0257 & 0.003420 & 0.003565 \\
    11 & 8,678 & HK & 88 & 27 & 0.001305 & 0.0268 & 0.003490 & 0.003528 \\
    12 & 11,220 & VD & 100 & 29 & 0.001186 & 0.0266 & 0.003220 & 0.003395 \\
    13 & 17,928 & HA & 126 & 31 & 0.001012 & 0.0265 & 0.003289 & 0.003462 \\
    14 & 18,913 & TP & 130 & 33 & 0.001001 & 0.0267 & 0.003161 & 0.003394 \\
    15 & 20,305 & SS & 134 & 33 & 0.000957 & 0.0261 & 0.003345 & 0.003383 \\
    16 & 58,758 & HA & 230 & 41 & 0.000689 & 0.0268 & 0.003246 & 0.003381 \\
    17 & 76,097 & VD & 261 & 45 & 0.000621 & 0.0263 & 0.003370 & 0.003441 \\
    18 & 78,438 & HK & 265 & 45 & 0.000614 & 0.0263 & 0.003273 & 0.003417 \\
    19 & 78,889 & TP & 266 & 45 & 0.000614 & 0.0263 & 0.003209 & 0.003412 \\
    20 & 84,410 & SS & 275 & 45 & 0.000601 & 0.0264 & 0.003318 & 0.003443 \\
    \addlinespace
    \multicolumn{5}{@{}l}{\textbf{Mean ($N \ge 10^4$, 9 targets)}} & 0.000810 & 0.0264 & 0.003270 & 0.003414 \\
    \bottomrule
  \end{tabular}

\end{table}

\section{Single-Axis Coaxial Rings vs.\ Multi-Axis Orthogonal Belts}
\label{app:multi-axis}

\begin{remark}[Single-Axis Rings vs.\ Multi-Axis Belts]
\label{rem:multi-axis}
The per-particle excess grows toward the poles, as $|\cot\theta|^{2/3}$, so one might instead cover $\Sbb$ with three orthogonal equatorial belts. This lowers the continuum $N^{2/3}$ constant by $60\%$ inside the belts, but the three ring families meet along non-coaxial seams where they cannot be phase-locked, and the seam cost more than cancels the gain: the total excess is $9$--$27\times$ that of \texttt{Terraced}\-\texttt{Zonal} for $N = 10^3$ to $3 \times 10^5$. A single coaxial family of rings is therefore essential. The rest of this appendix gives the details.
\end{remark}

\begin{figure}[!htbp]
  \centering
  \includegraphics[width=0.92\textwidth]{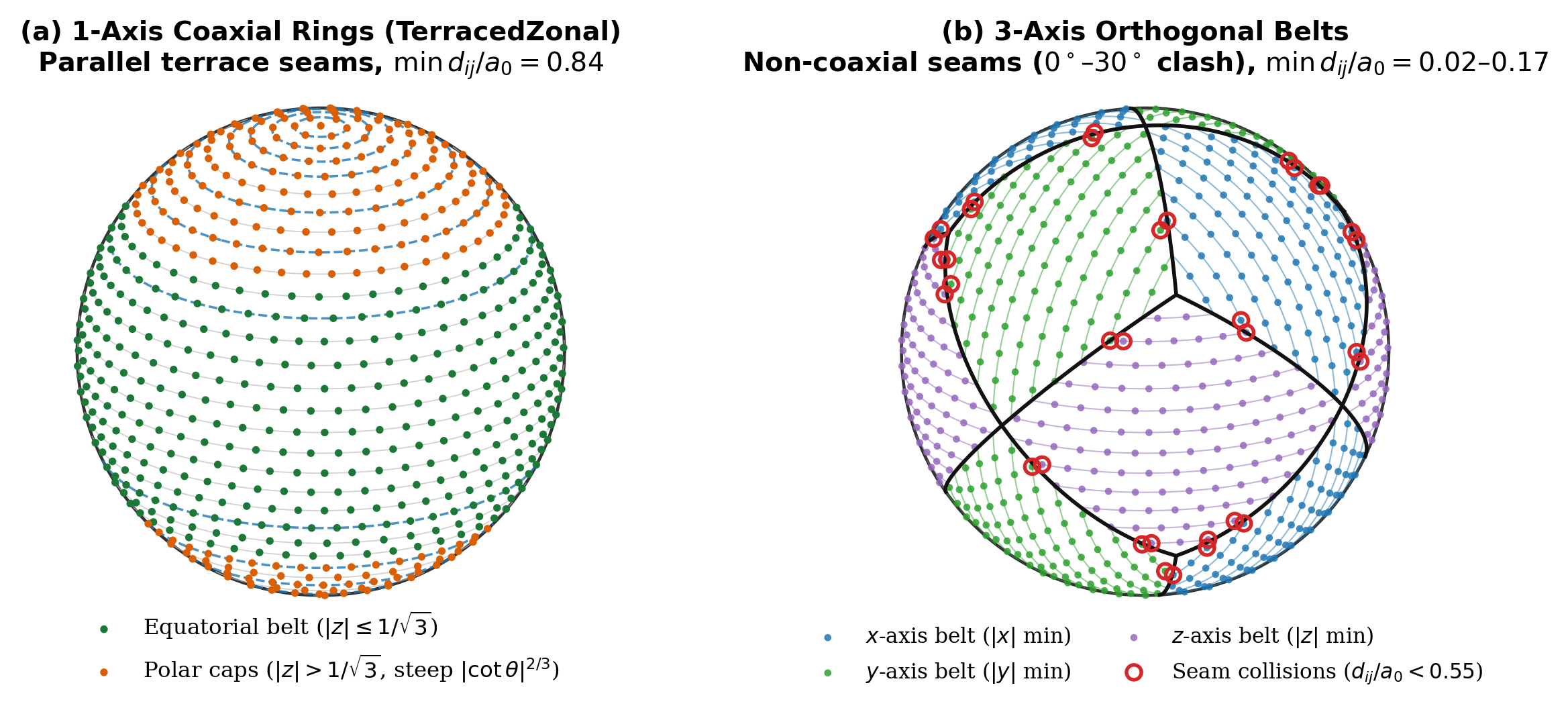}
  \caption{Seam geometry of single-axis and three-axis ring constructions. \textbf{(a)} \texttt{Terraced}\-\texttt{Zonal}: one coaxial family of latitude rings, whose terrace seams (dashed) are parallel circles; the polar caps $|z| > 1/\sqrt{3}$ (orange) are where the per-particle excess $\propto |\cot\theta|^{2/3}$ is largest. \textbf{(b)} Three orthogonal equatorial belts, each point assigned to the axis $a$ minimizing $|x_a|$: the three ring families meet along $12$ non-coaxial great-circle arcs (black), and points closer than $0.55\,a_0$ across the seams are circled in red.}
  \Description{Two spheres side by side. Left: points on horizontal rings, green in an equatorial belt and orange in the two polar caps, with dashed blue seam circles that are all parallel. Right: points in three colors, one per belt around the x, y, and z axes, each belt's rings running in a different direction; black curves mark where the belts meet, and red circles mark pairs of points that nearly collide along these curves.}
  \label{fig:multi-axis}
\end{figure}

Dividing the per-ring excess $f_\lambda(\theta)$ from the proof of Theorem~\ref{thm:energy} by the ring population $y(\theta) = \frac{1}{2}C_0 N^{1/2}\sin\theta$ shows that the per-particle excess density scales as $f_\lambda(\theta)/y(\theta) \propto |\cot\theta|^{2/3}$: near the equator ($\theta \approx \pi/2$), ring radii vary slowly so terraces are wide and cheap, whereas near the poles ($\theta \to 0,\pi$) ring radii shrink rapidly and terraces narrow (the polar caps $|z| \ge 0.9$, covering $10\%$ of $\Sbb$, carry $29\%$ of the $N^{2/3}$ excess). A natural idea is to avoid high latitudes altogether by covering $\Sbb$ with \emph{three orthogonal equatorial belts}---assigning each $\vx = (x, y, z) \in \Sbb$ to the coordinate axis $a \in \{x,y,z\}$ with smallest $|x_a| = \min(|x|,|y|,|z|) \le 1/\sqrt{3}$---and populating each belt with its own low-latitude terraced rings (Figure~\ref{fig:multi-axis}(b)). Inside the belts, restricting $|\cos\theta_a| \le 1/\sqrt{3}$ lowers the continuum $N^{2/3}$ constant by $60.1\%$ (from $C(c_{\mathrm{step}}) = 0.02638$ to $0.01052$; empirically $-66\%$ in the patch interiors at $N = 10^6$).

For a construction without a final global relaxation of the point positions, however, a single coaxial family of rings is essential. In \texttt{Terraced}\-\texttt{Zonal} (Figure~\ref{fig:multi-axis}(a)), all $K$ rings share one axis: every terrace seam is a parallel circle along which neighboring rings are staggered, well-separated ($\min d_{ij}/a_0 = 0.84$), and relaxed in $\mathcal{O}(N^{1/2})$ time by the 1D map~\eqref{eq:z-fixed-point}, while preserving the exact cyclotomic evaluation of Lemma~\ref{lem:cyclotomic}. In the 3-axis splice (Figure~\ref{fig:multi-axis}(b)), the three perpendicular ring families meet along $12$ great-circle arcs ($|x_a| = |x_b| \le |x_c|$, total length $L_{\mathrm{seam}} = 24\arcsin(\sqrt{2/3}) \approx 22.93$, or $6.02\,N^{1/2}$ lattice spacings) at lattice misorientations up to $30^\circ$. Because perpendicular rings cannot be phase-locked or relaxed across a shared seam by 1D ring shifts, points collide across the seams ($\min d_{ij}/a_0 \in [0.02, 0.17]$, red circles) and the unscreened seam penalty ($\gamma_{\mathrm{raw}} \in [0.14, 0.22]$ per boundary site, $>30\times c_{\mathrm{step}}$) overwhelms the interior savings, yielding $9.3\text{--}27.0\times$ higher total energy excess than single-axis \texttt{Terraced}\-\texttt{Zonal} ($9.4\text{--}16.1\times$ higher even after optimizing the three global axial rotation phases; Table~\ref{tab:multi-axis}).

\begin{table}[!htbp]
  \centering
  \caption{Single-axis \texttt{Terraced}\-\texttt{Zonal} vs.\ three orthogonal belts. The belts lower the continuum $N^{2/3}$ constant and the excess inside each patch, but their seams cost more than $30\times c_{\mathrm{step}}$ per boundary site, so the total excess $(E - \EBHS)/N$ is $9$--$27\times$ larger.}
  \label{tab:multi-axis}
  \footnotesize
  \setlength{\tabcolsep}{6pt}
  \begin{tabular}{@{}lcc@{}}
    \toprule
    Property / Metric & \textbf{1-Axis Coaxial (\texttt{Terraced}\-\texttt{Zonal})} & \textbf{3-Axis Orthogonal Belts} \\
    \midrule
    Axial latitude range $|\cos\theta_a|$ & $[0,\, 1]$ (pole to pole) & $[0,\, 1/\sqrt{3}]$ (equatorial belts only) \\
    Continuum bulk constant $C(c_{\mathrm{step}})$ & $0.02638$ ($1.00\times$) & $0.01052$ ($0.40\times$, $-60.1\%$) \\
    Patch-interior excess ($N = 10^6$) & $1.00\times$ & $0.34\times$ ($-66\%$) \\
    \addlinespace
    Seam geometry & Coaxial parallel circles & $12$ non-coaxial arcs ($0^\circ$--$30^\circ$ clash) \\
    Seam cost per boundary site & $c_{\mathrm{step}} \approx 0.0043$ & $\gamma_{\mathrm{raw}} \in [0.14,\, 0.22]$ ($>30\times c_{\mathrm{step}}$) \\
    Min.\ pair separation $\min d_{ij} / a_0$ & $0.84$ (well-separated) & $0.020$--$0.166$ (cross-seam collisions) \\
    Exact $\mathcal{O}(N)$ cyclotomic formula & Yes (Lemma~\ref{lem:cyclotomic}) & No (breaks coaxial symmetry) \\
    \addlinespace
    Exact $(E - \EBHS)/N$, $N = 10^3$ & $\mathbf{0.00243}$ & $0.06550$ ($27.0\times$ worse) \\
    Exact $(E - \EBHS)/N$, $N = 10^4$ & $\mathbf{0.00123}$ & $0.02666$ ($21.6\times$ worse) \\
    Exact $(E - \EBHS)/N$, $N = 3\times 10^4$ & $\mathbf{0.00086}$ & $0.01031$ ($12.0\times$ worse) \\
    Exact $(E - \EBHS)/N$, $N = 10^5$ & $\mathbf{0.00057}$ & $0.00530$ ($9.3\times$ worse) \\
    Exact $(E - \EBHS)/N$, $N = 3\times 10^5$ & $\mathbf{0.00039}$ & $0.00377$ ($9.6\times$ worse) \\
    \bottomrule
  \end{tabular}
\end{table}

\clearpage
\section{Spherical Integration Details}
\label{app:qmc}

\subsection{Minimum-Norm Gauss--Newton Legendre Projection}
Given Stage~1's ring populations $m_0,\dots,m_{K-1}$ ($\sum_{k=0}^{K-1} m_k = N - 2$) and Stage~2's relaxed ring heights $z^{(0)} \in (-1, 1)^K$ from~\eqref{eq:z-fixed-point}, line~7 of Algorithm~\ref{alg:terraced-zonal} solves the underdetermined moment system~\eqref{eq:zonal-moment-constraint} of order $Q' = \min(K - 1, Q)$ ($Q = 96$ by default) via up to $6$ damped minimum-norm Gauss--Newton iterations. At each iteration $t = 0,\dots,5$:
\begin{enumerate}[label=\arabic*.]
  \item Evaluate $P_0(z_k) = 1$, $P_1(z_k) = z_k$, and $(\ell + 1)P_{\ell+1}(z_k) = (2\ell + 1)z_k P_\ell(z_k) - \ell P_{\ell-1}(z_k)$ for $1 \le \ell < Q'$, together with the residual $r_\ell(z^{(t)}) = \frac{1}{N}(1 + (-1)^\ell) + \sum_{k=0}^{K-1} \frac{m_k}{N} P_\ell(z_k^{(t)})$ ($1 \le \ell \le Q'$). Stop if $\|r(z^{(t)})\|_\infty < 10^{-14}$.
  \item Form the $Q' \times K$ Jacobian $J_{\ell,k} = \frac{m_k}{N} P'_\ell(z_k^{(t)})$ via $P'_\ell(z) = \ell(z P_\ell(z) - P_{\ell-1}(z))/\min(-10^{-12}, z^2 - 1)$, and compute the minimum-norm step
  \begin{equation}
    \label{eq:gn-step}
    \Delta z \;=\; -J^\top \bigl(J J^\top + 10^{-14} I_{Q'}\bigr)^{-1} r\bigl(z^{(t)}\bigr).
  \end{equation}
  \item Rescale $\Delta z \leftarrow \Delta z \cdot \min\bigl(1,\, \frac{0.2/K}{\|\Delta z\|_\infty}\bigr)$ and clip $z^{\mathrm{cand}} = \operatorname{clip}(z^{(t)} + \Delta z, -1 + 10^{-9}, 1 - 10^{-9})$. If $z_0^{\mathrm{cand}} > z_1^{\mathrm{cand}} > \dots > z_{K-1}^{\mathrm{cand}}$, accept $z^{(t+1)} \leftarrow z^{\mathrm{cand}}$; otherwise terminate and return $z^{(t)}$.
\end{enumerate}
\par\smallskip\noindent
\begin{minipage}[c]{0.53\textwidth}
\paragraph{Choice of $Q = 96$.}
Figure~\ref{fig:qmc-sweep} sweeps the moment order $Q \in [0, 96]$ at $N = 10{,}242$ ($K = 95$ rings). Because Stage~2's heights $z^{\mathrm{zonal}}$ sit at a stationary point of $\Ezonal$ ($\nabla_z \Ezonal = 0$), projecting $z$ onto the first $Q$ Legendre moments perturbs the energy only at second order while $Q' = \min(K - 1, Q)$ leaves slack in the $K$ ring heights: $(E(X_N) - \EBHS(N))/N^{2/3}$ moves only from $0.02656$ ($Q = 0$) to $0.02685$ ($4 \le Q \le 92$), and ticks up to $0.02867$ only at $Q = K - 1 = 94$, where $Q' = 94$ pins all $K - 1$ interior heights. Meanwhile, matching $Q = 2$ moments removes the $\Theta(N^{-3/2})$ quadratic bottleneck ($110\times$ error drop on $f_{\mathrm{Franke}}$), and by $Q \approx 24$ the error saturates $3{,}800\times$ below \TZ{0}, where 1D Legendre truncation reaches machine precision and residual error is set by azimuthal ring aliasing. We fix $Q = 96$ to cover $Q \ge 24$ with a factor-of-$4$ margin while keeping $Q \ll K$ for $N > 10^4$. Table~\ref{tab:qmc-energy} compares \TZ{96} against \TZ{0} across $N = 10^3$ to $10^6$.
\end{minipage}\hfill
\begin{minipage}[c]{0.44\textwidth}
  \centering
  \setlength{\abovecaptionskip}{3pt}\setlength{\belowcaptionskip}{0pt}%
  \includegraphics[width=\linewidth]{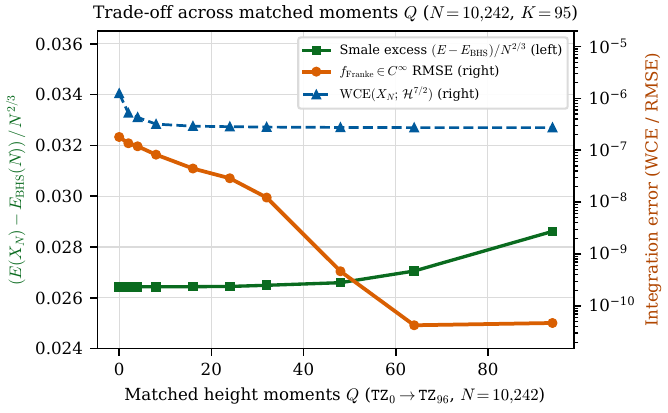}
  \captionof{figure}{Effect of matched height moments $Q$ at $N = 10{,}242$ ($K = 95$). Integration error on $f_{\mathrm{Franke}} \in C^\infty$ (orange, right axis) drops $110\times$ at $Q = 2$ and saturates $3{,}800\times$ below \TZ{0} by $Q \approx 24$, while $(E(X_N) - \EBHS(N))/N^{2/3}$ (green, left axis) stays flat at $\approx 0.0268$ until $Q = K - 1 = 94$.}
  \label{fig:qmc-sweep}
  \Description{Line plot against matched height moments Q from 0 to 96 at N = 10,242 (K = 95). Left axis (green squares): scaled energy excess (E - E_BHS)/N^{2/3} stays nearly flat around 0.0266 to 0.0269 for Q from 0 to 92 and rises slightly to 0.0287 at Q = 94. Right axis (log scale): integration error on f_Franke (orange circles) drops from 1.8 times 10^{-7} at Q = 0 to 1.6 times 10^{-9} at Q = 2 and plateaus at 4.7 times 10^{-11} for Q >= 24; worst-case error WCE in H^{7/2} (purple triangles) drops at Q = 2 and stays flat.}
\end{minipage}
\par\smallskip

\paragraph{Height shifts and energy.}
The height shifts produced by line~7 are tiny: $|\delta\theta_k|/h_0 < 2\%$ of the nominal ring spacing already at $N = 600$ (Figure~\ref{fig:qmc}(a)), falling to a median shift of $0.05\%$ (maximum $0.7\%$) at $N = 10^5$. Evaluated exactly via Lemma~\ref{lem:cyclotomic} (Table~\ref{tab:qmc-energy}), $(E(X_N) - \EBHS(N))/N^{2/3}$ is $0.0287$ at $N = 10^4$, $0.0265$ at $N = 10^5$, and $0.0261$ at $N = 10^6$---matching \texttt{Terraced}\-\texttt{Zonal} to four decimal places at $N = 10^6$ (an unscaled difference $\Delta E = +0.04$ across $10^6$ points) while driving $\max_{1 \le \ell \le Q'} |r_\ell(z)|$ below $4 \times 10^{-15}$.

\begin{table}[!htbp]
  \centering
  \caption{Exact logarithmic energy (Lemma~\ref{lem:cyclotomic}), ring-height shift, residual Legendre moment ($Q' = \min(K-1, 96)$), and runtime of \TZ{96} ($\TZs{96}$) vs.\ \TZ{0} ($\TZs{0}$). At $N = 10^7$, single-threaded Python/NumPy runtimes are $1.02\text{ s}$ ($\TZs{96}$) vs.\ $1.03\text{ s}$ ($\TZs{0}$), equal within timing noise.}
  \label{tab:qmc-energy}
  \footnotesize
  \setlength{\tabcolsep}{4.8pt}
  \begin{tabular}{@{}rrrrrrrrr@{}}
    \toprule
    $N$ & $K$ & $\frac{E(\TZs{0}) - \EBHS}{N^{2/3}}$ & $\frac{E(\TZs{96}) - \EBHS}{N^{2/3}}$ & $\Delta E$ & $K\|\delta z\|_\infty$ & $\max_{\ell \le Q'}|r_\ell(\TZs{0})|$ & $\max_{\ell \le Q'}|r_\ell(\TZs{96})|$ & Time (\TZs{96}) \\
    \midrule
    $10^3$ & $29$ & $0.02430$ & $0.02605$ & $+0.175$ & $0.0434$ & $1.5 \times 10^{-3}$ & $1.4 \times 10^{-16}$ & $5.3\text{ ms}$ \\
    $10^4$ & $94$ & $0.02656$ & $0.02867$ & $+0.976$ & $0.0274$ & $5.7 \times 10^{-4}$ & $3.5 \times 10^{-16}$ & $15.5\text{ ms}$ \\
    $3 \times 10^4$ & $164$ & $0.02680$ & $0.02715$ & $+0.342$ & $0.0107$ & $1.3 \times 10^{-4}$ & $1.2 \times 10^{-16}$ & $18.6\text{ ms}$ \\
    $10^5$ & $300$ & $0.02646$ & $0.02652$ & $+0.134$ & $0.0034$ & $1.7 \times 10^{-5}$ & $3.9 \times 10^{-15}$ & $30.9\text{ ms}$ \\
    $3 \times 10^5$ & $520$ & $0.02625$ & $0.02626$ & $+0.065$ & $0.0017$ & $5.0 \times 10^{-6}$ & $1.1 \times 10^{-16}$ & $69.7\text{ ms}$ \\
    $10^6$ & $951$ & $0.02612$ & $0.02613$ & $+0.038$ & $0.0007$ & $8.5 \times 10^{-7}$ & $7.2 \times 10^{-17}$ & $145.0\text{ ms}$ \\
    \bottomrule
  \end{tabular}
\end{table}

\subsection{Benchmark Definitions and Numerical Values}
In Figure~\ref{fig:qmc} and Table~\ref{tab:qmc-benchmarks}:
\begin{enumerate}[label=(\alph*)]
  \item \emph{Smooth global landscape~\cite{Renka1988}:} for $\vx = (x,y,z) \in \Sbb$,
  \begin{align*}
    f_{\mathrm{Franke}}(x,y,z) \;=\;&
      \tfrac{3}{4} e^{-((9x-2)^2 + (9y-2)^2 + (9z-2)^2)/4}
      + \tfrac{3}{4} e^{-(9x+1)^2/49 - (9y+1)/10 - (9z+1)/10} \\
      &+ \tfrac{1}{2} e^{-((9x-7)^2 + (9y-3)^2 + (9z-5)^2)/4}
      - \tfrac{1}{5} e^{-(9x-4)^2 - (9y-7)^2 - (9z-5)^2},
  \end{align*}
  with reference integral $\bar{f}_{\mathrm{Franke}} \approx 0.5329$, evaluated to $15$-digit precision by $240 \times 480$ Gauss--Legendre product quadrature.
  \item \emph{Localized $C^4$ bump~\cite{Wendland1995}:} $\psi(\vx) = \frac{1}{3}(1 - r)_+^6(35r^2 + 18r + 3)$ with chordal distance $r = \|\vx - \mathbf{u}\|$, exact sphere average $\bar{\psi} = \frac{1}{36}$. For both $f_{\mathrm{Franke}}$ and $\psi$, we report the root-mean-square error $\bigl(\frac{1}{M}\sum_{m=1}^M |I(f \circ R_m; X_N) - \bar{f}|^2\bigr)^{1/2}$ over $M = 500$ independent Haar-uniform rotations $R_m \in \mathrm{SO}(3)$ (cross-checked at $N = 3{,}242$ against the exact infinite-rotation $\mathrm{SO}(3)$ harmonic integral, agreeing within $1\%$). For \texttt{EqAreaIco}~\cite{HardinMichaelsSaff2016} we evaluate at the nearby pure Class-I icosahedral counts $N = 10b^2 + 2 \in \{362, 642, 1{,}002, 1{,}962, 3{,}242, 5{,}762, 10{,}242, 20{,}252\}$ (\texttt{FibLattice}~\cite{SwinbankPurser2006} uses odd counts $2M+1$).
  \item \emph{Worst-case error in $\mathcal{H}^{7/2}(\Sbb)$~\cite[eqs.~(3.15)--(3.18) and Thm.~3.5]{BrauchartSaffSloanWomersley2014}:} equipped with the generalized distance reproducing kernel at smoothness $s = 7/2$, the squared worst-case equal-weight integration error over $\|f\|_{\mathcal{H}^{7/2}(\Sbb)} \le 1$ has the exact closed form
  \begin{equation}
    \label{eq:bssw-wce}
    \mathrm{WCE}\bigl(X_N;\,\mathcal{H}^{7/2}(\Sbb)\bigr)^2
    \;=\; \frac{1}{N^2}\sum_{i=1}^N \sum_{j=1}^N \Bigl(\frac{64}{7} \;+\; \frac{3200}{231}\,P_2(\vx_i \cdot \vx_j) \;-\; \|\vx_i - \vx_j\|^5\Bigr),
  \end{equation}
  where $P_2(t) = \frac{1}{2}(3t^2 - 1)$. (We cross-checked~\eqref{eq:bssw-wce} against the truncated positive spherical-harmonic series of~\cite[eq.~(3.15)]{BrauchartSaffSloanWomersley2014} to eliminate catastrophic cancellation below $10^{-8}$.) Brauchart, Saff, Sloan, and Womersley introduced this measure to test logarithmic-energy point sets above the $s = 3$ barrier; for general $N$-point sets on $\Sbb$ the optimal rate in $\mathcal{H}^s(\Sbb)$ is $\Theta(N^{-s/2})$~\cite{HesseSloan2005}, here $N^{-7/4}$, and by reproducing-kernel duality~\cite[Thm.~3.5]{BrauchartSaffSloanWomersley2014} it requires no function sampling. At $N \approx 2 \times 10^4$, the scaled error $N^{7/4}\,\mathrm{WCE}$ is $17.3$ for \texttt{Terraced}\-\texttt{Zonal}, $17.6$ for \texttt{GenSpiral}, $18.8$ for Class-I \texttt{EqAreaIco}, $51.9$ for \texttt{FibLattice}, $161$ for \texttt{Hammersley}, and $2{,}350$ for \texttt{ZonalEQ}, whereas \TZ{96} stays in $[2.82, 3.06]$ across the whole range of $N$.
\end{enumerate}

\begin{table}[!htbp]
  \centering
  \caption{Equal-weight ($1/N$) spherical integration benchmarks at representative point counts $N$ (for \texttt{EqAreaIco} at $N \approx 10^3$, $N = 1{,}002$; for \texttt{FibLattice}, $N - 1$).}
  \label{tab:qmc-benchmarks}
  \footnotesize
  \setlength{\tabcolsep}{3.7pt}
  \begin{tabular}{@{}llrrrrrrr@{}}
    \toprule
    Metric & $N$ & \TZs{96} & \TZs{0} & \texttt{GenSpiral} & \texttt{EqAreaIco} & \texttt{FibLattice} & \texttt{Hammersley} & \texttt{ZonalEQ} \\
    \midrule
    $f_{\mathrm{Franke}}$ ($C^\infty$) & $974$ & $\mathbf{1.07\times 10^{-6}}$ & $1.13\times 10^{-5}$ & $5.95\times 10^{-6}$ & $6.49\times 10^{-6}$ & $1.80\times 10^{-5}$ & $5.95\times 10^{-5}$ & $1.42\times 10^{-4}$ \\
     & $3{,}242$ & $\mathbf{3.87\times 10^{-9}}$ & $1.22\times 10^{-6}$ & $9.69\times 10^{-7}$ & $1.78\times 10^{-6}$ & $2.77\times 10^{-6}$ & $8.60\times 10^{-6}$ & $4.43\times 10^{-5}$ \\
     & $10{,}242$ & $\mathbf{4.68\times 10^{-11}}$ & $1.81\times 10^{-7}$ & $1.68\times 10^{-7}$ & $5.46\times 10^{-7}$ & $5.28\times 10^{-7}$ & $1.51\times 10^{-6}$ & $1.37\times 10^{-5}$ \\
     & $20{,}252$ & $\mathbf{3.17\times 10^{-12}}$ & $6.74\times 10^{-8}$ & $6.28\times 10^{-8}$ & $2.74\times 10^{-7}$ & $1.65\times 10^{-7}$ & $5.78\times 10^{-7}$ & $7.00\times 10^{-6}$ \\
    \addlinespace
    $\psi$ ($C^4$ bump) & $974$ & $\mathbf{3.60\times 10^{-8}}$ & $8.50\times 10^{-6}$ & $2.85\times 10^{-6}$ & $4.54\times 10^{-6}$ & $8.61\times 10^{-6}$ & $2.75\times 10^{-5}$ & $4.44\times 10^{-5}$ \\
     & $3{,}242$ & $\mathbf{4.63\times 10^{-10}}$ & $7.63\times 10^{-7}$ & $4.62\times 10^{-7}$ & $1.30\times 10^{-6}$ & $1.38\times 10^{-6}$ & $3.71\times 10^{-6}$ & $1.38\times 10^{-5}$ \\
     & $10{,}242$ & $\mathbf{7.45\times 10^{-12}}$ & $1.76\times 10^{-7}$ & $8.01\times 10^{-8}$ & $4.04\times 10^{-7}$ & $2.40\times 10^{-7}$ & $6.31\times 10^{-7}$ & $4.30\times 10^{-6}$ \\
     & $20{,}252$ & $\mathbf{8.89\times 10^{-13}}$ & $5.75\times 10^{-8}$ & $2.88\times 10^{-8}$ & $2.03\times 10^{-7}$ & $9.43\times 10^{-8}$ & $2.50\times 10^{-7}$ & $2.20\times 10^{-6}$ \\
    \addlinespace
    $N^{7/4}\mathrm{WCE}$ ($\mathcal{H}^{7/2}$) & $974$ & $\mathbf{2.97}$ & $16.23$ & $8.73$ & $3.67$ & $23.70$ & $79.60$ & $235.9$ \\
     & $3{,}242$ & $\mathbf{2.89}$ & $14.67$ & $11.59$ & $5.67$ & $32.75$ & $95.27$ & $602.2$ \\
     & $10{,}242$ & $\mathbf{2.86}$ & $13.00$ & $14.90$ & $11.60$ & $42.47$ & $125.2$ & $1{,}396$ \\
     & $20{,}252$ & $\mathbf{2.82}$ & $17.32$ & $17.58$ & $18.84$ & $51.93$ & $161.4$ & $2{,}350$ \\
    \bottomrule
  \end{tabular}
\end{table}

\clearpage
\section{Glossary}
\label{app:glossary}

\begin{center}
  \small
  \renewcommand{\arraystretch}{1.06}
  \begin{tabular}{@{}>{\raggedright\arraybackslash}p{0.27\textwidth}>{\raggedright\arraybackslash}p{0.55\textwidth}>{\raggedright\arraybackslash}p{0.12\textwidth}@{}}
    \toprule
    Term, symbol & Meaning & Defined \\
    \midrule
    ring $k$; $K$; population $m_k$ & a latitude circle carrying $m_k$ equally spaced points at height $z_k = \cos\theta_k$ (colatitude $\theta_k$) and phase $\alpha_k$; the $K$ rings are numbered $k = 0,\dots,K-1$ from north to south & \S\ref{sec:terraces} \\
    ring configuration & the two poles plus $K$ coaxial latitude rings, ring $k$ carrying $m_k$ equally spaced points at height $z_k$ and phase $\alpha_k$; every output of \texttt{Terraced}\-\texttt{Zonal} is one & \S\ref{sec:cyclotomic} \\
    ideal population $y_k$; $C_0$ & the real-valued, equal-area population of ring $k$, $y_k \approx \frac{1}{2}C_0N^{1/2}\sin\theta_k$, with $C_0 = \sqrt{2\pi\sqrt{3}}$ & \eqref{eq:y-ideal} \\
    terrace $\mathcal{T}_t$; $T$ & a maximal run of consecutive rings with the same population; the $T$ terraces are numbered $t = 0,\dots,T-1$ from north to south & \eqref{eq:T-def} \\
    $M_t$, $L_t$ & the common population of the rings of terrace $t$, and their number; lowercase symbols ($m_k$, $y_k$) refer to rings, uppercase ($M_t$, $L_t$) to terraces & \S\ref{sec:terraces} \\
    optimal terrace length $L_*$ & the number of rings per terrace that balances stretching against seams, $\Theta(N^{1/3})$ & \S\ref{sec:terraces} \\
    seam & the boundary between two terraces, where the population jumps; rings on its two sides cannot interlock & \S\ref{sec:terraces} \\
    $A_2$ lattice; pitch $h_0$ & the equilateral triangular lattice; $h_0 = \sqrt{2\pi\sqrt{3}/N}$ is the ring spacing at which staggered equal rings form it & \S\ref{sec:terraces} \\
    stagger, phase locking & rotating adjacent rings against each other by half a point spacing, $\alpha_k - \alpha_{k-1} = 1/(2\operatorname{lcm}(m_{k-1}, m_k))$ & \eqref{eq:phase-locking} \\
    interlocking; $c_{\mathrm{step}}$ & two equal, staggered adjacent rings bind; $c_{\mathrm{step}} \approx 0.0043241$ is the binding energy per point & \eqref{eq:c-step} \\
    seam cost $c_{\mathrm{seam}}$ & energy per point lost at a seam, from all ring pairs that straddle it, $\approx 0.0042867$ & \eqref{eq:c-seam} \\
    aspect ratio $\beta_k$ & ring spacing relative to point spacing on ring $k$ in conformal coordinates; $\beta_k = \frac{\sqrt{3}}{2}(m_k/y_k)^2$, equilateral at $\beta_0 = \frac{\sqrt{3}}{2}$ & \S\ref{sec:kronecker} \\
    stretching cost $\psi_{\mathrm{stretch}}$, $c_{\mathrm{stretch}}$ & energy of ring $k$ due to $\beta_k \neq \beta_0$, quadratic in $m_k - y_k$ with stiffness $c_{\mathrm{stretch}} = \frac{1}{4}$ & \eqref{eq:stretch-cost} \\
    zonal energy $\Ezonal$ & the energy with every ring smeared into a uniform circle; depends only on populations and heights & Lemma~\ref{lem:cyclotomic} \\
    multipole energy $\Emult$ & the remainder, which depends on how the points of different rings line up; ring pairs $a < b$ couple through $\rho_{ab} = (\zeta_a/\zeta_b)^{\Lambda_{ab}}$, $\Lambda_{ab} = \operatorname{lcm}(m_a, m_b)$ & Lemma~\ref{lem:cyclotomic} \\
    stereographic radius $\zeta_k$ & $\zeta_k = \tan(\theta_k/2)$, the radius of ring $k$ after stereographic projection; $\xi_k = \log\zeta_k$ & Lemma~\ref{lem:cyclotomic} \\
    zonal heights $z^{\mathrm{zonal}}$ & the heights that exactly minimize $\Ezonal$ for given populations & \eqref{eq:z-zonal} \\
    BHS benchmark $\EBHS$; excess & the three-term expansion~\eqref{eq:bhs-expansion} with the conjectured constant $\CBHS$; the excess is $E(X_N) - \EBHS(N)$ & \S\ref{sec:intro} \\
    window $w$ & half-width of the range of candidate populations around $\operatorname{round}(y_k)$ in Stage~1 & \eqref{eq:window} \\
    residual $\mathrm{diff}_0$ & the point-count mismatch $(N-2) - \sum_k m_k$ before the repair loop & \S\ref{sec:stage1} \\
    \TZ{Q}; $Q$ & Algorithm~\ref{alg:terraced-zonal} with $Q \ge 0$ height moments matched ($Q = 0$ for Smale's problem, $Q = 96$ for integration) & \S\ref{sec:algorithm}, \eqref{eq:zonal-moment-constraint} \\
    \texttt{TZ-R} & a numerically optimized variant seeded by \texttt{Terraced}\-\texttt{Zonal}; a reference curve in Figure~\ref{fig:all-constructors}, with no runtime guarantee & App.~\ref{app:tables} \\
    \bottomrule
  \end{tabular}
\end{center}

\clearpage
\phantomsection
\section*{AI Methodology}
\label{sec:ai-methodology}
This work was produced using Gemini~4 Argon, which carried out most of the day-to-day technical, computational, and expository work. Specifically, this model:
\begin{itemize}[nosep,leftmargin=1.5em,topsep=3pt]
  \item explored candidate point constructions on $\Sbb$---including negative directions such as the multi-axis splice of Appendix~\ref{app:multi-axis}---and developed and refined the three-stage \texttt{Terraced}\-\texttt{Zonal} algorithm and its moment-matched integration variant \TZ{Q} (Sections~\ref{sec:algorithm} and~\ref{sec:qmc}, Appendix~\ref{app:algorithm});
  \item worked out the cyclotomic energy decomposition, the Kronecker--Epstein lattice expansion, the continuum variational problem and its closed-form $\Theta(N^{2/3})$ solution, and the discrete rounding and complexity analyses (Sections~\ref{sec:proof-complexity}--\ref{sec:proof-energy}, Appendices~\ref{app:proof-complexity}--\ref{app:proofs-energy});
  \item wrote the Python/NumPy reference implementation, fast cyclotomic evaluators, prior-construction baselines, adversarial stress tests, and spherical integration benchmarks, and generated the figures and numerical tables (Sections~\ref{sec:verification}--\ref{sec:qmc}, Appendices~\ref{app:tables} and~\ref{app:qmc}); and
  \item drafted, restructured, and revised the prose, bibliography, and glossary throughout the manuscript (Sections~\ref{sec:intro}--\ref{sec:conclusion}, Appendices~\ref{app:background}--\ref{app:related} and~\ref{app:glossary}).
\end{itemize}
\vspace{3pt}
\noindent The human authors posed the research problem, set high-level priorities, steered and critiqued the algorithmic and analytical directions, and audited intermediate outputs. In accordance with the STOC~2027 policy, the authors verified the correctness and originality of all content and take full responsibility for the submission.

\bibliographystyle{alphaurl}
\bibliography{references}

\end{document}